\documentclass[letterpaper,twocolumn,aps,rmp,superscriptaddress]{revtex4}
\usepackage{graphicx}
\usepackage{amsmath}
\usepackage{amsfonts}
\usepackage{amssymb}
\usepackage{epstopdf}

\begin{document}

\author{Yaroslav Tserkovnyak}
\affiliation{Lyman Laboratory of Physics, Harvard University, Cambridge,
Massachusetts 02138, USA}
\author{Arne Brataas}
\affiliation{Department of Physics, Norwegian University of
Science and Technology, N-7491 Trondheim, Norway}
\author{Gerrit E. W. Bauer}
\affiliation{Kavli Institute of NanoScience, Delft University of
Technology, 2628 CJ Delft, The Netherlands}
\author{Bertrand I. Halperin}
\affiliation{Lyman~Laboratory of Physics, Harvard University, Cambridge,
Massachusetts 02138, USA}

\title{Nonlocal magnetization dynamics in ferromagnetic heterostructures}

\begin{abstract}
Two complementary effects modify the GHz magnetization dynamics of nanoscale heterostructures of ferromagnetic and normal materials relative to those of the isolated magnetic constituents: On the one hand, a time-dependent ferromagnetic magnetization pumps a spin angular-momentum flow into adjacent materials and, on the other hand, spin angular momentum is transferred between ferromagnets by an applied bias, causing mutual torques on the magnetizations. These phenomena are manifestly nonlocal: they are governed by the entire spin-coherent region that is limited in size by spin-flip relaxation processes. We review recent progress in understanding the magnetization dynamics in ferromagnetic heterostructures from first principles, focusing on the role of spin pumping in layered structures. The main body of the theory is semiclassical and based on a mean-field Stoner or spin-density--functional picture, but quantum-size effects and the role of electron-electron correlations are also discussed. A growing number of experiments support the theoretical predictions. The formalism should be useful to understand the physics and to engineer the characteristics of small devices such as magnetic random-access memory elements.
\end{abstract}

\date{\today}
\maketitle
\tableofcontents
\section{Introduction}
\label{intro}

\subsection{Preliminaries}
\label{pre}

A ferromagnet is a symmetry-broken state in which a majority of electrons point their spin into a certain common direction below critical temperatures as high as 1000~K. The robustness of the magnetic order and the permanence of a given magnetization direction against elevated temperatures and external perturbations have been employed in applications as diverse as compass needles, refrigerator-door stickers, and memory devices.

In spite of its stability, ferromagnetism is neither rigid nor static. Depending on sample size and anisotropies due to crystal field and sample shape, a single-domain ferromagnet is often unstable with respect to a domain structure that lowers the macroscopic magnetic energy. Thermal fluctuations reduce the macroscopic moment until it completely vanishes at the critical temperature $T_c$. At temperatures sufficiently below $T_c$, the internal dynamics of the ferromagnet are dominated by low-energy transverse fluctuations of the magnetization, so-called spin waves or magnons, that are the magnetic equivalence of phonons in a lattice. Classical course-grain computer simulations of the detailed position- and time-dependent magnetization (``micromagnetism") describe these phenomena well \cite{Brown,Miltat}.

When magnetic grains become sufficiently small, the exchange stiffness renders domain structures energetically unfavorable and a single-domain picture is adequate. When the ferromagnet is exposed to a uniform driving field, the macroscopic magnetization dynamics may then be dominated by a collective precession of the entire ferromagnetic order parameter. Restricting the ferromagnetic degrees of freedom to this mode is often referred to as the ``macrospin model." In infinite ferromagnetic media, low-energy spin waves resemble symmetry-restoring Goldstone modes, but in real life, the spin-rotational symmetry is broken by magnetic anisotropies caused by magnetic dipolar fields or by crystal-field spin-orbit interactions. In thermodynamic equilibrium, the macrospin then points in a certain fixed direction with small thermal fluctuations around it. The ferromagnet can still be coerced into motion by applying external magnetic fields at a finite angle to the magnetization direction. The system then moves in response, trying to minimize its Zeeman energy. The compass needle, a freely suspended single-domain ferromagnet with a sufficiently high anisotropy (coercivity), does this by alignment of its lattice. In this review, we are interested in mechanically-fixed magnets whose magnetic moments move in the presence of external and anisotropy effective magnetic fields, as well as applied electric currents. Viscous damping processes are required to achieve a reorientation (switching) of the magnetization, if, for example, a magnetic-field direction is suddenly changed. Minimization of this finite switching time by engineering magnetic anisotropies and magnetization-damping rates is an important goal in the design of fast magnetic memories. When the applied magnetic fields are large enough to surmount the anisotropies, the magnetization can be reversed, often by large amplitude and complex trajectories, even in the simple macrospin model. At finite temperatures, the magnetization reorientation becomes probabilistic and is described by a Fokker-Planck equation on the unit sphere \cite{Brown}.

In the last two decades, a new subdiscipline in the field of magnetism has risen that is devoted to the studies of heterostructures of ferromagnets (F) with normal metals (N) and, to a lesser extent, semiconductors and superconductors. Especially magnetoelectronics, the science and technology directed at understanding and utilizing the transport properties of layered structures of ferromagnetic and normal metals, has grown into a mainstream topic of condensed-matter physics. Its attraction derives from large effects at room temperature that can be understood easily in terms of transparent physics and that have found already numerous applications. Two crucial discoveries in magnetic multilayers still reverberate in recent research, viz., the nonlocal oscillatory exchange coupling by \onlinecite{Grunberg:prl86} and the giant magnetoresistance (GMR) by \onlinecite{Baibich:prl88,Grunberg:prb89}. The exchange coupling through a metallic spacer favors an antiparallel coupling between ferromagnetic layers for certain spacer thicknesses, depending on the occupation of spin-polarized quantum-well states. It is therefore a quantum-interference effect sensitive to defect scattering, which vanishes exponentially with increasing spacer-layer thickness. GMR stands for a significant reduction of the resistance of multilayers when the magnetic configuration is forced from antiparallel to parallel by an applied magnetic field. In disordered multilayers, it is a semiclassical transport effect that can be understood in terms of a diffusion equation \cite{Camley:prl89,Valet:prb93}. In a configuration in which the currents are oriented perpendicular to the interface planes (CPP) \cite{Pratt:prl91,Gijs:prl93,Gijs:ap97}, electrical transport can be mapped on a two-channel (spin-up and spin-down) resistor model, in which interface and bulk resistances for a fixed spin are simply connected in series. The spin-relaxation processes are usually modeled by a local finite-resistance link connecting the spin-up and spin-down circuits.

Initially, the community focused its attention on stationary magnetic states, like those responsible for the magnetoresistance in metallic and tunneling structures with applied dc bias. This has changed drastically in recent years. The main catalyst was the experimental verifications of an earlier prediction by \onlinecite{Sloncz:mmm96} and \onlinecite{Berger:prb96} that electric currents can cause a reorientation of the ferromagnetic order in multilayer structures. \onlinecite{Tsoi:prl98} experimentally demonstrated magnetization precession in (Co$\mid$Cu)$_N$ multilayers by currents injected by a point contact, whereas \onlinecite{Myers:sc99} observed switching in the orientation of magnetic moments in Co$\mid$Cu$\mid$Co sandwich structures by perpendicular electric currents (``CPP spin valves"). Much earlier, a coupling between a dynamic ferromagnetic magnetization and spin accumulation in adjacent normal metals has been postulated by \onlinecite{Janossy:prl76,Silsbee:prb79}. These authors demonstrated that microwave transmission through normal-metal foils facilitated by conduction-electron spin transfer is significantly enhanced by ferromagnetic-layer coating.

This review covers the developments in the understanding of the magnetization dynamics in heterostructures of ferromagnets and normal conductors in the last five years or so. We believe that the time is ripe, since from a microscopic point of view, much of the basic physics is well understood. A consistent and coherent picture has evolved that is based on the diffusion equation for the bulk transport in metallic ferromagnets and normal conductors with quantum-mechanical boundary conditions at possibly sharp interfaces between them. Noncollinearity of the magnetization directions in structures with more than one magnet is an essential ingredient in order to understand the physics. We focus here on explicitly \textit{dynamic} effects, referring to a separate article \cite{Brataas:prp05} for the \textit{static} transport properties of magnetoelectronic circuits and devices. The present review follows a self-consistent approach extending the static magnetoelectronic circuit theory to time-dependent phenomena. It provides a framework to include on equal footing two physical effects that are two sides of one coin, viz., the spin-transfer torque induced by applied currents \cite{Sloncz:mmm96} and the spin pumping by moving ferromagnets into adjacent conductors \cite{Tserkovnyak:prl021}. The theory is derived from microscopic principles and the material-dependent input parameters are thus accessible to \textit{ab initio} calculations. We concentrate on quasi-one-dimensional models corresponding to, e.g., layered pillar structures. With few exceptions, we do not attempt accurate modeling of concrete device structures and deviations of the magnetization dynamics from the macrospin model, although the theory discussed in this review can be readily generalized to treat such situations. We focus on ``adiabatic" effects to lowest order in the characteristic Larmor frequency. In spite of these limitations, the agreement with various experiments is found to be gratifying.

Effects beyond such a model certainly may cause observable phenomena. For example, the quantum interference that leads to inversion of the magnetoresistance in high-quality tunnel junctions cannot be treated by the diffusion equation \cite{Yuasa:sc02}. Nonlinearities require numerical simulations or a stability analysis based on the theory of dynamic systems that are outside our scope \cite{Valet}. High temperatures and currents can best be treated by stochastic methods beyond the present review \cite{Li:prb041,Visscher:cm04}, but the input parameters of such approaches are provided here. The current-induced dynamics of domain walls (\onlinecite{Barnes:prl05,Tatara:prl04,Li:prl04} and references therein) are also beyond the macrospin considerations central to this review. Whereas the spin-transfer-torque--induced dynamics are a crucial ingredient, space constraints force us to abandon complete coverage of the numerous experiments published recently.

Throughout the review, we focus on self-consistent effects arising from the time-dependent ferromagnetic exchange field felt by itinerant carriers in the mean-field picture. We take the spin-orbit interaction into account only in terms of a phenomenological spin-flip relaxation time, Secs.~\ref{sm}-\ref{dec}, but consider it more seriously in Secs.~\ref{sdm} and \ref{mca}. Most results are not material specific, but, unless specified otherwise, we have heterostructures of transition-metal ferromagnets (and its alloys) with noble or other simple normal metals in mind.

The main body of this review is organized as follows: Sec.~\ref{intro} introduces several basic concepts that we rely on in the remainder of the article, which is primarily aimed at nonspecialists. Sec.~\ref{sm} is a brief but in-depth discussion of the magnetoelectronic circuit theory (see also \onlinecite{Brataas:prp05}), which is then generalized in Sec.~\ref{pump} to dynamic problems by means of the spin-pumping concept. Secs.~\ref{gde} and \ref{dec} respectively discuss Gilbert damping and dynamic ferromagnetic exchange in heterostructures, which are mediated by spin pumping and spin-transfer torques. Sec.~\ref{lra} is devoted to an alternative linear-response view of the nonlocal magnetization dynamics, and Sec.~\ref{misc} treats several special topics before we conclude the article with summary and outlook in Sec.~\ref{sum}.

\subsection{Nonlocal exchange coupling and giant magnetoresistance}
\label{rkky}

The discovery that the energy of magnetic multilayers made from alternating ferromagnetic and normal-metal films depends on the relative direction of the individual magnetizations \cite{Grunberg:prl86} is perhaps the most important in magnetoelectronics. The existence of the antiparallel ground-state configuration at certain spacer-layer thicknesses was essential for the subsequent discovery of the giant magnetoresistance \cite{Baibich:prl88,Grunberg:prb89}. Adjacent ferromagnetic layers in such structures are coupled by nonlocal and, as a function of normal-metal layer thickness, oscillatory \cite{Parkin:prl90} exchange interaction that can be qualitatively understood by perturbation theory analogous to the RKKY \cite{Ruderman:pr54,Kasuya:ptp56,Yosida:pr57} exchange coupling between magnetic impurities in a normal-metal host. Different oscillation periods, which can be resolved in measured magnetization configuration as a function of spacer thickness, are well explained in terms of the normal-metal Fermi surface calipers in the growth direction. The magnetic ground-state configuration is, at least in principle, accessible to first-principles electronic-structure calculations in the spin-density--functional theory formalism, and that is basically the end of the story. However, in order to make connection to the main topic of this review, we briefly discuss the formulation of the equilibrium exchange coupling in terms of scattering theory \cite{Sloncz:prb89,Erickson:prb93,Sloncz:mmm93}, that can also be formulated from first principles and calculated by density-functional theory \cite{Bruno:prb95,Stiles:mmm99}. Another advantage of a scattering-theory formulation is that effects of disorder can be understood employing the machinery of mesoscopic physics, such as random-matrix \cite{Beenakker:rmp97} or diagrammatic perturbation theory.

\begin{figure}[pth]
\includegraphics[width=0.95\linewidth,clip=]{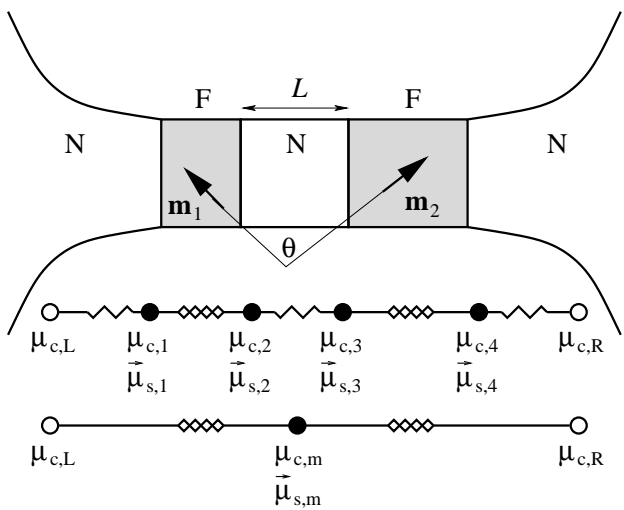}
\caption{Spin-valve schematic: two monodomain ferromagnets (F) separated by a normal (N) spacer and attached to normal leads. In this CPP configuration, the current flowing between two normal reservoirs sequentially traverses two magnetic layers. Also shown are two effective circuits discussed in Sec.~\ref{Mdc} for the semiclassical regime where the interlayer exchange coupling vanishes.}
\label{sv}
\end{figure}

Let us consider a layered N$\mid$F$\mid$N$\mid$F$\mid$N spin valve with angle $\theta$ between the magnetizations and an N-spacer with thickness $L$, see Fig.~\ref{sv} schematic. Suppose we can view the F$\mid$N$\mid$F trilayer as some spin-dependent scatterer embedded into a normal-metal medium. The trilayer gives a $\theta$-dependent contribution to the total ground-state energy, given by a standard formula \cite{Akkermans:prl91}:
\begin{equation}
E(L,\theta)=\frac{1}{2\pi i}\int_{-\infty}^{\varepsilon_F}\varepsilon\frac{\partial}{\partial\varepsilon}\ln\det\mathbf{s}(L,\theta,\varepsilon)d\varepsilon\,,
\label{TotEn}
\end{equation}
in terms of the energy-dependent scattering matrix $\mathbf{s}(L,\theta,\varepsilon)$ of the trilayer. The scattering matrix is made up from the matrices $\mathbf{r}$ and $\mathbf{t}$ of the reflection and transmission coefficients for a basis of spin-resolved incoming states at energy $\varepsilon$ from the right normal metal whereas the primed ones ($\mathbf{r}^{\prime}$ and $\mathbf{t}^{\prime}$) are defined for states coming from the left normal metal:
\begin{equation}
\mathbf{s}=\left(
\begin{array}[c]{cc}
\mathbf{r} & \mathbf{t}^{\prime}\\
\mathbf{t} & \mathbf{r}^{\prime}
\end{array}\right)\,.
\label{TotS}
\end{equation}
The scattering matrix of the total system can be composed out of the transmission and reflection coefficients of the F$\mid$N interfaces as well as of the bulk layers by well-known concatenation rules. Quantum-well states and resonances are formed by multiple reflections at interfaces caused by potential steps and electronic-structure mismatches. The angle and thickness dependence of the total energy (\ref{TotEn}) can be understood in terms of the variation of the interference pattern of the spin-dependent electron waves in and close to the normal-metal spacer. By varying $L$ and $\theta$, quantum-well states enter or leave the Fermi sea with abrupt changes in the total energy that can be large for small $L$. The minimum energy for a given $L$ is usually found at $\theta=0$ and $\pi$, i.e., either parallel or antiparallel configurations are favored. Though exponentially suppressed, the coupling between the magnetic layers persists when the insertion is an insulating barrier \cite{Sloncz:prb89,Bruno:prb95}.

Of special interest is the asymptotic dependence of energy $E(L,\theta)$ for large $L$. In this limit, the sharp jumps in energy by the population or depopulation of individual quantum-well states as a function of $L$ and $\theta$ become less prominent. The quasi-continuous angular dependence of the energy in this limit is well described by the lowest term in the expansion into Legendre polynomials:
\begin{equation}
E(L,\theta)\overset{L\rightarrow\infty}{=}\cos\theta\sum_\alpha\frac{J_\alpha}{L^2}\sin\left(q_{\alpha\perp}L+\phi_\alpha\right)\,,
\label{Easym}
\end{equation}
which is a sum of contributions from each critical caliper at the Fermi surface of the normal-metal spacer labeled by $\alpha$. The parameters $J_\alpha$ and $\phi_\alpha$ are model and material dependent \cite{Stiles:mmm99}. $q_{\alpha\perp}$ is the critical Fermi-surface--spanning wave vector in the layering direction, which determines the caliper in reciprocal space. Note that whereas we in principle require the scattering matrix for \textit{all} occupied states in Eq.~(\ref{TotEn}), Eq.~(\ref{Easym}) is governed by the scattering coefficients at the Fermi energy only, just as the transport properties at low temperature and bias. In practice, Eq.~(\ref{Easym}) can often be used for all but the most narrow spacer layers.

At configurations that are not at equilibrium, the derivative
\begin{equation}
\tau=-\frac{\partial}{\partial\theta}E(L,\theta)
\label{tdE}
\end{equation}
does not vanish. A finite $\tau$ is therefore interpreted as an exchange torque acting on the magnetizations, pulling them into an energetically-favorable configuration. Physically, this torque is an angular-momentum transfer that is carried by the electron spin. A spin valve that is ``strained" by a relative misalignment of the magnetization directions from their lowest-energy configuration therefore supports dissipationless spin currents. The situation is quite analogous to the Josephson junction in which a difference of the superconducting phase over a weak link induces a supercurrent. We note that at finite temperatures, the ground-state energy $E(L,\theta)$ should be replaced with the free energy $F(L,\theta)$ in Eq.~(\ref{tdE}).

Essential for the existence and the magnitude of the nonlocal exchange coupling and the corresponding persistent spin currents is the phase coherence of the wave functions in the normal spacer. An incoming electron in the spacer with information of the left magnetization direction has to be reflected at the right interface and interfere with itself at the left interface in order to convey the coupling information. This implies strong sensitivity to the effects of impurities, since the diffuse scattering destroys the regular interference pattern required by a sizable coupling. This qualitative notion has been formulated by \onlinecite{Waintal:prb00} in the scattering-theory formalism invoking the ``isotropy" condition for validity of the random-matrix theory. Isotropy requires diffuse transport, viz., that $L$ is larger than the mean free path due to bulk and interface scattering. It can then be shown rigorously that the equilibrium spin currents vanish on average with fluctuations that scale like $N^{-1}$, where $N$ stands for the number of transverse transport channels in the normal-metal spacer. In layered metallic structures, $N$ is large and the static exchange coupling and spin currents can safely be disregarded in the diffuse limit. On top of the suppression by disorder, the magnitude of the coupling scales like $L^{-2}$ even in ballistic samples, see Eq.~(\ref{Easym}). Experimentally, even the presumably best Co$\mid$Cu$\mid$Co samples indeed do not show an appreciable coupling beyond a spacer-layer thickness of about 20 atomic monolayers.

In section~\ref{dec}, we discuss the magnetization dynamics of multilayers and superlattices. We find that, on top of the equilibrium spin currents that communicate the nonlocal static exchange coupling, a dynamic exchange interaction with a much longer range becomes important. Any significant coupling at equilibrium in the dynamic studies can be represented approximately in terms of parametrized conservative forces that react to deviations from equilibrium without interference with nonequilibrium spin currents.

The GMR has been originally discovered in a configuration in which the current flows in the plane of the film (CIP). Multilayer pillar structures in which the current flows perpendicular to the planes (CPP) \cite{Pratt:prl91,Gijs:prl93,Gijs:ap97} are more relevant in the present context. Assuming diffusive transport, the CPP GMR is easily understood in terms of a two-channel series-resistor model \cite{Valet:prb93}: In the parallel configuration, the charge current is short-circuited by the low--electrical-resistance spin channel. The charge and spin transport in the intermediate configurations in which the magnetizations vary between parallel and antiparallel is important in the context of the present review due to the emergence of transverse spin currents that are absorbed by the magnetizations and contribute as a driving torque to the dynamics. This dissipative channel for transverse spins also modifies the angular magnetoresistance of spin valves. These and other noncollinear magnetoelectronic dc phenomena are reviewed by \onlinecite{Brataas:prp05}.

\subsection{Landau-Lifshitz-Gilbert phenomenology}
\label{llgt}

At temperatures well below the ferromagnetic critical temperature $T_c$, the equilibrium magnetization of a bulk ferromagnet saturates at some material-specific value $M_s$. Symmetry-restoring Goldstone modes of this broken-symmetry state are spin waves (magnons) mentioned in Sec.~\ref{pre}, which can condense into macroscopic transverse magnetization dynamics. In itinerant ferromagnets, there are also longitudinal spin excitations, the Stoner modes. However, since macroscopic variation of the magnetization magnitude at $T \ll T_c$ is very costly in free energy, we focus on the purely transverse motion of the position-dependent magnetization direction $\mathbf{m}=\mathbf{M}/M_s$, with a fixed magnitude $|\mathbf{M}|=M_s$.

A traditional starting point in studying the transverse magnetization dynamics in a ferromagnetic medium is based on the phenomenological Landau-Lifshitz (LL) equation \cite{Landau80}. The magnetization direction $\mathbf{m}(\mathbf{r},t)$ is treated in this approach as a classical position- and time-dependent variable obeying equations of motion which are determined by the free-energy functional $F[\mathbf{M}]$ for degrees of freedom coupled to the magnetization distribution $\mathbf{M}(\mathbf{r})$ (such as the electromagnetic field or itinerant electrons experiencing a ferromagnetic exchange field):
\begin{equation}
\partial_t\mathbf{m}(\mathbf{r},t)=-\gamma\mathbf{m}(\mathbf{r},t)\times\mathbf{H}_{\text{eff}}(\mathbf{r})\,,
\label{ll}
\end{equation}
where $\gamma$ is (minus) the gyromagnetic ratio and
\begin{equation}
\mathbf{H}_{\text{eff}}(\mathbf{r})=-\partial_\mathbf{M}F[\mathbf{M}]
\label{Heff}
\end{equation}
is the so-called ``effective magnetic field." Corresponding to the respective contributions to the free energy, the effective field can usually be decomposed into the applied, dipolar demagnetization, crystal-anisotropy, and exchange fields. In the case of free electrons, $\gamma=2\mu_B/\hbar>0$, in terms of Bohr magneton $\mu_B$ and Planck's constant $h=2\pi\hbar$, and it is usually close to this value in transition-metal ferromagnets.

It is easy to see that the LL equation (\ref{ll}), with effective field (\ref{Heff}), describes transverse-magnetization dynamics preserving the free energy $F[\mathbf{M}]$. Definition (\ref{Heff}) with the effective field depending on the instantaneous magnetic configuration assumes that the magnetization dynamics are very slow on the scale of the relevant microscopic relaxation processes. However, some slow degrees of freedom may not respond sufficiently fast to the magnetization motion, making the effective field dependent on the history of the magnetization dynamics $\mathbf{M}(\mathbf{r},t)$. This should be associated with dissipation of energy into the degrees of freedom that are coupled to the magnetization.

As a specific example, consider the magnetization dynamics (\ref{ll}) described by the effective field
\begin{equation}
\mathbf{H}_{\text{eff}}(\mathbf{r},t)=-\partial_{\mathbf{M}}\left\langle H(\mathbf{M})\right\rangle_t\,,
\label{HM}
\end{equation}
where $H(\mathbf{M})$ is the many-body Hamiltonian for itinerant electrons, parametrized by a mean-field magnetic configuration $\mathbf{M}(\mathbf{r},t)$ of, e.g., some localized magnetic orbitals (as in the $s-d$ model), and $\langle\rangle_t$ evaluates its expectation value for the many-body state (or ensemble) at time $t$. Setting the many-body ensemble at time $t$ to its thermal-equilibrium configuration determined by $\mathbf{M}(t)$ reproduces the Landau-Lifshitz (LL) definition (\ref{Heff}). In the opposite extreme, when electrons do not respond at all to the fast magnetic dynamics, $\langle\rangle_t\approx\langle\rangle_0$, the effective field is determined by the history-independent functional $\langle H(\mathbf{M})\rangle_0$ instead. In the intermediate regime, a finite time lag in the response of the itinerant electrons to the varying magnetization causes dissipation of the magnetic energy, as discussed in Sec.~\ref{sdm}. To lowest order in frequency (i.e., keeping only terms linear in $\partial_t$), such damping can be described by an additional torque in Eq.~(\ref{ll}) \cite{Gilbert:pr55,Gilbert:ieeem04}:
\begin{equation}
\partial_t\mathbf{m}=-\gamma\mathbf{m}\times\mathbf{H}_{\text{eff}}+\alpha\mathbf{m}\times\partial_t\mathbf{m}\,,
\label{llg}
\end{equation}
where $\alpha$ is the dimensionless Gilbert constant and $\mathbf{H}_{\text{eff}}$ is an effective field depending only on the instantaneous magnetic configuration. (Partial time derivatives imply here the possibility of spatial variations of the magnetization, as, e.g., in the case of spin waves; the full time derivatives are reserved for the monodomain dynamics.) The Gilbert term in Eq.~(\ref{llg}) has been obtained for various microscopic formulations of the magnetization dynamics, see, e.g., \onlinecite{Heinrich:pss67,Korenman:prb72,Lutovinov:zetf79,Safonov:prb00,Tserkovnyak:prl021,Kunes:prb02,Sinova:prb04}.

Energy dissipation implied by Eq.~(\ref{llg}) preserves the local magnitude of the magnetization. For example, for a constant $\mathbf{H}_{\text{eff}}$ obeying Eq.~(\ref{Heff}) and $\alpha=0$, $\mathbf{m}$ precesses around the field vector with frequency $\omega=\gamma H_{\text{eff}}$. When damping is switched on, $\alpha>0$ (assuming positive $\gamma$, as in the case of free electrons), the precession spirals down on a time scale of $(\alpha\omega)^{-1}$ to a time-independent magnetization along the field direction, i.e., the lowest--free-energy state. Close to an equilibrium axis with rotational symmetry, the Landau-Lifshitz-Gilbert (LLG) equation (\ref{llg}) is obeyed by a small-angle damped circular precession, while in the presence of anisotropies, small-angle trajectories are elliptic and the damping is in general a tensor. For most of our purposes, simple circular precession with a scalar damping $\alpha$ suffices (but see Sec.~\ref{mca}). It is sometimes convenient to work with a different Gilbert parameter
\begin{equation}
G=\alpha\gamma M_s\,.
\label{G}
\end{equation}

It can be made explicit that magnetization dynamics described by Eq.~(\ref{llg}) dissipate energy at a rate determined by $\alpha$. To this end, suppose for simplicity that $\gamma\mathbf{H}_{\text{eff}}=\omega_0\mathbf{\hat{z}}$, $\omega_0>0$, is uniform throughout a monodomain ferromagnetic sample, so that Eq.~(\ref{llg}) describes a damped macrospin circular precession around the $z$ axis. Small-angle dynamics around the $z$ axis can thus be resonantly excited by a (right-hand) circularly polarized rf field with a small amplitude $h_{\text{rf}}$ and frequency $\omega$ close to $\omega_0$, that is $h_-(t)=h_x(t)-ih_y(t)=h_{\text{rf}}\exp(-i\omega t)$. The magnetic response to such a field is $\delta M_-(\omega)=\chi_{-+}(\omega)h_-(\omega)$, where
\begin{equation}
\chi_{-+}(\omega)=\frac{\gamma M_s}{(\omega_0-\omega)-i\alpha\omega}\,
\end{equation}
is the transverse magnetic susceptibility. The linear-response expression for the energy-dissipation power per unit of volume
\begin{equation}
P=\omega\mbox{Im}\chi_{-+}(\omega)h^2_{\text{rf}}=\frac{\alpha\gamma M_s\omega^2h^2_{\text{rf}}}{(\omega_0-\omega)^2+(\alpha\omega)^2}
\label{P}
\end{equation}
does not depend on the microscopic origin of $\alpha$, as long as Eq.~(\ref{llg}) holds. For a steady precession, one can also show that $P=-\mathbf{h}(t)\cdot\mathbf{\dot{m}}(t)M_s$ is the work done by the rf field $\mathbf{h}(t)$. The stability of the system, $P>0$, requires that $\alpha\gamma>0$.

Eq.~(\ref{llg}) has been found to very successfully describe the dynamics of ultrathin ferromagnetic films as well as bulk materials in terms of a few material-specific parameters that are accessible to ferromagnetic-resonance (FMR) experiments \cite{Bhagat:prb74,Heinrich:ap93}. FMR spectra are obtained by placing the sample into a microwave cavity and sweeping the external dc field. $\gamma\mathbf{H}_{\text{eff}}$ then determines the position and $\alpha$ the width of the resonance absorption peak. The FMR linewidth can have an additional contribution due to inhomogeneities in $\mathbf{H}_{\text{eff}}$, loosely corresponding to a finite range of the resonance frequency $\omega_0$ in Eq.~(\ref{P}). For example, small disorder by surface roughness or a nonuniform surface field in exchange-biased thin films, contributes to the resonance broadening by (in quantum-mechanical terms) two-magnon scattering \cite{Mills03}. The inhomogeneous linewidth broadening is associated with dephasing of the global precession that in general does not conserve the magnitude of the magnetization. Whereas the Gilbert damping predicts a strictly linear dependence of FMR linewidths on frequency, the inhomogeneous broadening is usually associated with weaker frequency dependence as well as a zero-frequency contribution. Another common technique in studying long-wavelength spin waves is Brillouin light scattering (BLS) (see, e.g., \onlinecite{Demokritov:jp94}). Both FMR and BLS probe magnetic excitations close to the surface, i.e., within the corresponding skin depth of the order of 100~nm for FMR and 10~nm for BLS \cite{Mills03}. In contrast to FMR, BLS excites spin waves with finite wavelengths in the surface plane (in the range of that of the visible light), bearing consequences for the signal linewidths, see \onlinecite{Mills03} and Sec.~\ref{nmd}. In closing this subsection, we remark that ferromagnetic magnetization dynamics and, in particular, magnetization relaxation processes are collective many-body phenomena that continue to fascinate in spite of decades of theoretical and experimental efforts to understand them, see, e.g., \onlinecite{Dobin:prl03,Qian:prl02}.

\subsection{Current-induced magnetization dynamics}

It has been realized only relatively recently that in magnetic multilayers the magnets can be excited by other means than external magnetic fields. \onlinecite{Sloncz:mmm96} and \onlinecite{Berger:prb96} predicted that in CPP spin-valve structures a dc current in the right direction can excite and even reverse the magnetization of a magnetic layer. This can be observed by monitoring $dI/dV$ which depends on the magnetic configuration, as in the GMR. The predictions have by now been amply confirmed by many recent experiments \cite{Myers:sc99,Katine:prl00,Tsoi:nat00,Wegrowe:epl01,Myers:prl02,Ozyilmaz:prl03,Kiselev:nat03,Urazhdin:prl03,Ji:prl03,Puffal:prb04,Sun:jap05,Krivorotov:sc05}. The current-induced magnetic dynamics have been found to also affect current-noise spectra \cite{Covington:prb04}. The prediction \cite{Polianski:prl04,Stiles:prb04} that even a \textit{single} dc-current--driven ferromagnetic layer may undergo a resonant finite--wave-vector spin-wave excitation has been experimentally confirmed by \onlinecite{Ozyilmaz:prl04}, see also \onlinecite{Ji:prl03}. Consequently, higher--wave-vector spin-wave excitations may in some situations successfully compete \cite{Ozyilmaz:prb05,Brataas:cm05} with current-induced macrospin motion considered below. The current-induced magnetization dynamics pose a challenging physics problem that requires understanding of the coupling of nonequilibrium quasiparticles with the collective magnetization dynamics. It carries technological potential as well. In small structures, the current-induced magnetization reversal may be more power efficient to write information into magnetic random-access memories compared to switching by Ampere magnetic field. The generation of microwaves by exciting stable magnetization orbits with dc bias-dependent frequencies may also satisfy technological needs \cite{Kiselev:nat03,Rippard:prl04}. 

Current-induced magnetization dynamics are a consequence of the spin-dependent transport in F$\mid$N heterostructures. For example, Slonczewski's magnetization torque \cite{Sloncz:mmm96} is equivalent to absorption of an incident spin current with a polarization component perpendicular to the magnetization \cite{Brataas:prl00,Waintal:prb00,Stiles:prb02}. The component of the electron spin perpendicular to the magnetization is not a constant of the motion in a ferromagnet. On the other hand, neglecting the effects of spin-orbit coupling (other than the macroscopic anisotropy already included in $\mathbf{H}_{\text{eff}}$) and other spin-flip processes, the total spin angular momentum is conserved. The spin angular-momentum difference between an electron entering and leaving a ferromagnet is therefore transferred to the magnetization. Under a sufficiently large angular-momentum transfer, the magnetization starts to move. The component of the net spin angular-momentum flow \textit{out} of the ferromagnet $\mathbf{I}_s$ parallel to $\mathbf{m}$ vanishes, since the outflow cancels the inflow for the parallel component (assuming spin along the uniform magnetization direction is conserved). The \textit{spin-transfer torque} $\boldsymbol{\tau}=-\mathbf{I}_s$ should be accounted for as an additional source term in the equation motion of the magnetization. In the presence of spin-flip scattering, a component parallel to $\mathbf{m}$ must be projected out to represent the torque that drives the transverse magnetization dynamics:
\begin{equation}
\boldsymbol{\tau}=-\mathbf{m}\times\mathbf{I}_s\times\mathbf{m}\,.
\label{spintorque}
\end{equation}

An electron injected into a ferromagnet at the Fermi energy and transverse polarization is not in a momentum eigenstate, but should be described by a linear combination of majority and minority spin eigenstates associated with different Fermi wave vectors, $k_F^\uparrow$ and $k_F^\downarrow$. The linear coefficients of up and down spins carry out oscillations as a function of time and position, equivalent to a spin precession around the exchange magnetic field.  Fermi-level states entering the ferromagnet at different angles precess on different length scales perpendicular to the interface, depending on the perpendicular component of the spin-up and spin-down wave-vector difference. In ferromagnets with a large cross-section area, a large number of transverse modes with different spin-precession lengths contribute to the total spin current. The destructive interference of numerous states with different phases corresponds to the \textit{absorption of the transverse spin current} inside the ferromagnet on the scale of the so-called transverse spin-coherence length
\begin{equation}
\lambda_{\text{sc}}=\frac{\pi}{|k_F^\uparrow-k_F^\downarrow|}\,.
\label{lsc}
\end{equation}
$\lambda_{\text{sc}}\sim \lambda_F$ (the Fermi wavelength), an atomistic length scale for, e.g., transition-metal ferromagnets like Co, Ni, and Fe, or their alloys. The smallness of penetration depths $\lambda_{\text{sc}}$ \textit{a posteriori} justifies the implicitly assumed clean limit, $\lambda_{\text{sc}}\ll\lambda$ (the mean free path). It should be noted that $\lambda_{\text{sc}}$ sets a length scale of a \textit{power-law} \cite{Stiles:prb02}, not an exponential suppression of the transverse spin current.

After transmission through a ferromagnetic film much thicker than $\lambda_{\text{sc}}$, electrons are completely polarized along the magnetization direction. When reflection at the F$\mid$N boundary may be disregarded, $\mathbf{I}_s$ on the right-hand side of (\ref{spintorque}) is simply the negative of the transverse spin current incident on the ferromagnet. When reflection cannot be neglected, the transverse polarization of the reflected electrons should be taken into account: although the reflected electrons hardly penetrate the ferromagnet (over the Fermi wavelength), the strong exchange field can still induce a significant precession of the reflected component \cite{Stiles:prb02}. This can lead to a reaction torque on the ferromagnet as an effective magnetic field oriented parallel to the spin accumulation in the normal metal. However, at interfaces to transition-metal ferromagnets, positive and negative contributions to the effective field typically average out to be small \cite{Xia:prb02}.

The dynamics of a monodomain ferromagnet of volume $V$ and magnetization $M_s$ that is subject to the torque (\ref{spintorque}) are modified by an additional source term on the right-hand side of the LLG equation \cite{Sloncz:mmm96}: 
\begin{equation}
\left.\frac{\partial\mathbf{m}}{\partial t}\right\vert_{\text{torque}}=\frac{\gamma}{M_sV}\mathbf{m}\times\mathbf{I}_s\times\mathbf{m}\,.
\label{Gllg}
\end{equation}
For a fixed current density, Eq.~(\ref{Gllg}) is proportional to the interface area and inversely proportional to the volume of the ferromagnet. Current-induced magnetization dynamics are usually realized in perpendicular spin valves with one hard (highly coercive) ferromagnet that acts as a static polarizer and a second soft ferromagnet that responds sensitively to the spin-transfer torque.

\subsection{Spin emission by excited ferromagnets}

When seeking a consistent theory of the magnetization dynamics in heterostructures, the current-induced magnetization torque discussed above is only one side of the coin. A moving magnetization in a ferromagnet that is in electric contact with normal conductors emits (``pumps") a spin current into its nonmagnetic environment \cite{Tserkovnyak:prl021}, giving a contribution to $\mathbf{I}_s$ in Eq.~(\ref{Gllg}). The spin pumping thus leads to an additional source term in the LLG equation, also when the magnetization dynamics are induced by external magnetic fields and not by applied current bias. In typical biased systems with current-induced dynamics, the spin pumping is of the same order as the current-driven torque and should be treated on equal footing, as explained in Sec.~\ref{bsv}.

The spin pumping by a precessing ferromagnet is, in some sense, the reverse process to the current-induced magnetization dynamics. When the pumped spin angular momentum is not quickly dissipated to the normal-metal atomic lattice, a spin accumulation builds up and creates reaction torques due to transverse-spin backflow into ferromagnets. The interplay between the magnetization dynamics and the nonequilibrium spin-polarized transport in heterostructures is the central topic of this review. The conversion of magnetization movement into spin currents and \textit{vice versa} at a possibly different location is what we mean by the ``nonlocality of the magnetization dynamics" in our title. In the remainder of this subsection, we put this topic into a historic perspective.

Nonlocality of the magnetization dynamics can be interpreted as a nonlocal exchange coupling with explicit time dependence. A first step into this direction was carried out by \onlinecite{Barnes:jpf74} who generalized the RKKY theory for the static linear response of the electron gas to magnetic impurities  to dynamic phenomena in order to understand the electron-spin resonance of localized magnetic moments embedded in a conducting medium. He showed that the dynamic part of the RKKY interaction in diffuse media is limited by the spin-diffusion length. A related experimental observation of ``giant electron-spin--resonance transmission" through a Cu foil implanted with magnetic Mn ions on one or both sides \cite{Monod:prl72} showed that precessing impurity magnetic moments cause nonequilibrium spin diffusion.

Subsequently, \onlinecite{Silsbee:prb79} observed a strong enhancement of the microwave transmission through a Cu foil with a thin ferromagnetic layer evaporated on one side, when the ferromagnetic and Cu conduction-electron--spin resonances are tuned into a collective mode. This is related to the enhancement of ``Larmor waves" in nonresonant electron-spin transmission through normal-metal foils coated with a ferromagnetic layer \cite{Janossy:prl76}. The experiments were interpreted by postulating a phenomenological spin interdiffusion through the F$\mid$N interface by nonequilibrium components of the magnetization/spin accumulation on both sides. These authors concluded that the precessing magnetic moments can be a source of nonequilibrium spin accumulation diffusing through the nonmagnetic conducting medium. \textit{Vice versa}, the nonequilibrium spin accumulation can be transferred into the magnetization motion. This picture was investigated further by \onlinecite{Janossy:prb80,Parks:prb87} and was invoked later to qualitatively interpret the experiments by \onlinecite{Hurdequint:jdp88,Hurdequint:mmm91,Hurdequint:mmm912}.

The discussion of the dynamic coupling between a precessing magnetization and itinerant electrons in layered F$\mid$N structures has been (independently) revived by \onlinecite{Berger:prb96}. He predicted an enhanced Gilbert damping in thin ferromagnetic films in trilayer F$\mid$N$\mid$F configurations, relying on an elementary quantum process of magnon annihilation associated with electron spin flip. A very different approach to the problem was taken by \onlinecite{Tserkovnyak:prl021}. They used the formalism of parametric pumping \cite{Buttiker:zpb94,Brouwer:prb98} developed in the context of mesoscopic scattering problems in order to show that the time-dependent magnetization induces spin emission into the itinerant degrees of freedom, see Sec.~\ref{pump}. [A host of other mesoscopic spin pumps have been proposed in recent years, see, e.g., \onlinecite{Sharma:prl03} and references therein, at least one of which has been realized experimentally \cite{Watson:prl03}.] The spin-pumping picture enables us to discuss several topics of this review in a unified manner and is easily rendered quantitative. More recently, a linear-response formalism similar to that of \onlinecite{Barnes:jpf74} has been put forward by \onlinecite{Simanek:prb031}. This alternative point of view has the advantage to be more familiar to many in the magnetism community, but it is much less suited for quantitative comparison with experiments, as discussed in Sec.~\ref{lra}.

\section{Scattering-theory approach to magnetoelectronics}
\label{sm}

\subsection{Magnetoelectronic dc circuit theory}
\label{Mdc}

Electron spin and charge transport in F$\mid$N heterostructures with static magnetic configurations has attracted a considerable attention following the discovery of GMR. Most of the activities in recent years, including the work reviewed here, concentrated on the CPP geometry in which the electrons pass sequentially through magnetic and nonmagnetic elements of the circuit, see \onlinecite{Gijs:ap97} for a review. A systematic and quite general, yet easy-to-handle, semiclassical approach to this problem|the magnetoelectronic circuit theory|is reviewed by \onlinecite{Brataas:prp05}. In the following we want to give a brief account of that theory before extending it to the dynamic magnetic configurations in Sec.~\ref{pump}.

\begin{figure}[pth]
\includegraphics[width=\linewidth,clip=]{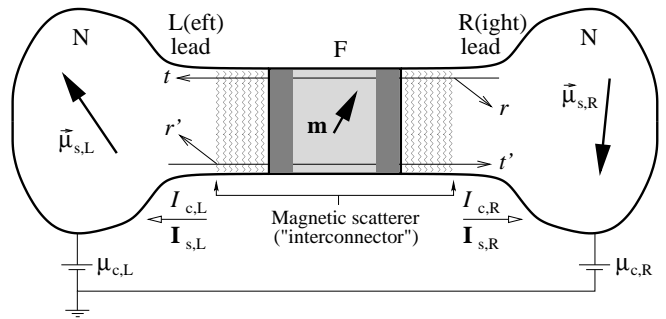}
\caption{Magnetic scatterer (interconnector) connecting two chaotic normal-metal (N) nodes via ballistic leads that support a quantized number of transverse channels at the Fermi energy. The scatterer includes a ferromagnetic (F) region characterized by a uniform magnetic direction $\mathbf{m}$ (but not necessarily uniform magnitude of the exchange spin splitting along $\mathbf{m}$), which is depicted as the gray box in the center. The dark-gray areas of the ferromagnetic region near both F$\mid$N interfaces mark the extent of the transverse-spin coherence characterized by $\lambda_{\text{sc}}$, Eq.~(\ref{lsc}). Each of the two normal-metal regions is divided into a reservoir characterized by the electrochemical potential $\mu_c$ and (vector) spin accumulation $\boldsymbol{\mu}_s$, a ballistic lead with a fixed number of transport channels, and possibly a disordered region incorporated in the interconnector (depicted by wavy lines), which accounts for relevant normal-metal scattering processes. The interconnector is described by spin-dependent reflection and transmission coefficients forming the scattering matrix, Eq.~(\ref{s}), for quantum channels in the normal leads. The purpose of the formalism is to calculate the nonequilibrium charge and spin flows in the leads, $I_c$ and $\mathbf{I}_s$, induced by spin accumulations and electrochemical-potential imbalance in the nodes.}
\label{sc}
\end{figure}

A basic element of the magnetoelectronic circuit theory is a magnetic ``interconnector" between two normal nodes or reservoirs attached to the ferromagnet via Ohmic leads, as shown in Fig.~\ref{sc}. Physically, this could be realized, e.g., as a layered pillar N$\mid$F$\mid$N structure. The normal nodes are chaotic to the extent that the nonequilibrium transport through the leads can be expressed in terms of the energy-dependent distribution functions $\hat{f}(\varepsilon)$ in each node (averaged over orbital states at energy $\varepsilon$), which are $2\times2$ energy-dependent matrices in spin space of spin-$1/2$ electrons. (We make a convention of denoting such $2\times2$ matrices in spin space by hats.) In thermal equilibrium $\hat{f}(\varepsilon)=f_{\text{FD}}(\varepsilon)\hat{\sigma}_0$, where $f_{\text{FD}}(\varepsilon)$ is the Fermi-Dirac distribution for electrons and $\hat{\sigma}_0$ is the $2\times2$ unit matrix. In linear response, it is convenient to define local electrochemical potentials $\mu_c$ and ``spin accumulations" $\boldsymbol{\mu}_s$ in all nodes: \cite{Brataas:prl00,Brataas:epjb01}
\begin{align}
\label{ca}
\mu_c=&\frac{1}{2}\int_{\varepsilon_0}^\infty d\varepsilon\mbox{Tr}[\hat{f}(\varepsilon )]\,,\\
\boldsymbol{\mu}_s=&\int_{\varepsilon_0}^\infty d\varepsilon\mbox{Tr}[\boldsymbol{\hat{\sigma}}\hat{f}(\varepsilon)]\,,
\label{sa}
\end{align}
choosing a reference energy $\varepsilon_0$ that lies below the Fermi energy by much more than the thermal energy and voltage biases, but which is arbitrary otherwise. $\boldsymbol{\hat{\sigma}}=(\hat{\sigma}_x,\hat{\sigma}_y,\hat{\sigma}_z)$ is a vector of the Pauli matrices. Notice that in our convention, the ``spin accumulation" $\boldsymbol{\mu}_s$ is a vector with the direction determined by the total nonequilibrium spin-imbalance density $\mathbf{s}$ and the magnitude given by the corresponding energy splitting of spins along this direction. In linear response and at low temperatures, these quantities are related through the Fermi-level density of states (per spin and unit volume) in the node, $\mathcal{N}(\varepsilon_F)$: $\mathbf{s}=(\hbar/2)\mathcal{N}(\varepsilon_F)\boldsymbol{\mu}_s$.

Here we wish to calculate the dc charge and spin angular-momentum currents, $I_c$ and $\mathbf{I}_s$, \textit{entering} the nodes through the leads, which are induced by the nonequilibrium spin accumulations \textit{in} the nodes and/or electrochemical imbalance \textit{between} the nodes. It is convenient to define the $2\times2$ tensor current
\begin{equation}
\hat{I}=\frac{1}{2}\hat{\sigma}_0I_c-\frac{e}{\hbar}\boldsymbol{\hat{\sigma}}\cdot\mathbf{I}_{s}\,,
\label{Ii}
\end{equation}
whose isotropic and traceless components determine respectively the charge and spin currents. Since, as discussed above, spin currents are not conserved at F$\mid$N interfaces, we use the convention that it is calculated on the normal side (unless specified otherwise).
The $2\times2$ current operator $\hat{I}_l$ for the $l$th lead ($l={\rm L,~R}$) can be expressed in terms of operators $a^\sigma_{n,l}(\varepsilon)$ [$b^\sigma_{n,l}(\varepsilon)$] that annihilate a spin-$\sigma$ electron with energy $\varepsilon$ leaving (entering) the $l$th node through the $n$th quantum channel of the lead \cite{Tserkovnyak:prb01}:
\begin{align}
I_l^{\sigma\sigma^\prime}=&\frac{e}{h}\sum_{m}\int d\varepsilon d\varepsilon^{\prime}\left[a^{\sigma^\prime}_{m,l}(\varepsilon)^\dagger a^{\sigma}_{m,l}(\varepsilon^{\prime})\right.\nonumber\\
&\left.-b^{\sigma^\prime}_{m,l}(\varepsilon)^\dagger b^\sigma_{m,l}(\varepsilon^{\prime })\right]\,.
\label{I}
\end{align}
Suppose, as a starting point, that the momentum-space distribution in each node is isotropic, i.e., $\hat{f}(\varepsilon)$ does not depend on orbital quantum numbers. This is true if the nonequilibrium currents do not cause significant drift contributions to the distribution function. (We will revisit and drop this assumption in Sec.~\ref{icm}.) For the $l$th lead then
\begin{equation}
\left\langle a^\sigma_{n,l}(\varepsilon)^\dagger a^{\sigma^\prime}_{n^\prime,l^\prime}(\varepsilon^{\prime})\right\rangle=f^{\sigma^\prime\sigma}_l(\varepsilon)\delta_{ll^{\prime}}\delta_{nn^\prime}\delta(\varepsilon-\varepsilon^{\prime})\,,
\label{exp}
\end{equation}
and it is now straightforward to evaluate the expectation value $\langle I_l^{\sigma\sigma^\prime}\rangle$ (also denoted simply by $I_l^{\sigma\sigma^\prime}$) of the current operator, after relating the scattered states to the incoming states via the scattering matrix of the magnetic interconnector:
\begin{equation}
b^\sigma_{n,l}(\varepsilon)=\sum_{\sigma^\prime n^\prime,l^\prime}s_{nn^\prime,ll^{\prime}}^{\sigma\sigma^\prime}(\varepsilon)a^{\sigma^\prime}_{n^\prime,l^\prime}(\varepsilon)\,.
\label{bsa}
\end{equation}
The scattering coefficients $s_{nn^\prime,ll^{\prime}}^{\sigma\sigma^\prime}$ characterize reflection if $l=l^\prime$ and transmission otherwise. Eq.~(\ref{bsa}) assumes that the entire interconnector is elastic, so that the electron energy is conserved upon scattering between the normal nodes. For a ferromagnet with an exchange spin splitting along unit vector $\mathbf{m}$ and vanishing spin-orbit interaction in the system \cite{Brataas:prl00},
\begin{equation}
\hat{s}_{nn^\prime,ll^{\prime}}=s_{nn^\prime,ll^{\prime}}^{\uparrow}\hat{u}^{\uparrow}+s_{nn^\prime,ll^{\prime}}^{\downarrow}\hat{u}^{\downarrow}
\label{s}
\end{equation}
in terms of the scattering coefficients for spins up (down) along $\mathbf{m}$, $s_{nn^\prime,ll^{\prime}}^{\uparrow(\downarrow)}$, and the projection matrices
\begin{equation}
\hat{u}^{\uparrow(\downarrow)}=\frac{1}{2}\left(\hat{\sigma}_0\pm\boldsymbol{\hat{\sigma}}\cdot\mathbf{m}\right)\,.
\label{u}
\end{equation}
A consequence of the elastic-scattering approximation is a rigid exchange field: there are no magnons excited by the electron transport. If we furthermore assume a sufficiently low temperature, voltage imbalance, and spin accumulations, so that the scattering-matrix variation on these energy scales is negligible, we can replace $\hat{s}_{nn^\prime,ll^{\prime}}(\varepsilon)$ by its value at the Fermi level $\varepsilon_F$.

It is then convenient to group the conductance parameters into two pairs of $2\times2$ matrices. For electrons incident from the right lead, we define
\begin{align}
\label{g}
g^{\sigma\sigma^\prime}=&\sum_{nn^\prime}\left[\delta_{nn^\prime}-r_{nn^\prime}^{\sigma}(r_{nn^\prime}^{\sigma^\prime})^\ast\right]\,,\\
t^{\sigma\sigma^\prime}=&\sum_{nn^\prime}t_{nn^\prime}^{\sigma}(t_{nn^\prime}^{\sigma^\prime})^\ast\,,
\label{t}
\end{align}
where the index $n^\prime$ is summed over the channels in the right lead, and $n$ runs over the channels in the right lead in Eq.~(\ref{g}) and left lead in Eq.~(\ref{t}). The coefficients $r_{nn^\prime}^{\sigma}$ and $t_{nn^\prime}^{\sigma}$ are reflection and transmission amplitudes, i.e., elements $s_{nn^\prime,ll^{\prime}}^{\sigma }$ of the scattering matrix (\ref{s}) with $l \ne l^{\prime}$ and $l = l^{\prime}$, respectively. The range of summation for $n^\prime$ in Eqs.~(\ref{g}) and (\ref{t}), i.e., the total number of transverse quantum channels in the right lead (for a given spin species at the Fermi level) is called the Sharvin conductance, $g^{\text{Sh}}$, a quantity which will be useful later. For electrons incident from the left lead, we denote the reflection and transmission amplitudes by primed quantities, and we similarly define primed matrices $g^{\prime\sigma\sigma^\prime}$ and $t^{\prime\sigma\sigma^\prime}$ in terms of the primed scattering amplitudes. We denote the Sharvin conductance of the left lead by $g^{\prime\text{Sh}}$. It should be understood that all scattering coefficients and corresponding conductance parameters are evaluated at the Fermi level. Putting above equations together \cite{Brataas:prl00,Brataas:epjb01,Brataas:prl03},
\begin{align}
\label{IcR}
I_{c,{\rm R}}^{(0)}=&\frac{e}{2h}\left\{2\left(g^{\uparrow\uparrow}+g^{\downarrow\downarrow}\right)\left(\mu_{c,{\rm R}}-\mu_{c,{\rm L}}\right)\right.\nonumber\\
&+\left.\left(g^{\uparrow\uparrow}-g^{\downarrow\downarrow}\right)\left(\boldsymbol{\mu}_{s,{\rm R}}-\boldsymbol{\mu}_{s,{\rm L}}\right)\cdot\mathbf{m}\right\}\,,\\
\mathbf{I}_{s,{\rm R}}^{(0)}=&-\frac{1}{8\pi}\left\{2\left(g^{\uparrow\uparrow}-g^{\downarrow\downarrow}\right)\left(\mu_{c,{\rm R}}-\mu_{c,{\rm L}}\right)\mathbf{m}\right.\nonumber\\
&+\left(g^{\uparrow\uparrow}+g^{\downarrow\downarrow}\right)\left[\left(\boldsymbol{\mu}_{s,{\rm R}}-\boldsymbol{\mu}_{s,{\rm L}}\right)\cdot\mathbf{m}\right]\mathbf{m}\nonumber\\
&+2g_r^{\uparrow\downarrow}\mathbf{m}\times\boldsymbol{\mu}_{s,{\rm R}}\times\mathbf{m}+2g_i^{\uparrow\downarrow}\boldsymbol{\mu}_{s,{\rm R}}\times\mathbf{m}\nonumber\\
&\left.-2t_r^{\prime\uparrow\downarrow}\mathbf{m}\times\boldsymbol{\mu}_{s,{\rm L}}\times\mathbf{m}-2t_i^{\prime\uparrow\downarrow}\boldsymbol{\mu}_{s,{\rm L}}\times\mathbf{m}\right\},
\label{IsR}
\end{align}
and the currents through the left lead are obtained by interchanging ${\rm L}\leftrightarrow{\rm R}$, $g^{\sigma\sigma^\prime}\leftrightarrow g^{\prime\sigma\sigma^\prime}$ and $t^{\prime\uparrow\downarrow}\leftrightarrow t^{\uparrow\downarrow}$. [The superscript $(0)$ introduced here denotes currents with static magnetizations.] By unitarity of the scattering matrix, $g^{\sigma\sigma}=g^{\prime\sigma\sigma}=t^{\sigma\sigma}=t^{\prime\sigma\sigma}$, if the spin component along the magnetization direction is conserved. The dc transport in the two-terminal geometry is then determined by two real-valued spin-dependent conductances $g^{\sigma\sigma}$ and four complex-valued (spin-)mixing parameters $g^{\uparrow\downarrow}=g^{\uparrow\downarrow}_r+ig^{\uparrow\downarrow}_i$, $t^{\uparrow\downarrow}=t^{\uparrow\downarrow}_r+it^{\uparrow\downarrow}_i$ (the subscripts $r$ and $i$ respectively denoting the real and imaginary parts), $g^{\prime\uparrow\downarrow}$ and $t^{\prime\uparrow\downarrow}$. For a mirror-symmetric structure, $g^{\uparrow\downarrow}=g^{\prime\uparrow\downarrow}$ and $t^{\uparrow\downarrow}=t^{\prime\uparrow\downarrow}$. 

We have now also access to microscopic expressions for the spin-transfer torques (\ref{spintorque}). The torque on the right surface of the ferromagnet  $\boldsymbol{\tau}_{\rm R}=-\mathbf{m}\times\mathbf{I}_{s,{\rm R}}^{(0)}\times\mathbf{m}$:
\begin{align}
\boldsymbol{\tau}_{\rm R}=&\frac{1}{4\pi}\left(g_r^{\uparrow\downarrow}\mathbf{m}\times\boldsymbol{\mu}_{s,{\rm R}}\times\mathbf{m}+g_i^{\uparrow\downarrow}\boldsymbol{\mu}_{s,{\rm R}}\times\mathbf{m}\right.\nonumber\\
&\left.-t_r^{\prime\uparrow\downarrow}\mathbf{m}\times\boldsymbol{\mu}_{s,{\rm L}}\times\mathbf{m}-t_i^{\prime\uparrow\downarrow}\boldsymbol{\mu}_{s,{\rm L}}\times\mathbf{m}\right)\,,
\label{tR}
\end{align}
is proportional to the spin-mixing (i.e., off-diagonal) components of the conductance matrices (\ref{g}) and (\ref{t}). The first two terms in $\boldsymbol{\tau}_{\rm R}$ involve reflection at the right F$\mid$N junction and the last two terms|transmission through the entire N$\mid$F$\mid$N structure. The latter terms can thus be disregarded when the ferromagnet is much thicker than the transverse-spin coherence length $\lambda_{\text{sc}}$, Eq.~(\ref{lsc}), since transmitted electrons accumulate phases differing by more than $\pi$ for opposite spins (along $\mathbf{m}$). In that limit, the first term, proportional to $\mathbf{m}\times\boldsymbol{\mu}_{s,{\rm R}}\times\mathbf{m}$, is similar to the torque introduced by \onlinecite{Sloncz:mmm96} that is responsible for instability leading to magnetization precession or reversal. The second term, proportional to $\boldsymbol{\mu}_{s,{\rm R}}\times\mathbf{m}$, acts as an effective magnetic field collinear with the spin accumulation in the right normal node. In transition-metal ferromagnets, $g_i^{\uparrow\downarrow}\lesssim0.1g_r^{\uparrow\downarrow}$, see, e.g., Table~\ref{tabmix} and Sec.~\ref{uml}, so that the effective magnetic field can be disregarded in many practical situations.

\begin{figure}[pth]
\includegraphics[width=0.75\linewidth,clip=]{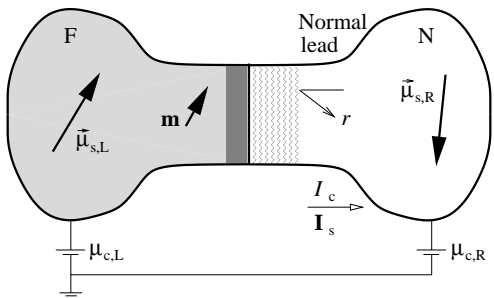}
\caption{Contact between a ferromagnetic and a normal node. The notation is analogous to Fig.~\ref{sc}. Here, the charge and spin currents in the normal lead depend on the conductance matrix $g^{\sigma\sigma^\prime}$ defined in terms of the spin-dependent reflection coefficients from the normal-metal side, as before, see Eq.~(\ref{g}). The nonequilibrium spin accumulation in the ferromagnetic node is collinear with the  magnetization $\mathbf{m}$.}
\label{scfn}
\end{figure}

For ferromagnetic films much thicker than $\lambda_{\text{sc}}$, the remaining mixing conductances $g^{\uparrow\downarrow}$ and $g^{\prime\uparrow\downarrow}$ are insensitive to scattering processes deep inside the ferromagnet (i.e., in the light-gray area of the ferromagnet in Fig.~\ref{sc}) and are determined by the scattering potential of a thin slice of the ferromagnet near the interfaces (the dark-gray areas) and eventually the normal metal (the wavy-line area). The \textit{mixing conductance is then a property of the isolated {\rm F}$\mid${\rm N} junction rather than the entire {\rm N}$\mid${\rm F}$\mid${\rm N} scatterer}. In this limit, we may introduce a ferromagnetic node at a sufficient distance from the interfaces and consider without lack of generality the two interface separately. This is allowed, since in the bulk of the ferromagnet the spin accumulation becomes a well-defined semiclassical distribution function collinear with the magnetization. An analysis entirely analogous to the one leading to Eqs.~(\ref{IcR}) and (\ref{IsR}) results in the same equations for the currents $I_c$ and $\mathbf{I}_s$ (on the normal-metal side of the right interface). Only the terms proportional to $t^{\prime\uparrow\downarrow}_r$ and $t^{\prime\uparrow\downarrow}_i$ in Eq.~(\ref{IsR}) drop out now because of the collinearity of $\boldsymbol{\mu_{s,{\rm L}}}$ and $\mathbf{m}$, left node now being assumed ferromagnetic, see Fig.~\ref{scfn} \cite{Brataas:prl00,Brataas:epjb01}. Naturally, when the F-layer thickness is much larger than $\lambda_{\text{sc}}$, the mixing conductance for the normal lead $g^{\uparrow\downarrow}$ in the F$\mid$N structure of Fig.~\ref{scfn} is the same as $g^{\uparrow\downarrow}$ for the right lead in Fig.~\ref{sc} (assuming same F$\mid$N contacts including dark-gray and wavy-line areas, at the right lead). On the other hand, the spin-up and spin-down conductances of the N$\mid$F$\mid$N structure are not identical to the conductance parameters of a single F$\mid$N interface, being dependent on two F$\mid$N junctions and the bulk F layer.

So far, we focused on the spin and charge flow through a single resistive element, i.e., a single N$\mid$F$\mid$N or F$\mid$N junction. Equations~(\ref{IcR}) and (\ref{IsR}), which can be viewed as a generalization of the Landauer-B\"{u}ttiker formulas (see, e.g., \onlinecite{Imry97}) to describe charge and spin currents in a two-terminal geometry with possibly nonequilibrium spin accumulations in the nodes. These are basic building blocks of the magnetoelectronic circuit theory. We also need the Kirchhoff laws generalized to consider the spin currents and spin accumulations on equal footing with the usual charge currents and voltage biases \cite{Brataas:prp05}.  The properties of a given device or circuit can be calculated by first prudently separating it into \textit{reservoirs}, \textit{nodes}, and \textit{resistors} (interconnectors), where the latter are the current-limiting elements. The nodes are supposed to have a negligibly small resistance and their choice may depend on the problem at hand. In a disordered multilayer, for example, it is convenient to imagine (fictitious) nodes at both sides of an interface, treating the latter as a separate resistive element (whose conductance parameters may need to be redefined, however, as explained in Sec.~\ref{icm}). Reservoirs represent the ``battery poles," i.e., large thermodynamic baths at thermal equilibrium with a constant electrochemical potential. The electrochemical potentials and spin accumulations in the nodes are not known \textit{a priori}, but determined from the generalized Kirchhoff laws based on the conservation of spin and charge. For example, disregarding spin-flip scattering, in the stationary dc state, all spin and charge flows into a normal-metal node must vanish:
\begin{equation}
\sum_lI_{c,l}=0\,,\,\,\,\sum_l\mathbf{I}_{s,l}=0\,,
\label{IcIs}
\end{equation}
summing over all leads attached to the node. The net spin flow into a ferromagnetic node projected onto its magnetization also vanishes, whereas the transverse currents are absorbed at the interface, as discussed above. We have also seen that the spin and charge currents in each lead can be calculated as a function of the distributions on adjacent nodes/reservoirs in terms of well-defined conductance parameters. The spin- and charge-current conservation laws (\ref{IcIs}) then allow computation of the circuit properties as a function of, e.g., the voltages applied to the reservoirs. The protocol for calculating the current-voltage curves may be summarized as:
\begin{enumerate}
\item Divide the circuit into resistors, nodes, and reservoirs.
\item Specify the control variables, e.g., the voltages (electrochemical potentials) applied to the reservoirs. Parametrize the electrochemical potential and spin accumulation of each node.
\item Compute the currents through the resistors as a function of the distributions in the adjacent nodes, which requires the spin-dependent and spin-mixing conductances defined earlier.
\item Use the spin- and charge-current conservation laws (\ref{IcIs}) at each node. If there is spin decoherence, use a modified continuity equation with spin relaxation.
\item Solve the resulting system of linear equations to obtain all currents as a function of the electrochemical potentials of the reservoirs.
\end{enumerate}
The charge and spin currents through the resistors, the net spin torques on the ferromagnets, and the spin accumulations anywhere in the circuit can be computed this way.

As a specific application, consider the layered spin-valve structure and its effective circuit models sketched in Fig.~\ref{sv}. The reservoirs are described by electrochemical potentials $\mu_{c,{\rm L}}$ and $\mu_{c,{\rm R}}$. We can define four nodes on the normal side of each F$\mid$N interface, with electrochemical potentials $\mu_{c,i}$ and spin accumulations $\boldsymbol{\mu}_{s,i}$. There are correspondingly three normal resistors in the problem, each described by a single real-valued conductance parameter $g$, and two magnetic resistors, corresponding to two ferromagnets, each described by two real-valued spin-dependent conductances $g^{\uparrow\uparrow}$ and $g^{\downarrow\downarrow}$ and four complex-valued spin-mixing conductances $g^{\uparrow\downarrow}$, $g^{\prime\uparrow\downarrow}$, and $t^{\uparrow\downarrow}$, $t^{\prime\uparrow\downarrow}$. In the absence of spin-orbit coupling, the charge and spin currents in each node depend on the electrochemical potentials and spin accumulations on each side of the respective ferromagnet according to Eqs.~(\ref{IcR}) and (\ref{IsR}). Transport across normal resistors is described by simpler equations that could be easily obtained by setting $g^{\sigma\sigma^\prime},~g^{\prime\sigma\sigma^\prime},~t^{\sigma\sigma^\prime},~t^{\prime\sigma\sigma^\prime}\to g$ in Eqs.~(\ref{IcR}) and (\ref{IsR}). By following the steps outlined above, these equations can be used to self-consistently determine $\mu_{c,i}$ and $\boldsymbol{\mu}_{s,i}$ and then charge and spin currents into each node, as a function of the applied bias $\mu_{c,{\rm L}}-\mu_{c,{\rm R}}$. Obviously, the same procedure can be carried out for current-biased instead of voltage-biased systems. When the conductance parameters are to be evaluated from first principles, the definitions (\ref{g}) and (\ref{t}) have to be corrected for kinetic effects when the interfaces are highly transmitting, as discussed in the next subsection. Placing of the nodes is to some extent arbitrary but different choices should not lead to contradictory results. For the system in Fig.~\ref{sv}, for example, it might be more convenient to replace nodes 2 and 3 with a single one somewhere in the middle of the normal spacer and eliminate nodes 1 and 4 altogether, while redefining the conductance matrices of the magnetic regions to include scattering on the normal-metal sides. In such a case, the entire structure would consist of two magnetic scatterers connecting each reservoir with the middle of the normal spacer. On the other hand, for sufficiently thick ferromagnets, we may introduce four additional nodes on the ferromagnetic side of the interfaces, and define pure F$\mid$N-interface resistances and ferromagnetic bulk resistances.

Circuit theory assumes momentum scattering in each node but not necessarily inelastic scattering. It is only required that the chemical-potential gradients/drops are small enough such that linear-response theory holds, and the energy dependence of the scattering matrix may be disregarded. Inelastic scattering in the nodes per definition does not affect the transport properties. It is, however, often stated that the scattering approach to transport is valid only when the orbital-dephasing length $\lambda_\phi$ is sufficiently longer than the dimensions of the scattering region. This is a relevant statement only when phase coherence is essential for the physics under consideration. In magnetoelectronic systems, we have to worry about orbital interference only at sharp interfaces. The spin-mixing conductances $g^{\uparrow\downarrow}$, $g^{\prime\uparrow\downarrow}$ and $t^{\uparrow\downarrow}$, $t^{\prime\uparrow\downarrow}$ in Eq.~(\ref{IsR}) govern transverse spin currents under the condition that $\lambda_\phi\gg\lambda_{\text{sc}}$. Also, the description of conventional spin-dependent conductances by first-principles band-structure calculations assumes wave-function coherence on atomistic length scales as well. These conditions are obviously hardly restrictive and assumed to be valid up to high (room) temperature. When spin-flip scattering is strong in diffusive regions, the spin-diffusion equation should be solved their, for which Eqs.~(\ref{IcR}) and (\ref{IsR}) provide the boundary conditions. Sec.~\ref{ds} demonstrates how this is carried out in practice. Although we illustrated here circuit theory using simple linear structures, it can be easily applied to more general multiterminal devices, such as spin transistors.

\subsection{Interfacial and thin-film conductance matrices}
\label{icm}

In the previous subsection, the electron states in the nodes were assumed to be occupied according to energy and spin, but without any regard to their momentum. Physically, this ``isotropy" in momentum space implies that net currents in the nodes may be disregarded. This is allowed only when the incoming and outgoing currents do not significantly disturb the isotropic momentum space distribution. When the contacts to the nodes are relatively small (point contacts) or highly resistive (tunnel junctions), this approximation holds. Highly-conductive metallic multilayers that are the main subject of this review, do not satisfy such a condition. The node interconnectors are then intermetallic interfaces or thin films. The isotropy of the distribution functions in the nodes can be significantly distorted by the current induced by a given voltage bias. In that limit, simple Kirchhoff laws with Landauer-B\"{u}ttiker conductances parameters, Eqs.~(\ref{g}) and (\ref{t}), do not apply. This subsection summarizes how to rescue circuit-theory by only modifying conductance parameters \cite{Schep:prb97, Bauer:prb03}.

We first illustrate the issue for a nonmagnetic metallic pillar with a uniform cross section connecting two reservoirs. Transport through a ballistic pillar is governed by the Sharvin conductance $g^{\text{Sh}}$, i.e., the number of propagating transport channels. Let us introduce $M$ interfaces (e.g., grain boundaries) in series that scramble the transverse-momentum distribution of incident electrons without any significant backscattering. The total conductance must then still amount to $g^{\text{Sh}}$, since w excluded any reflection \cite{Imry97}. If we naively carried out the circuit-theory protocol, we could assign a conductance $g_i=g^{\text{Sh}}$ to each interface and insert $M-1$ nodes between them. We would trivially obtain that the total resistance $1/g$ is the sum of the individual ones:
\begin{equation}
\frac{1}{g}=\sum_{i=1}^M\frac{1}{g_i}=\frac{M}{g^{\text{Sh}}}\,,
\end{equation}
which is obviously wrong. This breakdown of the circuit theory can be ``fixed" by renormalizing the individual interface resistance from scattering theory $1/g_i$ by subtracting the Sharvin resistance $1/g^{\text{Sh}}$:
\begin{equation}
\frac{1}{\tilde{g}_i}=\frac{1}{g_i}-\frac{1}{g^{\text{Sh}}}\,,
\end{equation}
At the same time, the contact resistances between pillar and reservoirs are assigned half of their Sharvin resistance. We then arrive at the total resistance
\begin{equation}
\frac{1}{g}=\frac{1}{2g^{\text{Sh}}}+\sum_{i=1}^M\frac{1}{\tilde{g}_i}+\frac{1}{2g^{\text{Sh}}}=\frac{1}{g^{\text{Sh}}}\,.
\end{equation}
In this way, we enforced a vanishing local voltage drop over each interface, corresponding to $\tilde{g}_i=\infty$, and a finite Sharvin resistance for the entire structure that is governed by the geometrical sample cross section. The total conductance is $g^{\text{Sh}}$, as it should.

\onlinecite{Schep:prb97} justified such ``renormalized" resistances, deriving a more general result for interfaces between different materials:
\begin{equation}
\frac{1}{\tilde{g}_i}=\frac{1}{g_i}-\frac{1}{2}\left(\frac{1}{g_{i,L}^{\text{Sh}}}+\frac{1}{g_{i,R}^{\text{Sh}}}\right)\,,
\label{schep}
\end{equation}
where $g_{i,L}^{\text{Sh}}$ and $g_{i,R}^{\text{Sh}}$ are Sharvin conductances on two sides of the interface, thus allowing to compare first-principles calculations of interface resistances with experimental results on diffuse multilayer systems that access $\tilde{g}_i$. Eq.~(\ref{schep}) requires global diffusivity that separately randomizes the momentum distributions for right- and left-moving electrons and destroys quantum interference between consecutive contacts.

The renormalization of resistances by subtracting Sharvin contributions is closely related to early controversies about resistance in mesoscopic systems. The ``bare" Landauer-B\"{u}ttiker conductance $g=\sum_{nn^\prime}|t_{nn^\prime}|^2$ is suitable for the description of two-point measurement in which reservoirs are current as well as voltage sources (see, e.g., \onlinecite{Imry97}). In an idealized four-point measurement where the voltage drop is measured directly across the scatterer (e.g., an interface), the conductance should be renormalized since there is no geometric contribution. In a single-mode quantum wire with $g^{\rm Sh}=1$, the renormalized conductance reduces to the ``old" Landauer formula $\tilde{g}=g/(1-g)$ \cite{Landauer:pm70}. When the scatterer is embedded in a diffuse environment and circuit theory applies, the situation is analogous to the four-point measurement and Eq.~(\ref{schep}) should be used. In collinear magnetic structures, Eq.~(\ref{schep}) should be applied for each spin channel separately \cite{Schep:prb97}.

The ``kinetic" corrections for magnetic structures with noncollinear magnetizations require additional thought. As a starting point we consider a single F$\mid$N interface as sketched in Fig.~\ref{scfn} with conductance matrices $g^{\sigma\sigma^{\prime}}$ and disregard spin-flip processes. In the presence of nonequilibrium currents, the effective electrochemical potentials and spin accumulations, Eqs.~(\ref{ca}) and (\ref{sa}), on either side of the interface are different for left- and right-moving modes. The discussion of Sec.~\ref{Mdc} must thus be modified to account for this asymmetry in the left- and right-moving distributions (``drift"). Like in \onlinecite{Schep:prb97}, the central underlying assumption that allows an easy implementation of this program \cite{Bauer:prb03,Bauer03} is a globally diffuse or chaotic system. In that case, the distribution of electrons incident on each interface from either side is isotropic in momentum space. Such a randomization is likely to be provided in realistic structures by interfacial disorder and bulk scattering. By extension of either the random matrix theory of \onlinecite{Waintal:prb00} or circuit theory, the analysis \cite{Bauer:prb03,Bauer03} then boils down to a set of simple rules. Eqs.~(\ref{IcR}) and (\ref{IsR}) remain unchanged but the interfacial conductances are renormalized as
\begin{equation}
\frac{1}{\tilde{g}^{\sigma\sigma^\prime}}=\frac{1}{g^{\sigma\sigma^\prime}}-\frac{1}{2}\left(\frac{1}{g_{\rm N}^{\text{Sh}}}+\frac{\delta_{\sigma\sigma^\prime}}{g_{{\rm F}\sigma}^{\text{Sh}}}\right)\,,
\label{gr}
\end{equation}
whereas the \textit{average} electrochemical potentials and spin accumulations in the nodes have to be found self-consistently as before. (We recall that the conductance parameter $t^{\prime\uparrow\downarrow}$ drops out of the discussion for a single F$\mid$N interface.) Additionally, the leads to real reservoirs must be assigned half of the respective Sharvin resistances, for each spin $\sigma$. 

The renormalization (\ref{gr}) is highly significant for understanding of the measured resistances of metallic multilayers in the CPP configuration \cite{Bass:mmm99}, as well as other measurements, by \textit{ab initio} calculations \cite{Schep:prb97,Xia:prb01,Xia:prb02,Bauer:prb03,Zwierzycki:prb05}. In Table~\ref{tabmix} we quote the theoretical results \cite{Zwierzycki:prb05} for two representative N$\mid$F material combinations: Au$\mid$Fe(001) and Cu$\mid$Co(111), the former routinely used by the Simon-Fraser group (e.g., \onlinecite{Urban:prl01,Heinrich:jap02,Heinrich:jap03}) and the latter by the Cornell group (e.g., \onlinecite{Myers:sc99,Katine:prl00}). The large difference between the $\tilde{g}$'s and $g$'s is evident, when using Eq.~(\ref{gr}), as the bare interfacial conductances are comparable with the Sharvin conductances.

\begin{table}[pth]
\begin{center}
\begin{tabular}
[c]{ccccccccc}\hline\hline
System & Interface & $g^{\uparrow\uparrow}$ & $g^{\downarrow\downarrow}$ & $g_{r}^{\uparrow\downarrow}$ & $g_{i}^{\uparrow\downarrow}$ & $g^{\text{Sh}}_{\rm N}$ & $g^{\text{Sh}}_{{\rm F}\uparrow}$ & $g^{\text{Sh}}_{{\rm F}\downarrow}$\\\hline
Au$_{\text{fcc}}$$\mid$Fe$_{\text{bcc}}$ & clean & 10.3 & 2.1 & 12.0 & 0.1 & 11.9 & 21.4 & 11.9\\
(001) & alloy & 10.1 & 4.6 & 11.9 & 0.1 &  &  & \\
Cu$_{\text{fcc}}$$\mid$Co$_{\text{fcc}}$ & clean & 10.8 & 9.8 & 14.1 & 0.4 & 15.0 & 11.9 & 27.9\\
(111) & alloy & 10.8 & 8.5 & 14.6 & -1.1 &  &  & \\\hline\hline
\end{tabular}
\end{center}
\caption{Calculated interface conductances (in units of quantum channels per nm$^2$). The results are shown for clean and disordered interfaces. The latter are modeled by two atomic monolayers of 50\% alloy. From \onlinecite{Zwierzycki:prb05}.}
\label{tabmix}
\end{table}

Eq.~(\ref{gr}) can be used only when the distribution function in the ferromagnetic layers is well defined. This is not the case anymore when the magnetic-film thickness is of the order or smaller than the transverse-spin coherence length (\ref{lsc}). In the latter situation, calculations \cite{Bauer:prb03,Bauer03} that lead to Eq.~(\ref{gr}) have to be repeated for two normal-metal layers separated by a thin magnetic film as in Fig.~\ref{sc}. We find that the basic circuit-theory equations (\ref{IcR}) and (\ref{IsR}) hold for all film thicknesses after replacing the conductance parameters for the N$\mid$F$\mid$N sandwiches with the renormalized ones, $\tilde{g}^{\sigma\sigma^\prime}$ and $\tilde{t}^{\sigma\sigma^\prime}$:
\begin{align}
\frac{1}{\tilde{g}^{\sigma\sigma}}=&\frac{1}{g^{\sigma\sigma}}-\frac{1}{2}\left(\frac{1}{g_{\rm N}^{\text{Sh}}}+\frac{1}{g_{\rm N}^{\prime\text{Sh}}}\right)\,,\nonumber\\
\frac{1}{\tilde{g}^{\uparrow\downarrow}}=&\frac{1}{g^{\uparrow\downarrow}+t^{\uparrow\downarrow}t^{\prime\uparrow\downarrow}/(2g_{\rm N}^{\prime\text{Sh}}-g^{\prime\uparrow\downarrow})}-\frac{1}{2g_{\rm N}^{\text{Sh}}}\,,\nonumber\\
\frac{1}{\tilde{t}^{\uparrow\downarrow}}=&\frac{(2g_{\rm N}^{\text{Sh}}-g^{\uparrow\downarrow})(2g_{\rm N}^{\prime\text{Sh}}-g^{\prime\uparrow\downarrow})/t^{\uparrow\downarrow}-t^{\prime\uparrow\downarrow}}{4g_{\rm N}^{\text{Sh}}g_{\rm N}^{\prime\text{Sh}}}\,,
\label{tr}
\end{align}
and the same after interchanging $g\leftrightarrow g^{\prime}$ and $t\leftrightarrow t^{\prime}$.

In the presence of weak (compared to momentum scattering) spin-flip scattering during bulk diffusion, the same renormalizations, Eqs.~(\ref{gr}), (\ref{tr}), still hold for interfacial transport. In the opposite limit of high spin-flip rates, the layers can act as ideal sinks for spin currents. In particular, in the regime of pure spin transport, such layers are fully equivalent to reservoirs, bearing all consequences and rules outlined in this subsection for the reservoirs. The regime of  intermediate-strength spin dephasing cannot be characterized by such simple statements and is omitted from our discussion.

\subsection{Time-dependent theory}
\label{tdt}

In this review we are mostly interested in the dynamic phenomena stemming from a slow variation in the magnetization direction $\mathbf{m}$ of ferromagnets that are part of an Ohmic circuitry. We focus on the adiabatic response of the itinerant carriers to the time-dependent $\mathbf{m}$ that is driven by external magnetic fields or applied (spin) currents. It turns out that the magnetoelectronic (dc) circuit theory discussed above contains already all the necessary parameters in the adiabatic regime. An adiabatic approximation is applicable when the frequency of the magnetization modulation is much smaller than the characteristic ferromagnetic exchange spin splitting, which is safely fulfilled for transition-metal--based structures. The total current is a sum of the currents induced by the bias applied via the reservoirs (including the spin-transfer torques), viz., Eqs.~(\ref{IcR}), (\ref{IsR}), and the ``pumping" component proportional to the rate of variation of the scattering potentials \cite{Buttiker:zpb94}:
\begin{align}
\label{Ict}
I_c=&I_c^{(0)}+I_c^{\text{pump}}\,,\\
\mathbf{I}_s=&\mathbf{I}_{s}^{(0)}+\mathbf{I}_{s}^{\text{pump}}\,.
\label{Ist}
\end{align}
We will make a convention to include in $I_c^{(0)}$ and $\mathbf{I}_{s}^{(0)}$ the currents driven by any charge and spin imbalance brought about by the pumping, as well as by the applied bias.

In Sec.~\ref{pump}, we derive the currents pumped by a time-dependent magnetization direction $\mathbf{m}$ into adjacent normal nodes:
\begin{align}
I_{c,{\rm R}}^{\text{pump}}=&0\,,\\
\mathbf{I}_{s,{\rm R}}^{\text{pump}}=&\frac{\hbar}{4\pi}\left(\mathcal{A}^{\uparrow\downarrow}_r\mathbf{m}\times\frac{d\mathbf{m}}{dt}+\mathcal{A}^{\uparrow\downarrow}_i\frac{d\mathbf{m}}{dt}\right)\,,
\label{Is}
\end{align}
introducing a new complex-valued parameter
\begin{equation}
\mathcal{A}^{\uparrow\downarrow}=\mathcal{A}^{\uparrow\downarrow}_r+i\mathcal{A}^{\uparrow\downarrow}_i=g^{\uparrow\downarrow}-t^{\prime\uparrow\downarrow}
\label{A}
\end{equation}
that determines the magnitude of the spin pumping as a function of device parameters \cite{Tserkovnyak:prl021}. Magnetism (or another source of spin-dependent scattering) is essential for a nonvanishing pumping parameter $\mathcal{A}^{\uparrow\downarrow}$. The spin current into the left lead is similar to Eq.~(\ref{Is}) and governed by $\mathcal{A}^{\prime\uparrow\downarrow}=g^{\prime\uparrow\downarrow}-t^{\uparrow\downarrow}$. When the ferromagnet is much thicker than the coherence length (\ref{lsc}), the quantities $t^{\uparrow\downarrow}$ and $t^{\prime\uparrow\downarrow}$ vanish and the mixing conductances $g^{\uparrow\downarrow}$ and $g^{\prime\uparrow\downarrow}$ do not depend on the thickness of the ferromagnet; the spin pumping currents originate from the interfaces. The spin pumping (\ref{Is}) thus does not depend on spin-flip processes in the ferromagnet when far from the F$\mid$N interface on the scale of the coherence length (\ref{lsc}) or, in other words, when the spin-flip scattering rate is small compared to the exchange splitting. This is typically the case in real materials. The spin (de)coherence in the attached normal conductors is crucial, however, since it affects the self-consistent reaction torque exerted on a slowly-precessing monodomain ferromagnet, as explained in Sec.~\ref{gde}. 

It can be shown that the renormalizations (\ref{tr}) introduced in the dc circuit theory for layered structures must be also applied to the mixing conductances in $\mathcal{A}^{\uparrow\downarrow}$, Eq.~(\ref{A}), for multilayers with diffuse spin backscattering, but not for contacts to spin-sink reservoirs. Using Eqs.~(\ref{tr}), we find for $\tilde{\mathcal{A}}^{\uparrow\downarrow}=\tilde{g}^{\uparrow\downarrow}-\tilde{t}^{\prime\uparrow\downarrow}$
\begin{equation}
\frac{\tilde{\mathcal{A}}^{\uparrow\downarrow}}{2g^{\text{Sh}}_{\rm N}}=\frac{g^{\uparrow\downarrow}(2g^{\prime\text{Sh}}_{\rm N}-g^{\prime\uparrow\downarrow})-t^{\uparrow\downarrow}(2g^{\prime\text{Sh}}_{\rm N}-t^{\prime\uparrow\downarrow})}{(2g^{\text{Sh}}_{\rm N}-g^{\uparrow\downarrow})(2g^{\prime\text{Sh}}_{\rm N}-g^{\prime\uparrow\downarrow})-t^{\uparrow\downarrow}t^{\prime\uparrow\downarrow}}\,,
\label{rA}
\end{equation}
and the same after interchanging primed and unprimed quantities. For mirror-symmetric structures, Eq.~(\ref{rA}) reduces to simply
\begin{equation}
\frac{1}{\tilde{\mathcal{A}}^{\uparrow\downarrow}}=\frac{1}{\mathcal{A}^{\uparrow\downarrow}}-\frac{1}{2g^{\text{Sh}}_{\rm N}}\,.
\end{equation}

The spin-pumping expression (\ref{Is}) sets the stage for several interesting developments reviewed in the following. We therefore devote the entire section \ref{pump} to derive Eq.~(\ref{Is}) and discuss its physical content, see also \onlinecite{Tserkovnyak:prl021,Tserkovnyak:prb022,Tserkovnyak:jap03}.

\section{Spin emission by coherently-precessing ferromagnets}
\label{pump}

\subsection{Parametric spin pumping}
\label{psp}

The adiabatic spin-pumping expression (\ref{Is}) can be derived by time-dependent scattering theory \cite{Tserkovnyak:prl021}. Consider an N$\mid$F$\mid$N junction as in Fig.~\ref{sc}, where the two normal nodes are now assumed to be large reservoirs in a common thermal equilibrium. Without a voltage bias, no spin or charge currents flow when the magnetization of the ferromagnet is static. When it is moving, however, the time dependence of the scattering matrix in spin space can induce nonequilibrium spin currents in the nonmagnetic leads. The current $\hat{I}(t)$ pumped by the precession of the magnetization into the right and left paramagnetic reservoirs can be calculated in the adiabatic approximation, since typical precession frequencies are several orders of magnitude smaller than the ferromagnetic exchange field that sets the relevant energy scale for spin-dependent transport. The adiabatic charge-current response in nonmagnetic systems has been derived by \onlinecite{Buttiker:zpb94}. The generalization to the $2\times2$ spin- and charge-current matrix (\ref{Ii}) by \onlinecite{Tserkovnyak:prl021,Tserkovnyak:prb022} is explained in the following.

The $2\times2$ current operator $\hat{I}_l$ for the $l$th lead ($l={\rm L,~R}$) is in general given by Eq.~(\ref{I}). When the scattering matrix $s_{nn^\prime,ll^{\prime }}^{\sigma\sigma^\prime}(t)$ of the ferromagnetic layer varies slowly compared to the relevant microscopic time scales of the system, an adiabatic approximation may be used, meaning that the energy of the scattered states is assumed to be weakly modulated with respect to the energy of the incoming states by the oscillating part of the scattering matrix: The state annihilated by $a^{\sigma^\prime}_{n^\prime,l^\prime}(\varepsilon,t)=a^{\sigma^\prime}_{n^\prime,l^\prime}(\varepsilon)e^{-i\varepsilon t/\hbar}$ is partitioned into states in the $m$th channel of lead $l^\prime$ with energies determined by the time dependence of $s_{nn^\prime,ll^{\prime}}^{\sigma\sigma^\prime}(\varepsilon,t)a^{\sigma^\prime}_{n^\prime,l^{\prime}}(\varepsilon,t)$, see \onlinecite{Buttiker:zpb94}. The scattering amplitude at a given energy shift is determined by the Fourier transform of $s_{nn^\prime,ll^{\prime}}^{\sigma\sigma^\prime}(\varepsilon,t)$ in time space. The expectation value of the current operator $\hat{I}_l$ is evaluated similarly to the dc limit discussed in Sec.~\ref{Mdc}. When the scattering matrix depends on the real-valued parameter $X(t)$, the Fourier transform of the current expectation value $\hat{I}_{l}(\omega)=\int dte^{i\omega t}\hat{I}_l(t)$ can be written as
\begin{equation}
\hat{I}_{l}(\omega )=\hat{g}_{X,l}(\omega )X(\omega)
\label{Il}
\end{equation}
in terms of the frequency $\omega$ and $X$-dependent parameter $\hat{g}_{X,l}$:
\begin{align}
\hat{g}_{X,l}(\omega)=&\frac{e\omega }{4\pi}\int d\varepsilon\frac{d f_{\text{FD}}(\varepsilon)}{d\varepsilon}\sum_{nn^\prime l^\prime}\frac{\partial\hat{s}_{nn^\prime,ll^{\prime}}(\varepsilon)}{\partial X}\hat{s}_{nn^\prime,ll^{\prime}}^{\dagger}(\varepsilon)\nonumber\\
&-\text{H.c.}\,.
\label{gXl}
\end{align}
Eq.~(\ref{Il}) is the first-order (in frequency) correction to the dc theory of Sec.~\ref{Mdc}. At sufficiently low temperatures, one can approximate $-\partial_\varepsilon f_{\text{FD}}(\varepsilon)$ by a $\delta$ function centered at the Fermi energy. The expectation value of the $2\times2$ particle-number operator $\hat{Q}_{l}(\omega )$ [defined by $\hat{I}_{l}(t)=d\hat{Q}_{l}(t)/dt$ in time or by $\hat{I}_{l}(\omega )=-i\omega \hat{Q}_{l}(\omega )$ in frequency domain] for the $l$th reservoir is then given by 
\begin{equation}
i\hat{Q}_{l}(\omega )=\left( \frac{e}{4\pi}\sum_{nn^\prime l^{\prime }}\frac{\partial \hat{s}_{nn^\prime,ll^{\prime }}}{\partial X}\hat{s}_{nn^\prime,ll^{\prime}}^{\dagger }-\text{H.c.}\right) X(\omega )\,,
\label{Q}
\end{equation}
where the scattering matrix is evaluated at the Fermi energy. Because the prefactor on the right-hand side of Eq.~(\ref{Q}) does not depend on frequency $\omega $, the equation is also valid in time domain. The change in particle number $\delta\hat{Q}_l(t)$ is thus proportional to the modulation $\delta X(t)$ of parameter $X$ and the $2\times 2$ matrix current reads 
\begin{equation}
\hat{I}_{l}(t)=e\frac{d\hat{n}_{l}}{dX}\frac{dX(t)}{dt}\,,
\label{Ip}
\end{equation}
introducing the matrix ``emissivity" into lead $l$:
\begin{equation}
\frac{d\hat{n}_{l}}{dX}=\left(\frac{1}{4\pi i}\sum_{nn^\prime l^{\prime}}\frac{\partial\hat{s}_{nn^\prime,ll^{\prime}}}{\partial X}\hat{s}_{nn^\prime,ll^{\prime}}^{\dagger}\right)+\text{H.c.}\,.
\label{em}
\end{equation}
The right-hand side of Eq.~(\ref{Ip}) should be in general summed over all parameters, if any, that modulate the scattering matrix. When the spin-flip scattering is disregarded, the scattering matrix $\hat{s}$ can be written in terms of the spin-up and spin-down scattering coefficients $s^{\uparrow(\downarrow)}$ using the projection matrices, see Eqs.~(\ref{s}) and (\ref{u}). The spin current pumped by a magnetization precession is obtained by identifying $X(t)=\varphi(t)$, where $\varphi$ is the azimuthal angle of the magnetization direction in the plane perpendicular to the instantaneous rotation axis. For example, suppose the magnetization rotates around the $z$ axis: $\mathbf{m}=(\cos\varphi,\sin\varphi,0)$. Using Eq.~(\ref{s}), it is then easy to calculate the emissivity (\ref{em}) for this process: 
\begin{equation}
\frac{d\hat{n}_{\rm R}}{d\varphi}=\frac{1}{4\pi}\left[\mathcal{A}^{\uparrow\downarrow}_i(\sigma_x\sin\varphi-\sigma_y\cos\varphi)-\mathcal{A}^{\uparrow\downarrow}_r\sigma_z\right]\,.
\label{nl}
\end{equation}
Inserting this into Eq.~(\ref{Ip}) and comparing the result with the definition (\ref{Ii}), one finally finds that the charge current vanishes, $I_{c,{\rm R}}^{\text{pump}}=0$, and the spin current
\begin{equation}
\mathbf{I}_{s,{\rm R}}^{\text{pump}}=\left(-\mathcal{A}^{\uparrow\downarrow}_i\sin\varphi,\mathcal{A}^{\uparrow\downarrow}_i\cos\varphi,\mathcal{A}^{\uparrow\downarrow}_r\right)\frac{\hbar}{4\pi}\frac{d\varphi}{dt}
\end{equation}
can be rewritten as Eq.~(\ref{Is}). Since the spin current transforms as a vector in spin space, Eq.~(\ref{Is}) is also valid in the case of a general (slow) motion of the magnetization direction.

Even though the mathematics of the scattering approach to adiabatic spin pumping is entirely analogous to the charge-pumping theory developed by \onlinecite{Buttiker:zpb94,Brouwer:prb98}, there is an important difference in the physics. In the case of a nonmagnetic scattering region, the average pumped charge current has the same direction in the two leads, because charge cannot be accumulated or depleted for long; the net charge entering the scattering region through one lead must leave it via the other within a period of the external-gate modulation. On the other hand, the total conduction-electron spin angular momentum does not have to be conserved since the magnetization can act as a sink or a source. The preceding analysis shows that a precessing ferromagnet polarizes adjacent nonmagnetic conductors. The phenomenon can be called as well a ``spin well" or ``spin fountain" rather than spin pump: An excited ferromagnet emits spins into all adjacent conductors, spending its ``own" angular momentum rather than pumping spins from one lead into the other. If the lost angular momentum is not replenished by an external magnetic field, the precession invariably will die out.

Per revolution, the precession pumps an angular momentum into an adjacent normal-metal layer which is proportional to $\mathcal{A}^{\uparrow\downarrow}_r$, in the direction of the precession axis. At first sight, it may be surprising that a net dc pumping can be achieved by a single parameter varying in time, whereas it can be shown \cite{Brouwer:prb98} that at least two periodic parameters are required for that. However, there are actually two periodic parameters (out of phase by $\pi/2$) hidden behind $\varphi(t)$, viz., the projections of the unit vector defined by $\varphi$ in the plane perpendicular to the axis of precession.

\subsection{Rotating-frame analysis}
\label{rfa}

It turns out that Eqs.~(\ref{Is}), (\ref {A}) for the nonequilibrium spin currents pumped out of a precessing ferromagnet can be understood as an equilibrium property by a transformation to a rotating frame. We prove this for a ferromagnetic film sandwiched between two normal-metal nodes at a common thermal equilibrium, $\mu_{c,{\rm L}}=\mu_{c,{\rm R}}$, as shown in Fig.~\ref{sc}. Let
\begin{equation}
H(t)=H_0+H^\prime(t)
\label{Ht}
\end{equation}
be the Hamiltonian experienced by conduction electrons, where $H_0$ is the sum of the kinetic and potential energies and $H^\prime(t)$ is a time-dependent exchange Hamiltonian in the ferromagnet:
\begin{equation}
H^\prime(t)=\int d\mathbf{r}V_x(\mathbf{r})\sum_{\sigma\sigma^\prime}\Psi_\sigma^\dagger(\mathbf{r})\left[\boldsymbol{\sigma}^{\sigma\sigma^\prime}\cdot\mathbf{m}(\mathbf{r},t)\right]\Psi_{\sigma^\prime}(\mathbf{r})\,,
\label{Hf}
\end{equation}
where $V_x$ is the local exchange interaction along the magnetization direction $\mathbf{m} (\mathbf{r},t )=\mathbf{m} (t)$ that is assumed to be uniform in the ferromagnet. $\Psi_\sigma$ is the spin-$\sigma$ electron field operator.

At time $t<0$, the entire system is in a common--thermal-equilibrium state corresponding to a time-independent magnetization direction $\mathbf{m}(0)$. At $t=0$, $\mathbf{m}(t)$ starts to rotate with frequency $\omega$ about an axis denoted $\mathbf{\hat{z}}$ that is perpendicular to $\mathbf{m}$, $\mathbf{\dot{m}} = \omega \mathbf{\hat{z}}\times\mathbf{m}$. The time-dependent magnetization drives the electron system out of equilibrium.

Let us introduce a (many-body) state $\Phi^\prime(t)$ of the electrons in the rotating frame of reference for the spin variables, which is related to the solution of the Schr\"{o}dinger equation $i\hbar\dot{\Phi}(t)=H(t)\Phi(t)$ by the unitary transformation $\Phi(t)=U(t)\Phi^\prime(t)$. Here, $U=\exp(-iS_z\omega t/\hbar)$ is a spin-rotation operator in terms of the total-spin z-axis projection $S_z$. Since $H^\prime(t)=U(t)H^\prime(0)U^\dagger(t)$, $\Phi^\prime(t)$ satisfies a Schr\"{o}dinger equation \cite{Slichter90}:
\begin{equation}
i\hbar\frac{d\Phi^\prime(t)}{dt}=\left[H(0)-S_z\omega\right]\Phi^\prime(t)\,,
\label{phi}
\end{equation}
provided $[H_0,S_z]=0$, i.e., spin-orbit coupling or other spin-flip processes are disregarded. This is the equation of motion for electrons subjected to a static magnetic field $\omega/\gamma$ in the $z$ direction, in addition to a static exchange interaction (\ref{Hf}) along $\mathbf{m}(0)$ in the ferromagnet. A steady-state solution for the system in the rotating frame is thus characterized by spin-polarized normal-metal nodes with $\langle s_z\rangle=\chi\omega/\gamma$ and $\langle s_x\rangle=\langle s_y\rangle=0$, where $\chi$ is their static (isotropic) spin susceptibility.

In the laboratory frame, the spin imbalance must be identical to the polarization in the rotating frame along the $z$ axis. Since there is no magnetic field in the laboratory frame, this spin imbalance can only be accounted for by a nonequilibrium spin accumulation, i.e., difference between the chemical potentials for spins parallel and antiparallel to $\mathbf{\hat{z}}$, of magnitude
\begin{equation}
\mu_s=\hbar\omega\,.
\label{mu0}
\end{equation}
According to the magnetoelectronic dc circuit theory, the spin accumulation in turn drives a spin current
\begin{equation}
\mathbf{I}_{s,{\rm R}}^{\text{back}}=\frac{1}{4\pi}\left(\mathcal{A}^{\uparrow\downarrow}_r\boldsymbol{\mu}_s+\mathcal{A}^{\uparrow\downarrow}_i\boldsymbol{\mu}_s\times\mathbf{m}\right)
\label{II}
\end{equation}
from a normal-metal node back into the ferromagnet, where the coefficients $\mathcal{A}^{\uparrow\downarrow}$ have been defined in Eq.~(\ref{A}). Here, we used Eq.~(\ref{IsR}) with $\mu_{c,{\rm L}}-\mu_{c,{\rm R}}=0$ and $\boldsymbol{\mu}_{s,{\rm L}}=\boldsymbol{\mu}_{s,{\rm R}}=\boldsymbol{\mu}_s$. According to Eq.~(\ref{mu0}), $\boldsymbol{\mu}_s=\hbar\omega\mathbf{\hat{z}}=\hbar\mathbf{m}\times\mathbf{\dot{m}}$. Since in the steady state the backflow spin current (\ref{II}) must cancel the spin current pumped by the ferromagnet into the normal metal, above arguments thus provide an alternative derivation of the spin pumping (\ref{Is}). For finite-size nodes and slow rotation, equilibrium may be established in the rotating frame in a time short compared to the period of rotation. In the adiabatic limit, the equality of the spin currents into and out of the ferromagnet thus holds instantaneously, and not just over the average over a period. The assumption of a fixed rotation axis is also not essential, as long as this axis moves slowly in time. In the limit of vanishing exchange splitting, $\mathcal{A}^{\uparrow\downarrow}=0$ and the spin current (\ref{Is}) vanishes, as it should.

\begin{figure}[pth]
\includegraphics[width=0.8\linewidth,clip=]{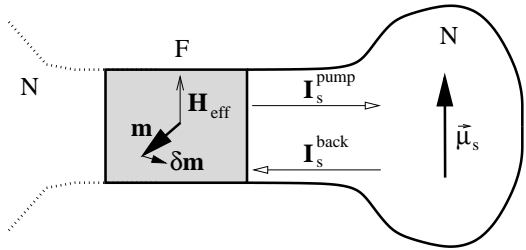}
\caption{The ferromagnetic magnetization $\mathbf{m}$ rotates around the effective field $\mathbf{H}_{\text{eff}}$. In the steady state, the spin pumping $\mathbf{I}_s^{\text{pump}}$ cancels the spin backflow $\mathbf{I}_s^{\text{back}}$ that accompanies the nonequilibrium spin accumulation $\boldsymbol{\mu}_s=\hbar\gamma\mathbf{H}_{\text{eff}}$.}\label{res}
\end{figure}

It is instructive to scrutinize the energy and spin--angular-momentum conservation laws of the combined F$\mid$N system in the laboratory frame of reference \cite{Tserkovnyak:prb022}. Suppose the rotation axis $z$ is oriented along the static effective magnetic field $\mathbf{H}_{\text{eff}}$, and let us focus on the nonequilibrium dynamics in the normal metal node, see Fig.~\ref{res}. After starting the magnetization rotation at $t=0$, net spin currents flow into the normal mode in a finite time interval before reaching the steady state. We wish to account for the energy and angular-momentum transfer between the ferromagnet and normal metal in this transient regime. If the normal node is sufficiently small, the steady state is reached after a vanishingly small transfer of spins, with only little effect on the ferromagnet. $N_s$ spins oriented along the $z$ axis transferred into the normal metal correspond to an excess energy $\Delta E_N=N_s\mu_{s}/2$ and angular momentum $\Delta L_N=N_s\hbar/2$. The conservation laws dictate $\Delta E_F=-\Delta E_N$ and $\Delta L_F=-\Delta L_N$ for the ferromagnet. Using Eq.~(\ref{Heff}), the magnetic energy $\Delta E_F=\gamma\Delta L_FH_{\text{eff}}$ and we find $N_s\mu_{s}/2=\gamma N_s(\hbar/2)H_{\text{eff}}$. It follows that $\mu_{s}=\hbar\gamma H_{\text{eff}}=\hbar\omega$, where $\omega=\gamma H_{\text{eff}}$ is the Larmor precession frequency in the effective field. Using energy and momentum conservation, we thus derived a nonequilibrium spin accumulation that agrees with the result of the rotating-frame analysis.

\subsection{FMR-operated spin battery}
\label{sb}

In the preceding subsection, we showed that precessing ferromagnets inject a spin current into adjacent conductors via Ohmic contacts. In this subsection, we discuss how this opens the way to create a pure spin source (``spin battery") by ferromagnetic resonance (FMR) and estimate the spin current and spin bias for different material combinations \cite{Brataas:prb02}. The spin source can be realized as a ferromagnetic layer at resonance with an rf field. Pure spin-current injection into low-density conductors should allow experimental studies of spintronic phenomena in mesoscopic, ballistic, and nanoscale systems. The combination of a ferromagnet at the FMR in Ohmic contact with a conductor can be interpreted as a spin battery, with analogies and differences with charge batteries. For example, charge-current conservation dictates that a charge battery has two poles, plus and minus. A spin battery requires only one pole, since the spin current does not need to be conserved. Furthermore, the polarity is not a binary, but a three-dimensional vector. 

Devices made from metallic layered systems displaying the giant \cite{Baibich:prl88} and tunnel \cite{Miyazaki:mmm95,Moodera:prl95} magnetoresistance have been proven useful for read-head sensors and magnetic random-access memories. Integration of such devices with semiconductor electronics is desirable but difficult because a large conductivity mismatch between magnetic and normal materials is detrimental to spin injection \cite{Schmidt:prb00}. Spin injection into bulk semiconductors has so far been reported only in optical pump-and-probe experiments \cite{Kikkawa:nat99} and with high-resistance ferromagnetic injectors \cite{Fiederling:nat99,Ohno:nat99} or Schottky/tunnel barriers \cite{Monsma:sc98,Zhu:prl01}. In these cases, the injected spin-polarized carriers are hot and currents are small, however. Desirable are semiconductor devices with an efficient all-electrical cold-electron spin injection and detection via Ohmic contacts at the Fermi energy, just as has been realized by \onlinecite{Jedema:nat01} for metallic devices. The spin battery discussed in this subsection is an alternative conceptual approach to accomplish such electrical spin injection.

In the absence of spin-orbit scattering, the spin-battery idea can be understood in terms of the rotating-frame analysis of the previous subsection. A magnetization undergoing a circular precession around the $z$ axis with frequency $\omega$ induces a spin-imbalance density $\mathbf{s}=\chi\omega\mathbf{\hat{z}}/\gamma$ in the normal metal or semiconductor adjacent to the ferromagnet. The spin susceptibility $\chi$ includes possible effects of electron-electron interactions in the normal metal/semiconductor. The chemical-potential difference (\ref{mu0}) between spin-up and spin-down electrons along the $z$ axis, on the other hand, is universal for the FMR-operated spin battery with vanishing spin-orbit coupling. Spin-orbit scattering limits the efficiency of the spin battery in real structures, as discussed in the following.

The important parameters of a charge battery are the maximum voltage in the absence of a load, as well as the maximum charge current that can be drawn from it. In the following we discuss estimates for the analogous characteristics of the spin battery, closely following \onlinecite{Brataas:prb02}. To this end, one should solve the dynamic problem of spin pumping (\ref{Is}) at the F$\mid$N contact as a boundary condition for the spin diffusion equations in the normal conductor. When the ferromagnet is thicker than the ferromagnetic coherence length (\ref{lsc}) (a few \AA ngstr\o ms in transition metals such as Co, Ni or Fe), the spin current (\ref{Is}) emitted into the normal conductor is determined by the mixing conductance $g^{\uparrow\downarrow}$, since $t^{\uparrow\downarrow}$ vanishes. With some exceptions such as ferromagnetic insulators \cite{Huertas:prl02}, the imaginary part of the mixing conductance is small \cite{Xia:prb02}, see also Table~\ref{tabmix}, and will be disregarded in the following. The spin current emitted into the normal metal is then
\begin{equation}
\mathbf{I}_{s}^{\text{pump}}=\frac{\hbar}{4\pi}g_r^{\uparrow\downarrow}\mathbf{m}\times\frac{d\mathbf{m}}{dt}\,.
\label{spinemission}
\end{equation}
When the spin current (\ref{spinemission}) flows freely into the normal metal, the corresponding loss of angular momentum increases the (Gilbert) damping of the magnetization dynamics, as discussed in detail in the next section. Eq.~(\ref{spinemission}) is the maximum spin current that can be drawn from the spin battery. The rotating-frame analysis indicates that a nonequilibrium spin accumulation builds up in the normal conductor when the spin-flip relaxation rate is smaller than the spin-injection rate. This in turn induces a ``backflow" spin current $\mathbf{I}_{s}^{(0)}$ that opposes the total spin current $\mathbf{I}_{s}=\mathbf{I}_s^{\text{pump}}+\mathbf{I}_s^{(0)}$. The component of the backflow spin current $\mathbf{I}_{s}^{(0)}$ parallel to the instantaneous magnetization direction $\mathbf{m}$ is canceled by an opposite flow from the ferromagnet when the FMR frequency and spin-flip scattering rate in the ferromagnet are much smaller than the characteristic spin-injection rate into the ferromagnet. The surviving component of $\mathbf{I}_{s}^{(0)}$, Eq.~(\ref{IsR}), is perpendicular to $\mathbf{m}$:
\begin{equation}
\mathbf{I}_{s}^{(0)}=-\frac{g^{\uparrow\downarrow}_r}{4\pi}\mathbf{m}\times\boldsymbol{\mu}_s\times\mathbf{m}\,.
\label{spinback}
\end{equation}
We note that the mixing conductance in Eqs.~(\ref{spinemission}) and ~(\ref{spinback}) ought to be renormalized in layered structures, as discussed in Sec.~\ref{icm}.

The relation between the spin accumulation $\boldsymbol{\mu}_s$ and the total spin current $\mathbf{I}_{s}$ in a normal diffuse conductor is governed by the spin-diffusion equation, see Sec.~\ref{ds}. Let us consider an F$\mid$N bilayer structure with cross section $S$ and thickness $L$, diffusion coefficient $D$ and characteristic spin-flip time $\tau_{\text{sf}}$ for the N-layer. The solution for the spin accumulation is simple when $\sqrt{D/\omega}\ll L\ll\sqrt{D\tau_{\text{sf}}}=\lambda_{\text{sd}}$ , requiring $\omega\gg\tau_{\text{sf}}^{-1}$ \cite{Brataas:prb02}. The spin accumulation in the normal layer is then nearly uniform and time-independent for a steady ferromagnetic precession cone with angle with angle $\theta$:
\begin{equation}
\boldsymbol{\mu}_s=\hbar\omega\frac{\sin^2\theta}{\sin^2\theta+\eta}\mathbf{\hat{z}}\,.
\label{mutheta}
\end{equation}
$\eta=\tau_{\text{i}}/\tau_{\text{sf}}$ is a reduction factor expressed in terms of the ``injection rate" $\tau_{\text{i}}^{-1}$. When the conductance of a normal-layer slab of thickness $\sqrt{D/\omega}$ is much larger than the contact conductance $g_r^{\uparrow\downarrow}$, $\tau_{\text{i}}^{-1}=g_r^{\uparrow\downarrow}/(h\mathcal{N}SL)$, where $\mathcal{N}$ is the normal-metal one-spin density of states per unit volume. In the opposite limit,  $\tau_{\text{i}}^{-1}=\sqrt{D\omega}/L$. The former regime is relevant for an opaque interface, such as a Schottky or tunnel barrier. In the latter regime, $\eta\ll1$ since we have assumed $\omega\tau_{\text{sf}}\gg1$ and $L\ll\sqrt{D\tau_{\text{sf}}}$. Large systems have a smaller injection rate since more states have to be filled. When $\tau_{\text{sf}}\rightarrow\infty$, $\eta=0$ and Eq.~(\ref{mutheta}) reduces to the result of the rotating-frame analysis (\ref{mu0}). \onlinecite{Schmidt:prb00} realized that efficient spin injection into semiconductors by Ohmic contacts is difficult with transition-metal ferromagnets since virtually all of the applied potential drops over the nonmagnetic part and is unavailable for spin injection. The present mechanism of spin injection is current- rather then bias-driven and thus does not suffer from this conductivity-mismatch problem.

When the spin-relaxation rate is smaller than the spin-injection rate and the precession-cone angle is sufficiently large, $\sin^2\theta>\eta$, the spin bias (\ref{mutheta}) saturates at its maximum value (\ref{mu0}), that does not depend on the material parameters. At resonance, $\sin\theta=h_{\text{rf}}/(\alpha H_{\text{eff}})$ (assuming for simplicity a constant effective field $\mathbf{H}_{\text{eff}}$). For a dc field of $H_{\text{eff}}=1$~T, rf field $h_{\text{rf}}=1$~mT and damping $\alpha=0.01$, $\sin^2\theta=0.01$. In order to realize a battery with maximum spin bias (\ref{mu0}), a suppression factor $\eta\lesssim0.01$ is thus required. Epitaxially-grown clean samples with low spin-flip rates are needed for spin batteries operating at small precession angles. The precession-cone angle $\theta$ in FMR experiments is typically small, but large angles could in principle be achieved by sufficiently intense rf fields and a soft ferromagnet such as permalloy or, for transient dynamics, by transverse magnetic-field pulses. 

The dc component of the maximum spin-current bias (\ref{spinemission}) is suppressed by $\sin^2\theta$ upon averaging over one precession period. For operation as a spin current source, a large-angle precession is required as well. The current can be increased by a larger F$\mid$N interface area. 

A potential practical problem of large-angle dynamics is the FMR energy dissipation. For a total spin $S_F=M_sV/\gamma$ of a ferromagnet of volume $V$, dissipation power $P=\alpha S_F\omega^2\sin^2\theta$ has to be efficiently drained in order to avoid excessive heating. On the other hand, possible undesirable spin-precession and energy generation in the nonmagnetic parts of the system is of no concern for material combinations with different $g$ factors, as, e.g., Fe ($g=2.1$) and GaAs ($g=-0.4$), or when the magnetic anisotropy shifts the resonance frequency with respect to electrons in the normal metal.

Purified elemental metals such as Al and Cu are suitable materials, since their spin-diffusion lengths are very long: In Cu, $\lambda_{\text{sd}}\sim1~\mu$m at low temperatures that is reduced by a factor of 3 at room temperature \cite{Jedema:nat01}, and similar numbers hold for Al \cite{Jedema:nat02}. Indirect proof of a spin accumulation pumped into Cu is provided by the FMR measurements of permalloy$\mid$Cu$\mid$Pt structures \cite{Mizukami:mmm02,Mizukami:prb02} discussed in Sec.~\ref{ds}. Semiconductors have the advantage of a larger ratio of spin bias to Fermi energy. In lightly $n$-doped GaAs, the spin-flip relaxation time can be very long: $\tau_{\text{s}}=10^{-7}~\text{s}$ at $n=10^{16}$~cm$^{-3}$ carrier density \cite{Kikkawa:prl98}. This favorable number is offset by the difficulty to form Ohmic contacts to GaAs, however. Large Schottky barriers exponentially suppress the interface mixing conductance. InAs has the advantage of a natural charge accumulation layer at the surface that avoids Schottky barriers when covered by high-density metals. However, the spin-orbit interaction in a narrow-gap semiconductor such as InAs is substantial, which reduces $\tau_{\text{sf}}$. In asymmetric-confinement structures, the spin-flip relaxation rate is governed by the Rashba-type spin-orbit interaction that vanishes in symmetric quantum wells \cite{Nitta:prl97,Engels:prb97}. The remaining D'yakonov-Perel scattering rate \cite{Dyakonov:jetp71} is reduced in narrow quasi-one-dimensional channels of width $w$ due to waveguide diffusion modes by a factor of $(L_s/w)^2$, where $L_s$ is the spin-precession length \cite{Malshukov:prb00}, which makes InAs and its alloys potentially interesting materials for a spin battery as well. In pure Si, the spin-flip relaxation time should be very long, since spin-orbit interaction is weak. Furthermore, the possibility of heavy doping allows control of Schottky barriers. Si therefore also appears to be a good candidate for spin injection into semiconductors.

The FMR-generated spin bias can be detected noninvasively via a ferromagnet that is connected to the normal metal by a tunnel junction. A voltage difference $p\mu_s$ should be measurable between parallel and antiparallel configurations of the analyzing magnetization with respect to the spin accumulation in the normal metal. Here $p=(G_{\uparrow}-G_{\downarrow})/(G_{\uparrow}+G_{\downarrow})$ is the relative polarization of the spin-dependent tunnel conductance $G^\sigma$ of the contact. For a fixed analyzing magnetization, the same voltage difference can be generated by reversing the static magnetic field. The spin current, on the other hand, can be measured via the drop of spin bias over a known resistive element with weak spin-flip scattering.

Spin pumping into the normal metal can also have consequences for the nuclei via the hyperfine interaction between electrons and nuclear spins \cite{Kawakami:sc01}. An initially unpolarized collection of nuclear spins can be oriented by a spin-polarized electron current, which transfers part of its angular momentum by spin-flop scattering due to the hyperfine interaction. A ferromagnetically-ordered nuclear-spin system, in turn, is felt by the electrons as the Overhauser magnetic field which induces an equilibrium spin density $\mathbf{s}_0$ in the normal metal \cite{Overhauser:pr53}. The spin-flop scattering can be included in electron-spin dynamics by adding an additional term to the total spin current, $\mathbf{I}_s=\mathbf{I}_s^{\text{pump}}+\mathbf{I}_s^{(0)}+\mathbf{I}_s^{\text{nuc}}$, to account for the exchange of angular momentum between electrons and nuclei \cite{Brataas:prb02}. The nuclear spin polarization increases with the spin accumulation and decreasing temperature. In bulk GaAs, the nuclear magnetic field is $H_n=5.3$~T when the nuclei are fully spin polarized \cite{Paget:prb77}, which should occur at thermal energies sufficiently  smaller than the FMR frequency, as can be understood from the rotating-frame analysis.

In conclusion, the spin battery is a source of spin, just as a conventional battery is a source of charge. Estimates of its performance for bilayers of metallic ferromagnets with either normal metals or doped semiconductors suggest that it is a feasible concept.

\section{Gilbert-damping enhancement}
\label{gde}

\subsection{Ideal spin sinks}
\label{pss}

In section~\ref{pump}, we showed that a moving magnetization emits spins into adjacent nonmagnetic conductors. This effect is necessarily associated with an energy loss for the ferromagnet, as explained in Sec.~\ref{rfa}. We now discuss spin pumping as a source of viscous damping of the magnetization dynamics in thin films or small particles. Under quite general conditions, this damping is consistent with the Gilbert phenomenology (\ref{llg}).

Consider a ferromagnet sandwiched between two normal metals labeled by $l={\rm L,~R}$. By conservation of angular momentum, spins ejected out of the ferromagnet exert a transverse reaction torque (\ref{spintorque}). The total spin current
\begin{equation}
\mathbf{I}_{s}=\sum_{l} ( \mathbf{I}_{s,l}^{(0)}+\mathbf{I}_{s,l}^{\text{pump}})
\label{Ipb}
\end{equation}
that determines the spin torque, i.e., the additional term (\ref{Gllg}) in the LLG equation, contains bias- and spin-pumping--induced contributions. Suppose initially that the normal metals act as ideal reservoirs in a common thermal equilibrium that are perfect spin sinks for the pumped spin currents, so that $\mathbf{I}_{s,l}^{(0)}=0$. This model is valid when the spin current $\mathbf{I}_{s,l}^{\text{pump}}$ is completely drained by massive and highly conductive reservoirs or by a material with effective spin-flip processes that prevent any spin-accumulation build-up and backscattering into the ferromagnet. Using Eq.~(\ref{Is}),
\begin{align}
\mathbf{I}_{s}=&\sum_{l}\mathbf{I}_{s,l}^{\text{pump}}=\frac{\hbar}{4\pi}\left[\left(\mathcal{A}^{\uparrow\downarrow}_r+\mathcal{A}^{\prime\uparrow\downarrow}_r\right)\mathbf{m}\times\frac{d\mathbf{m}}{dt}\right.\nonumber\\
&\left.+\left(\mathcal{A}^{\uparrow\downarrow}_i+\mathcal{A}^{\prime\uparrow\downarrow}_i\right)\frac{d\mathbf{m}}{dt}\right]\,.
\end{align}
Since this spin current is polarized perpendicularly to $\mathbf{m}$, the torque (\ref{spintorque}) becomes simply $\boldsymbol{\tau}=-\mathbf{m}\times\mathbf{I}_s\times\mathbf{m}=-\mathbf{I}_s$. Substituting this into equation (\ref{Gllg}), we recover the LLG equation (\ref{llg}) with a new effective damping constant, $\alpha _{\text{eff}}$, and gyromagnetic ratio, $\gamma_{\text{eff}}$, defined by
\begin{align}
\frac{\gamma}{\gamma_{\text{eff}}}=&1-\frac{\hbar\gamma}{4\pi M_sV}\left(\mathcal{A}_i^{\uparrow\downarrow}+\mathcal{A}_i^{\prime\uparrow\downarrow}\right)\,,\label{geff}\\
\alpha_{\text{eff}}\frac{\gamma}{\gamma_{\text{eff}}}=&\alpha+\frac{\hbar\gamma}{4\pi M_sV}\left(\mathcal{A}_r^{\uparrow\downarrow}+\mathcal{A}_r^{\prime\uparrow\downarrow}\right)\,.
\label{aeff}
\end{align}
The spin pumping thus affects FMR experiments as a shift of the resonance magnetic field via $\mathcal{A}_i^{\uparrow\downarrow}+\mathcal{A}_i^{\prime\uparrow\downarrow}$, whereas $\mathcal{A}_r^{\uparrow\downarrow}+\mathcal{A}_r^{\prime\uparrow\downarrow}$ increases the relative resonance linewidth. $\mathcal{A}_r^{\uparrow\downarrow}$ defined by Eq.~(\ref{A}) is positive, since, by unitarity of the scattering matrix, it can be rewritten as
\begin{equation}
\mathcal{A}_r^{\uparrow\downarrow}=\frac{1}{2}\sum_{nn^\prime}\left(\mid r_{nn^\prime}^\uparrow-r_{nn^\prime}^\downarrow\mid^2+\mid t_{nn^\prime}^{\prime\uparrow}-t_{nn^\prime}^{\prime\downarrow}\mid^2\right).
\end{equation}
The spin pumping by a moving magnetization therefore cannot reverse the sign of the effective damping parameter (without reversing the sign of the gyromagnetic ratio), as is required by the LLG phenomenology, see Sec.~\ref{llgt}.

Enhanced Gilbert damping leads to a broader resonance absorption peak (\ref{P}) of an rf magnetic field. By the fluctuation-dissipation theorem, this should be manifest in increased fluctuations of the magnetization in thermodynamic equilibrium. This additional magnetization noise can arise from the torques exerted by Johnson-Nyquist spin-current fluctuations that are exchanged between the ferromagnet and reservoirs or other spin sinks. Indeed, \onlinecite{Foros:prl05} proved explicitly that magnetic noise caused by the spin-current fluctuations is consistent with the dissipation power predicted by the spin-pumping theory for a monodomain ferromagnet in contact with ideal spin sinks.

\begin{figure}[ptb]
\includegraphics[width=0.95\linewidth,clip=]{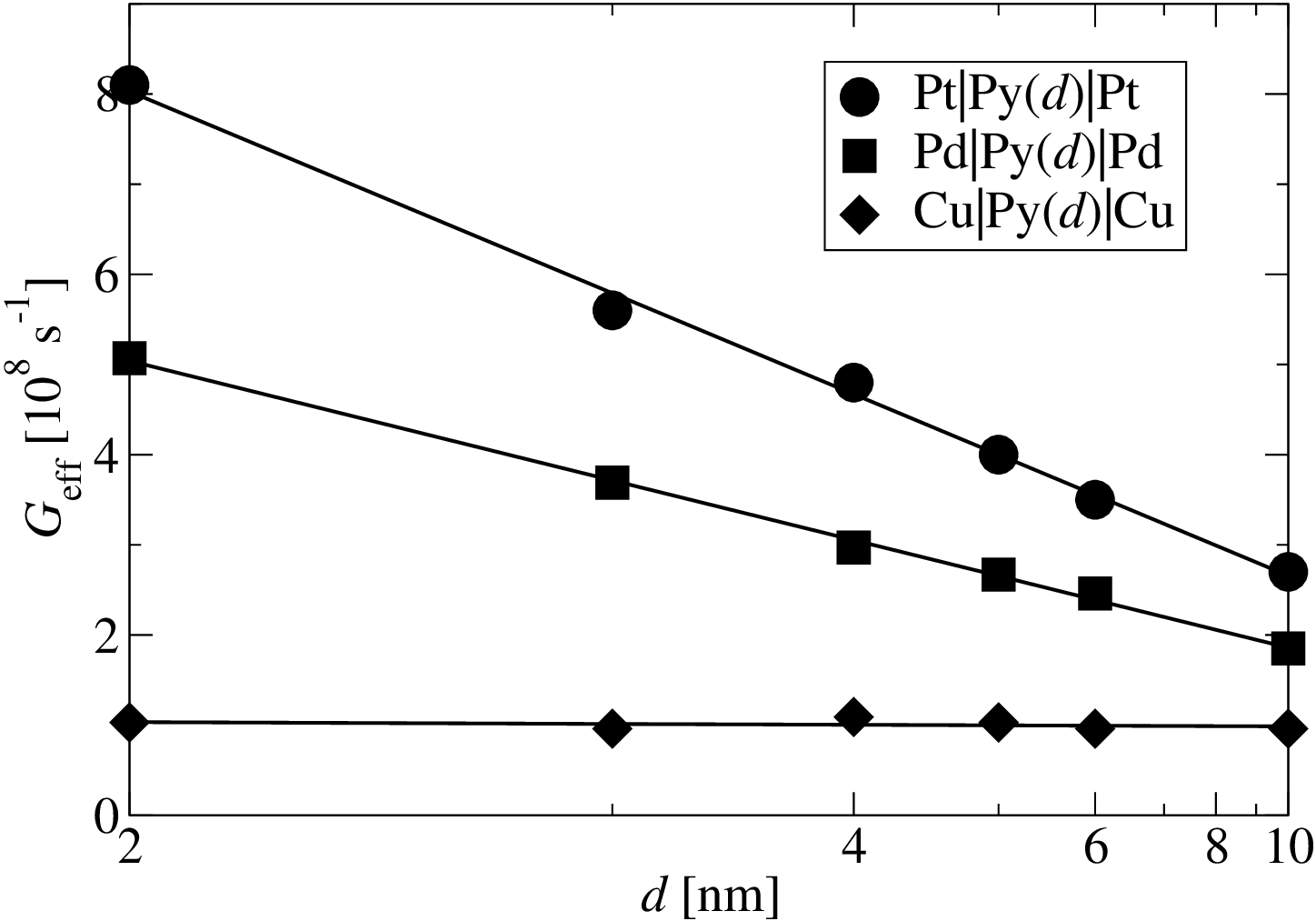}
\caption{Symbols represent the data points derived from FMR experiments on N$\mid$Py$\mid$N sandwiches with N=Pt, Pd, or Cu \cite{Mizukami:jjap01,Mizukami:mmm01}. The lines show fits by $G+a/d$. $G=0.97$, $1.06$, and $1.31\times10^{8}$~s$^{-1}$, where following Eq.~(\ref{aest}), $a$ corresponds to $g_r^{\uparrow\downarrow}/S=0.2$, $15.3$, and $25.8$~nm$^{-2}$ for Cu, Pd, and Pt, respectively, taking a \textit{g}-factor of $2.1$ for Py films \cite{Mizukami:jjap01,Mizukami:mmm01}. The horizontal axis uses a reciprocal scale.}
\label{miz1}
\end{figure}

\onlinecite{Mizukami:jjap01,Mizukami:mmm01} measured room-temperature FMR linewidths of sputtered N$\mid$permalloy (Ni$_{80}$Fe$_{20},$ Py)$\mid$N sandwiches, and discovered systematic trends in the damping parameter as a function of Py-film thickness $d$ for different normal metals N. Their data for the Gilbert parameter $G$, Eq.~(\ref{G}), are shown by symbols in Fig.~\ref{miz1}. In the following, we wish to compare these measurements with theoretical predictions based on Eq.~(\ref{aeff}). For the Py films, $d\geq2~{\rm nm}\gg\lambda_{\text{sc}}$, so that $t^{\uparrow\downarrow},\,t^{\prime\uparrow\downarrow}\approx0$ and $\mathcal{A}^{\uparrow\downarrow}$, Eq.~(\ref{A}), is simply the mixing conductance $g^{\uparrow\downarrow}$ of the Py$\mid$N interface. $g_i^{\uparrow\downarrow}\ll g_r^{\uparrow\downarrow}$ for the interfaces listed in Table~\ref{tabmix}, so that
\begin{equation}
g^{\uparrow\downarrow}\approx g_r^{\uparrow\downarrow}\,.
\label{gg1}
\end{equation}
Eq.~(\ref{geff}) then reduces to $\gamma_{\text{eff}}\approx\gamma$ and Eq.~(\ref{aeff}) becomes
\begin{equation}
G_{\text{eff}}\approx G+\frac{\hbar\gamma^2g_r^{\uparrow\downarrow}}{2\pi V}\,.
\label{aest}
\end{equation}
Moreover, for the parameters in Table~\ref{tabmix},
\begin{equation}
g_r^{\uparrow\downarrow}\approx g^{\rm Sh}_N\,.
\label{gg2}
\end{equation}
As explained by \onlinecite{Zwierzycki:prb05}, Eqs.~(\ref{gg1}) and (\ref{gg2}) are good approximations for intermetallic interfaces because of the large phase differences between spin-up and spin-down reflection coefficients (see, however, Sec.~\ref{uns} for ultrathin magnetic films, $d\lesssim\lambda_{\text{sc}}$). We will assume the validity of these approximations in much of the review. Sharvin conductances for different normal metals \cite{Zwierzycki:prb05} are listed in Table~\ref{stone}. The data in Fig.~\ref{miz1} are used to extract $g_r^{\uparrow\downarrow}$ according to Eq.~(\ref{aest}). A possible $d$-dependence of the bulk $G$ is disregarded in our analysis. The $G$'s in Fig.~\ref{miz1} are close to the previously measured bulk Gilbert damping constants of permalloy \cite{Patton:jap75,Bastian:pssa76}. We see that the extracted mixing conductance for the Pd$\mid$Py interface is similar to that expected from the Sharvin conductance of Pd, while that of Pt$\mid$Py is about $1.5$ times larger. \onlinecite{Mizukami:jjap01,Mizukami:mmm01} also reported FMR measurements on Ta$\mid$Py sandwiches that had only a small damping enhancement with respect to the bulk value.

\begin{table}[pth]
\begin{center}
\begin{tabular}
[c]{ccccc}\hline\hline
& Cu & Ta & Pd & Pt\\\hline
$g^{\rm Sh}$ & 15.0 & 25.0 & 16.0 & 17.6 \\
$\mathcal{N}$ & 2 & 10 & 15 & 12 \\
$(1-\mathcal{N}I_{\text{xc}})^{-1}$ & 1.1 & 1.9 & 4.4 & 2.2 \\\hline\hline
\end{tabular}
\end{center}
\caption{Sharvin conductances (in units of quantum channels per nm$^2$) for bulk fcc Cu, Pd, Pt and bcc Ta, density of states $\mathcal{N}$ at the Fermi level  (in units of states per Ry, atom, spin), and Stoner enhancement factor. From \onlinecite{Zwierzycki:prb05}. Typical values of the exchange-correlation integral, $I_{\text{xc}}$, were taken from \onlinecite{Gunnarsson:jpf76,Janak:prb77}.}
\label{stone}
\end{table}

According to the earlier arguments, the absence of a significant thickness dependence of the damping in the Cu$\mid$Py system could correspond to an unrealistically small conductance $g_r^{\uparrow\downarrow}$. The explanation should however be sought in the long spin-flip relaxation times in clean Cu \cite{Meservey:prl78}: Since the accumulated spins drive a diffusion spin current opposite to the pumping current, the additional damping vanishes in clean Cu. On the other hand, Pd and especially Pt (which is below Pd in the periodic table) have much larger spin-relaxation rates, that can be rationalized by the higher atomic number and complex Fermi surfaces (Pd appears to also have an additional spin-decoherence mechanism due to spin-flip scattering at magnetization fluctuations, see \onlinecite{Foros:jap05}). We note that the spin-flip efficiency of a dirty normal metal is determined by defects and impurities as well. \onlinecite{Lubitz:jap03} reported a significant damping contribution that scaled as $1/d$ as a function of the Fe-layer thickness $d$ in polycrystalline Cu$\mid$Fe$\mid$Cu sandwiches in contrast to the experiments by \onlinecite{Mizukami:jjap01} on Cu$\mid$Py$\mid$Cu. This was interpreted in terms of a larger Cu spin-relaxation rate in the Fe$\mid$Cu as compared to the Cu$\mid$Py system. The effect of spin flip in the normal metal on the excess damping in multilayers is treated quantitatively in the next subsection.

\onlinecite{Ingvarsson:prb02} also carried out FMR studies on Py films sandwiched by the normal metals Pt, Nb (which is above Ta in the periodic table), and Cu, as well as by insulators. Pt$\mid$Py combinations displayed a significantly stronger damping than other structures. They also identified an additional thickness-dependent damping correlated with the disorder in the Py films that was not reported by \onlinecite{Mizukami:jjap01,Mizukami:mmm01}. Damping processes that are intrinsic to the ferromagnetic layer cannot be addressed by the present theory, however.
 
\onlinecite{Lagae:mmm05} studied pulsed dynamics of a thin Py layer in contact with Cu or Ta layers. In addition, they investigated the role of a second, pinned ferromagnetic layer, attached to the free layer via a Cu spacer or a tunnel barrier. The measured Gilbert damping was found to be consistent with the spin-pumping picture. We will discuss the effect of the second ferromagnetic layer in Sec.~\ref{abi}.

Finally, we comment on the role of electron correlations in the spin-pumping formalism, see also Sec.~\ref{eei}. In order to calculate the mixing conductance, the scattering matrix has been obtained by \onlinecite{Xia:prb02,Zwierzycki:prb05} for the effective potential from Kohn-Sham density-functional theory. This potential is calculated self-consistently and includes electron-electron interaction effects via the Hartree and an exchange-correlation potential in the local--spin-density approximation. The interface parameters when computed self-consistently and nonperturbatively include the magnetic moments induced in the normal metal by the proximity to the ferromagnet (also discussed by \onlinecite{Simanek:prb032}). For Cu and Au normal-metal films, this effect is very small due to the small densities of states at the Fermi energy. The question arises whether larger effects may be expected for materials such as Pd and Pt with a large Fermi-level density of states (see Table~\ref{stone}). These metals are known to be ``almost ferromagnetic" with a Stoner-enhanced spin susceptibility $\chi/\chi_0=(1-\mathcal{N}I_{\text{xc}})^{-1}$, also listed in Table~\ref{stone}. So, can the increased damping of the magnetization dynamics in contact with thin layers of Ta, Pd and Pt as compared to Cu \cite{Mizukami:jjap01,Mizukami:mmm01} connected to their larger Fermi-level densities of states? In the spin-pumping theory, the quantity that governs the damping enhancement is the mixing conductance. For metallic structures, the band-structure calculations give results that are close to the Sharvin conductance of the normal metal, Eq.~(\ref{gg2}), thus not sensitive to the magnetic moments induced in the proximity to the ferromagnet. The Sharvin conductance for Cu, Ta, Pd and Pt  is seen to change less than $\mathcal{N}$. More significantly, the trend does not correspond to that observed experimentally for the damping enhancement: The Sharvin conductance is maximal for Ta, not Pd with the highest spin susceptibility, supporting our earlier conclusion that the spin dissipation in the normal conductors is crucial for the enhanced magnetization damping, as is explored in more detail below.

\subsection{Diffuse systems}
\label{ds}

In the previous subsection, we concentrated on the extreme situation in which the normal-metal layer is an ideal spin sink that immediately absorbs the injected spin current $\mathbf{I}_s^{\text{pump}}$, either by relaxation through spin-flip processes or the absence of back scattering. Disregarding  $\mathbf{I}_s^{(0)}$, the total spin current through the contact, Eq.~( \ref{Ipb}), reduces to $\mathbf{I}_s= \mathbf{I}_s^{\text{pump}}$. In general, we have to take into account the spin accumulation in a diffuse normal metal that drives a spin current $\mathbf{I}_s^{(0)}$ back into the ferromagnet. As a technical note, we recall that in this regime the drift correction (\ref{tr}) should be applied to the conductance parameters entering Eqs.~(\ref{IcR}), (\ref{IsR}), and (\ref{Is}).

\begin{figure}[pth]
\includegraphics[width=0.7\linewidth,clip=]{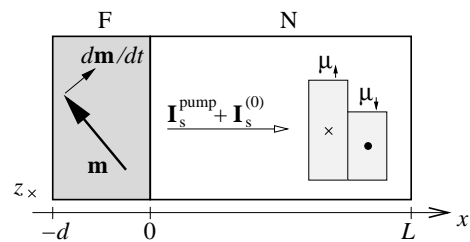}
\caption{Schematic view of an F$\mid$N bilayer in which the magnetization direction $\mathbf{m}(t)$ of the ferromagnet F rotates in an easy plane, pumping a spin current $\mathbf{I}_s^{\text{pump}}$ into the adjacent normal-metal layer N. The N layer here is not an ideal reservoir but rather a film of the same cross section as the magnetic layer F. The spin pumping builds up a position ($x$) dependent spin accumulation in N that either relaxes by spin-flip scattering or drives a spin current back into the ferromagnet as $\mathbf{I}_s^{(0)}$.}
\label{fn}
\end{figure}

As the simplest example, we first discuss an F$\mid$N bilayer, in which a magnetization precession generates a pure spin but no charge current. The spin accumulation or nonequilibrium chemical potential imbalance $\boldsymbol{\mu}_s(x)$ [similar to Eq.~(\ref{sa}), but with spatial dependence] in the normal metal is a vector with magnitude that depends on the distance from the interface $x$, $0<x<L$, where $L$ is the thickness of the normal-metal film, see Fig.~\ref{fn}. When the ferromagnetic magnetization steadily precesses (in an easy plane) around the $z$ axis, $\mathbf{m}\times\mathbf{\dot{m}}$ and the normal-metal spin accumulation $\boldsymbol{\mu}_s(x)$ are oriented along $z$, as depicted in Fig.~\ref{fn}. We will see in the following that the instantaneous $\boldsymbol{\mu}_s$ is always perpendicular to $\mathbf{m}$ even in the case of a \textit{precessing} ferromagnet with time-dependent rotation axis, as long as the precession frequency $\omega$ is smaller than the spin-flip rate $\tau _{\text{sf}}^{-1}$ in the normal metal.

In frequency domain, a spin accumulation diffuses in a normal metal according to \cite{Johnson:prb88a}
\begin{equation}
i\omega\boldsymbol{\mu}_s=D\partial_x^2\boldsymbol{\mu}_s-\frac{\boldsymbol{\mu}_s}{\tau_{\text{sf}}}\,,
\label{de}
\end{equation}
where $D=v_F^{2}\tau/3$ is the diffusion coefficient (in three dimensions) in terms of the Fermi velocity $v_F$, and the transport mean free time $\tau$. It is assumed here that the frequency $\omega$ is smaller than the scattering rate $\tau ^{-1}$. The ratio of the momentum to spin-flip scattering time is an important parameter:
\begin{equation}
\epsilon=\frac{\tau}{\tau_{\text{sf}}}\, .
\end{equation}
Eq.~(\ref{de}) holds when $\epsilon\ll1$, i.e., the spin-flip relaxation may be treated perturbatively. The boundary conditions are given by the spin current $\mathbf{I}_s$ at the F$\mid$N interface $x=0$ and vanishing of the spin current at the outer boundary $x=L$:
\begin{align}
x=&0:~\partial_x\boldsymbol{\mu}_s=-\frac{2}{\hbar\mathcal{N}SD}\mathbf{I}_s\,,\nonumber\\
x=&L:~\partial_x\boldsymbol{\mu}_{s}=0\,,
\label{bc}
\end{align}
where $S$ is the interface area and $\mathcal{N}$ the normal-metal one-spin density of states per unit volume. The solution of Eqs.~(\ref{de}), (\ref{bc}) is 
\begin{equation}
\boldsymbol{\mu}_s(x)=\frac{\cosh\left[\kappa(x-L)\right]}{\sinh\kappa L}\frac{2}{\hbar\mathcal{N}SD\kappa}\mathbf{I}_s\,,
\label{sn}
\end{equation}
where $\kappa =\lambda_{\text{sd}}^{-1}\sqrt{1+i\omega\tau_{\text{sf}}}$, recalling that $\lambda_{\text{sd}}=\sqrt{D\tau_{\text{sf}}}$ is the spin-diffusion length in the normal metal. In Sec.~\ref{sb}, we used similar arguments to calculate the spin accumulation generated by the precessing magnetization. Whereas in Sec.~\ref{sb} the magnitude of the spin accumulation and its relevance for spintronics are discussed, we focus here on the effect of the spin accumulation on the ferromagnetic magnetization dynamics.

We assume in the following that the precession frequency $\omega$ is smaller than the spin-flip relaxation rate, $\omega\ll\tau_{\text{sf}}^{-1}$, so that $\kappa\approx\lambda_{\text{sd}}^{-1}$. For a typical effective field of 1~T, $\omega \sim10^{11}$~s$^{-1}$. The scattering rate corresponding to a mean free path of $\lambda\sim10$~nm is $\tau^{-1}\sim 10^{14}$~s$^{-1}$. Consequently, for such metals, $\omega\ll\tau_{\text{sf}}^{-1}$ requires $\epsilon\gtrsim10^{-3}$.  This condition is easily satisfied \cite{Meservey:prl78} for impurities with higher atomic numbers $Z$ [since $\epsilon$ scales as $Z^{4}$ \cite{Abrikosov:zetf62}]. The high-frequency limit $\omega\gtrsim\tau_{\text{sf}}^{-1}$, on the other hand, is relevant for systems with weak spin-flip scattering in the normal metal, and has been discussed in the context of the spin-battery concept in Sec.~\ref{sb}. We will see that a sizable Gilbert-damping enhancement requires a large spin-flip probability $\epsilon\gtrsim10^{-1}$, thus $\omega\ll\tau_{\text{sf}}^{-1}$, unless the frequency is comparable with the momentum scattering rate in the normal metal. [The latter regime lies below our treatment based on Eqs.~(\ref{de}), (\ref{bc}).]

It is convenient to define an effective energy-level spacing of the states participating in the spin-flip scattering events in a thick (i.e., $L\gg\lambda_{\text{sd}}$)  normal-metal film as
\begin{equation}
\delta_{\text{sd}}=\frac{1}{\mathcal{N}S\lambda_{\text{sd}}}\,.
\label{td}
\end{equation}
The spin-diffusion length (which sets the scale for spin-accumulation penetration into the normal layer) written in terms of the relevant scattering times is
\begin{equation}
\lambda_{\text{sd}}=v_F\sqrt{\frac{\tau\tau_{\text{sf}}}{3}}\,.
\label{sd}
\end{equation}
The spin-accumulation--driven spin current $\mathbf{I}_{s}^{(0)}$ through the interface is obtained by substituting $\boldsymbol{\mu}_s(x=0)$ from Eq.~(\ref{sn}) into Eq.~(\ref{IsR}) to give
\begin{equation}
\mathbf{I}_s=\mathbf{I}_s^{\text{pump}}+\mathbf{I}_s^{(0)}=\mathbf{I}_s^{\text{pump}}-\beta\left(\tilde{g}_r^{\uparrow\downarrow}\mathbf{I}_s+\tilde{g}_i^{\uparrow\downarrow}\mathbf{I}_s\times\mathbf{m}\right)\,,
\label{sc1}
\end{equation}
where the spin current returning into the ferromagnet is governed by the ``backflow factor" $\beta$: 
\begin{equation}
\beta=\frac{\tau_{\text{sf}}\delta_{\text{sd}}/h}{\tanh(L/\lambda_{\text{sd}})}.
\label{beta}
\end{equation}
The last equality in Eq.~(\ref{sc1}) is obtained from Eq.~(\ref{IsR}) by assuming that the nonequilibrium transport is limited to spin currents that are polarized normal to the magnetization. This is allowed in the adiabatic regime when $\omega\ll\tau_{\text{sf}}$ where the average angular momentum of the pumped and backscattered spins is at a given instant perpendicular to the magnetization. Consequently, as long as the ferromagnetic spin-relaxation length is larger than the transverse spin-coherence length $\lambda_{\text{sc}}$, we may disregard spin relaxation inside the ferromagnet. $\beta$ is given by the ratio between the energy-level spacing of the normal-metal film with a thickness $L_{\text{sf}}=\min(L,\lambda _{\text{sd}})$ and the spin-flip rate. When the normal metal is shorter than its spin-flip diffusion length, $L\ll\lambda_{\text{sd}}$, $\beta\rightarrow\tau_{\text{sf}}\delta/h$, where $\delta=(\mathcal{N}SL)^{-1}$ is the energy-level splitting. In the opposite regime of thick normal metals, $L\gg\lambda_{\text{sd}}$, $\beta\rightarrow\tau _{\text{sf}}\delta_{\text{sd}}/h$. 

By solving Eq.~(\ref{sc1}), the total spin current $\mathbf{I}_{s}$ can be expressed in terms of the pumped spin current $\mathbf{I}_{s}^{\text{pump}}$, Eq.~(\ref{Is}), to finally obtain
\begin{align}
\mathbf{I}_s=&\frac{\hbar}{4\pi}\left(\mbox{Re}\mathcal{A}_{\text{eff}}^{\uparrow\downarrow}\mathbf{m}\times\frac{d\mathbf{m}}{dt}+\mbox{Im}\mathcal{A}_{\text{eff}}^{\uparrow\downarrow}\frac{d\mathbf{m}}{dt}\right)\,,
\label{sc2}
\end{align}
which has the same form as the original Eq.~(\ref{Is}), but with effective ``spin-pumping efficiencies" $\mathcal{A}_{\text{eff}}^{\uparrow\downarrow}$:
\begin{equation}
\frac{1}{\mathcal{A}_{\text{eff}}^{\uparrow\downarrow}}=\frac{1}{\tilde{g}^{\uparrow\downarrow}}+\frac{R_{\text{sd}}}{\tanh(L/\lambda_{\text{sd}})}\,.
\label{tA}
\end{equation}
Here, $R_{\text{sd}}=\tau_{\text{sf}}\delta_{\text{sd}}/h$ is the dimensionless resistance [in units of $(2e^2/h)^{-1}$] of the normal-metal layer of thickness $\lambda_{\text{sd}}$, which follows from Eq.~(\ref{td}) and the Einstein relation $\sigma=e^2D\mathcal{N}$ between conductivity $\sigma$ and diffusion coefficient $D$. When $L\gg\lambda_{\text{sd}}$, the effective spin pumping out of the ferromagnet is governed by the mixing conductance of the F$\mid$N interface in series with a diffusive normal-metal film of thickness $\lambda_{\text{sd}}$.

By accepting the approximation (\ref{gg1}), we disregard $\tilde{g}_i^{\uparrow\downarrow}$ in the remainder of this section. $\mbox{Im}\mathcal{A}^{\uparrow\downarrow}_{\text{eff}}$ then also vanishes and upon substitution into Eq.~(\ref{Gllg}) the damping torque due to the spin current $\mathbf{I}_s$ obeys the Gilbert phenomenology. The effective Gilbert-damping parameter in the diffuse model reads [cf. Eq.(\ref{aest})]:
\begin{equation}
G_{\text{eff}}-G=\left[1+\tilde{g}_r^{\uparrow\downarrow}\frac{R_{\text{sd}}}{\tanh(L/\lambda_{\text{sd}})}\right]^{-1}\frac{\hbar\gamma^2\tilde{g}_r^{\uparrow\downarrow}}{4\pi V}\,.
\label{afn}
\end{equation}
The prefactor on the right-hand side of Eq.~(\ref{afn}) reflects the reduction effect on the Gilbert damping that is caused by the spin-diffusion back into the ferromagnet. This correction has been disregarded in Sec.~\ref{pss} where the normal metals were considered to be ideal spin sinks. Because spins accumulate in the normal metal polarized transversely to the ferromagnetic magnetization, the spin-accumulation--driven transport across the F$\mid$N contact is governed by the spin-mixing conductance (just like the spin-pumping current). The absence of spin accumulations or currents polarized collinear to the magnetization (in the limit $\omega\ll\tau_{\text{sf}}^{-1}$) explains why the diagonal components of the conductance matrix $\tilde{g}^{\sigma\sigma^\prime}$ do not enter Eq.~(\ref{afn}).

The numerical values of the parameters in Eq.~(\ref{afn}) can be estimated by the free-electron model for the normal metal, for which $\mathcal{N}=k_F^{2}/(\pi hv_F)$. Using Eqs.~(\ref{sd}) and (\ref{td}), we find $R_{\text{sd}}^{-1}=h/(\delta_{\text{sd}}\tau_{\text{sf}})=4\sqrt{\epsilon/3}g^{\text{Sh}}_{\rm N}$, where $g^{\text{Sh}}_{\rm N}=Sk_F^{2}/(4\pi)$ is the dimensionless Sharvin conductance (number of transport channels) of the normal metal. Using approximation (\ref{gg2}) for the mixing conductance $g_r^{\uparrow\downarrow}$, and Eq.~(\ref{gr}) to correct for the drift contribution, we obtain $\tilde{g}_r^{\uparrow\downarrow }\approx 2g^{\text{Sh}}_{\rm N}$. We thus estimate 
\begin{equation}
G_{\text{eff}}-G\sim\frac{\hbar\gamma^2\tilde{g}_r^{\uparrow\downarrow}/(4\pi V)}{1+\left[\sqrt{\epsilon}\tanh(L/\lambda_{\text{sd}})\right]^{-1}}\,.
\label{sp11}
\end{equation}
A significantly increased damping therefore requires a high spin-flip to spin-conserving scattering probability, $\epsilon\gtrsim 10^{-2}$, and sufficiently thick normal layers, $L\gtrsim\lambda_{\text{sd}}$ ($L\gtrsim\lambda_{\text{sd}}$ alone is not a sufficient condition, however, contrary to what has sometimes been assumed in the literature). Poor spin sinks that do not modify the magnetization dynamics are characterized by a large denominator in Eq.~(\ref{sp11}). Clean light metals with atomic number $Z\lesssim50$, such as Al, Cr, and Cu, as well as heavier metals with only $s$ electrons in the conduction band, such as Ag, are less effective spin sinks because of a relatively small intrinsic spin-orbit coupling, typically in the range of $\epsilon\lesssim10^{-2}$ \cite{Meservey:prl78,Bergmann:zpb82,Yang:prl94}. Heavier elements with $Z\gtrsim50$ and $p$ or $d$ character of the conduction electrons, such as Pd, Pt, and Pb, are effective spin sinks with much larger $\epsilon\gtrsim 10^{-1}$ \cite{Meservey:prl78}. Au is intermediate in $\epsilon$ \cite{Chiang:prb04}, presumably due to the $s$ character of its outer orbital. These conclusions explain the hierarchy of the observed Gilbert damping enhancement observed by \onlinecite{Mizukami:jjap01,Mizukami:mmm01}, see Sec.~\ref{pss}. By doping a small-$\epsilon$ matrix with high-$Z$ or magnetic impurities (e.g., Cu lattice with Pt atoms), a good spin sink can be created. We conclude that the damping parameter (and thus the switching speed) of thin ferromagnetic films can be engineered flexibly by coating them with suitable normal metals.

In the limit of a large ratio of the spin-flip to momentum scattering, $\epsilon\sim1$, the spin-diffusion equation and, consequently, Eq.~(\ref{sp11}) do not hold. In this regime, the normal metal very efficiently relaxes the spins injected from the ferromagnet. \onlinecite{Foros:jap05} investigated this scenario for Pd (in contact with Fe) that appears to have a spin-relaxation rate, arguably by paramagnons, that is faster than the bulk momentum-scattering rate. When thicker than the spin-decoherence length $v_F \tau_{\text{sf}}$, otherwise ballistic Pd films should thus be good spin sinks. The ferromagnetic relaxation by ideal spin sinks is determined by the bare spin-mixing conductance $g^{\uparrow\downarrow}$, as discussed in Sec.~\ref{pss}. The latter generally provides an upper bound for the magnitude of the additional Gilbert damping. However, in the present theory, spin-orbit coupling is treated phenomenologically in the diffuse-transport regime only. Strong spin-orbit coupling immediately at interfaces, for example, requires generalization of spin-pumping and circuit theories beyond the scope of this review.

Infinite vs vanishing spin-flip rates in the normal metal are two extremes for the F$\mid$N bilayer  dynamics. In the former case, the damping parameter $G_{\text{eff}}$ is enhanced, and in the latter case unmodified. Both limits are experimentally accessible by using Pt as a good or Cu as a poor spin sink, as shown by \onlinecite{Mizukami:jjap01,Mizukami:mmm01} for N$\mid$Py$\mid$N sandwiches. 

\begin{figure}[pth]
\includegraphics[width=0.95\linewidth,clip=]{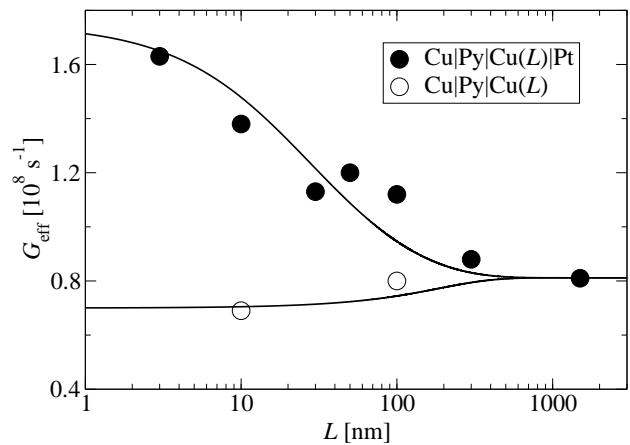}
\caption{Gilbert-damping constants (circles) in Cu$\mid$Py(3~nm)$\mid$Cu($L$)$\mid$Pt(5~nm) and Cu$\mid$Py(3~nm)$\mid$Cu($L$) structures measured by \onlinecite{Mizukami:mmm02,Mizukami:prb02} as a function of Cu-layer thickness $L$. Solid lines are the theoretical results according to Eq.~(\ref{afnn}) with parameters discussed in the text. The horizontal axis uses a logarithmic scale.}
\label{miz2}
\end{figure}

\begin{figure}[pth]
\includegraphics[width=0.9\linewidth,clip=]{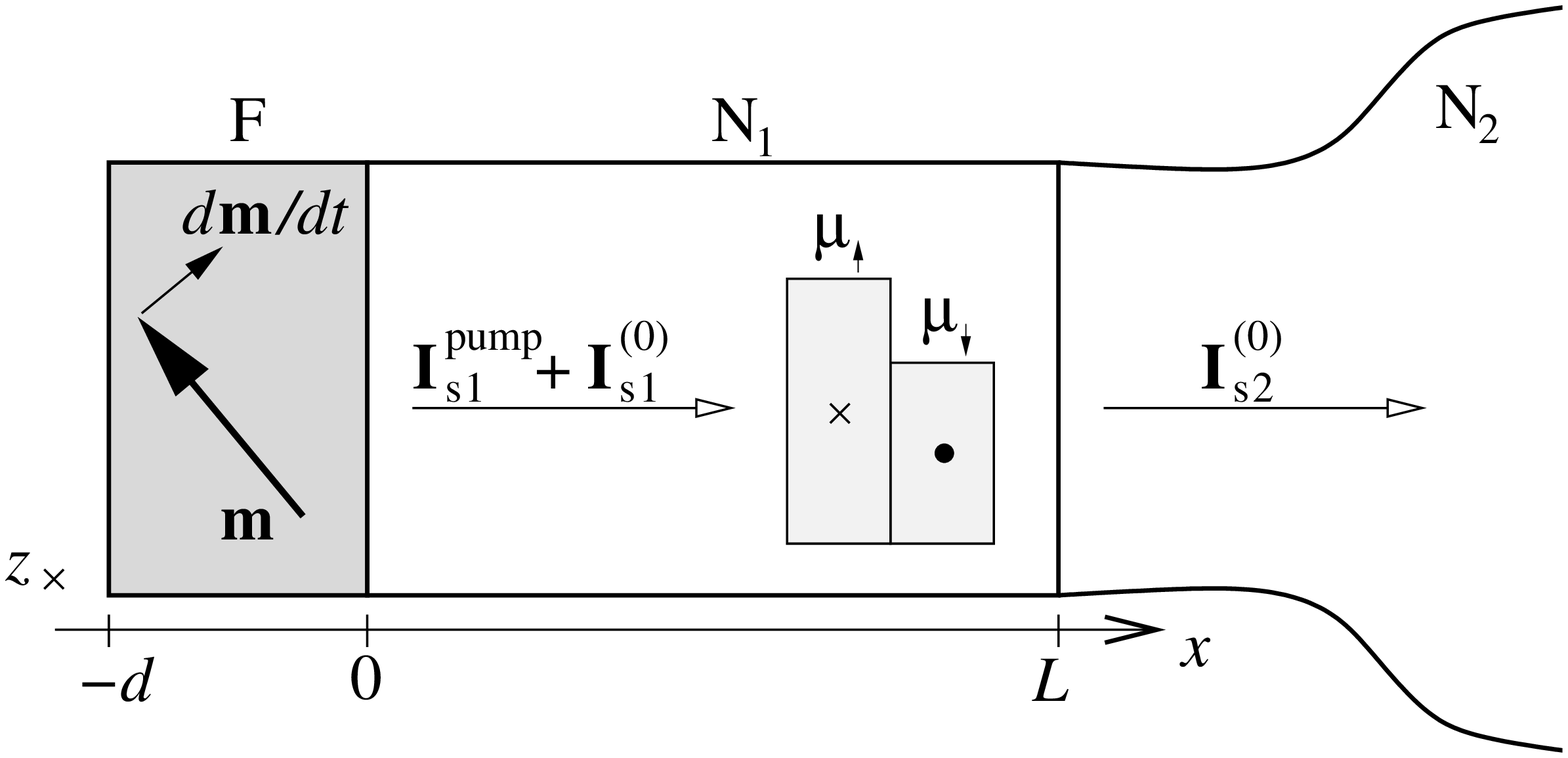}
\caption{Similar to Fig.~\ref{fn}, but the normal-metal system is an N$_1$$\mid$N$_2$ bilayer. Ferromagnetic magnetization precession pumps spins into the first normal-metal layer N$_1$. The spin accumulation in N$_1$ either flows back into the ferromagnet F as spin current $\mathbf{I}_{s1}^{(0)}$, relaxes in N$_1$, or leaves via the second normal-metal layer N$_2$ as spin current $\mathbf{I}_{s2}^{(0)}$. The spin accumulation in the ideal spin sink N$_2$ vanishes.}
\label{fnn}
\end{figure}

Experiments on Py where both Cu and Pt were combined in a Py$\mid$Cu$\mid$Pt heterostructure \cite{Mizukami:mmm02,Mizukami:prb02} provide a novel method to study spin-diffusion in the central Cu layer. The measured room-temperature magnetization damping in Cu$\mid$Py$\mid$Cu($L$) and Cu$\mid$Py$\mid$Cu($L$)$\mid$Pt structures as a function of Cu-film thickness $L$ is shown by circles in Fig.~\ref{miz2}. The experiments can be understood by a slight extension of the diffusion-theory--based discussion for F$\mid$N bilayers. The damping is governed by the angular-momentum loss of the ferromagnet through the normal-metal compound N$_1$$\mid$N$_2$, as schematically shown in Fig.~\ref{fnn}. Once injected into N$_1$, spins are either scattered back into the ferromagnet or relax in N$_1$ or N$_2$. When there is only a very weak spin-flip scattering in N$_1$ compared to N$_2$, the spins have to diffuse through N$_1$ before they can relax in the second normal-metal layer N$_2$. For simplicity, we model N$_2$ as an ideal sink that instantaneously relaxes incoming spins, as is appropriate for Pt. The analysis below illustrates that the ferromagnetic magnetization damping as a function of $L$ provides important information about spin transport through N$_1$ and the N$_1$$\mid$N$_2$ interface. The Cu substrate on the other side of the permalloy film \cite{Mizukami:mmm02,Mizukami:prb02} can be disregarded since it is a poor spin sink.

Similar to Eqs.~(\ref{bc}), the boundary conditions for the spin accumulation  in the normal metal N$_1$ are now: 
\begin{align}
x=0:~\partial_x\boldsymbol{\mu}_s=&-\frac{2}{\hbar\mathcal{N}SD} 
\mathbf{I}_{s1}\,,\nonumber\\
x=L:~\partial_x\boldsymbol{\mu}_s=&-\frac{2}{\hbar\mathcal{N}SD}
\mathbf{I}_{s2}\,.
\label{bc1}
\end{align}
$\mathbf{I}_{s1}$ and $\mathbf{I}_{s2}$ are the total spin currents through the left ($x=0$) and right ($x=L$) interfaces, respectively. $\mathbf{I}_{s1}=\mathbf{I}_{s1}^{\text{pump}}+\mathbf{I}_{s1}^{(0)}$, cf. Eq.~(\ref{sc1}) in the previous subsection, is the sum of the pumped (\ref{Is}) and the spin-accumulation--driven (\ref{IsR}) spin currents. $\mathbf{I}_{s2}$, on the other hand, is entirely governed by the N$_1$$\rightarrow$N$_2$ diffusion 
\begin{equation}
\mathbf{I}_{s2}=\frac{\tilde{g}}{4\pi}\text{\boldmath$\mu$}_{s}(x=L)\,,
\end{equation}
where $\tilde{g}$ is the effective spin conductance of the N$_1$$\mid$N$_2$ interface:
\begin{equation}
\frac{1}{\tilde{g}}=\frac{1}{g_{{\rm N}_1\mid{\rm N}_2}^{\sigma\sigma}}-\frac{1}{2g^{\text{Sh}}_{{\rm N}_1}}\,.
\label{rg}
\end{equation}
Here, we corrected the bare single-spin resistance $1/g_{{\rm N}_1\mid{\rm N}_2}^{\sigma\sigma}$ of the all-normal interface for the drift effect by subtracting the Sharvin contribution on the N$_1$ side only, because the layer N$_2$ is assumed to be an ideal spin bath. Solving the diffusion equation (\ref{de}) with the boundary conditions (\ref{bc1}), we can find the spin current $\mathbf{I}_{s1}$, which is transferred to the magnetization as a torque, as discussed in the preceding discussion for the bilayer. The Gilbert damping enhancement due to the spin relaxation in the trilayer is then found to be
\begin{equation}
G_{\text{eff}}-G=\left[ 1+\tilde{g}_r^{\uparrow\downarrow}R_{\text{sd}}\frac{1+\tanh(L/\lambda_{\text{sd}})\tilde{g}R_{\text{sd}}}{\tanh(L/\lambda_{\text{sd}})+\tilde{g}R_{\text{sd}}}\right]^{-1}\frac{\hbar\gamma^2\tilde{g}_r^{\uparrow\downarrow}}{4\pi V}\,,
\label{afnn}
\end{equation}
where $\lambda_{\text{sd}}$ and $R_{\text{sd}}$ parametrize the spin-diffusion in N$_1$, $\tilde{g}_r^{\uparrow\downarrow}$ is the renormalized mixing conductance (\ref{gr}) of the F$\mid$N$_1$ interface, and $\tilde{g}$ is the spin-transfer conductance (\ref{rg}) of the N$_1$$\mid$N$_2$ interface.

Setting $\tilde{g}=0$ decouples the two normal-metal systems and reduces Eq.~(\ref{afnn}) to Eq.~(\ref{afn}), the damping coefficient of the F$\mid$N$_1$ bilayer. In the experiment \cite{Mizukami:mmm02,Mizukami:prb02}, the permalloy thickness $d=3$~nm is fixed and the Cu film thickness $L$ is varied between 3 and 1500 nm as shown by the circles in Fig.~\ref{miz2}. The theoretical result (\ref{afnn}) is plotted in Fig.~\ref{miz2} for comparison, using the following parameters: the bulk damping $G=0.7\times 10^{8}$~s$^{-1}$ \cite{Patton:jap75,Bastian:pssa76}, the spin-flip probability $\epsilon=1/700$ and the spin-diffusion length $\lambda_{\text{sd}}=250$~nm for Cu (which correspond to the mean free path $\lambda=\sqrt{3\epsilon}\lambda_{\text{sd}}=16$~nm), in satisfactory agreement with values reported in literature \cite{Meservey:prl78,Yang:prl94,Jedema:nat01}, $\tilde{g}_r^{\uparrow\downarrow}/S=16$~nm$^{-2}$ extracted from the experimental angular magnetoresistance of Py$\mid$Cu \cite{Bauer:prb03}, and $\tilde{g}/S=35$~nm$^{-2}$ for the Cu$\mid$Pt contact. This $\tilde{g}$ corresponds to the bare one-spin conductance $g^{\sigma\sigma}/S=16$~nm$^{-2}$, which is close to the Sharvin conductance of Cu, see Table~\ref{tabmix}. Figure~\ref{miz2} shows a satisfactory agreement (within the experimental error) between experiments and theory. This proves the diffusive nature of spin transfer in the Cu spacer. Whereas the detailed mechanism for spin injection (relaxation) at the Py$\mid$Cu (Cu$\mid$Pt) interface cannot be deduced directly, the agreement on absolute scale obtained with parameters taken from other sources strongly supports the spin-pumping picture.

It is illuminating to discuss Eq.~(\ref{afnn}) in the limit of vanishing spin-flip in the spacer layer N$_1$. Recalling the definitions for $\lambda_{\text{sd}}$, Eq.~(\ref{sd}), and $\delta_{\text{sd}}$, Eq.~(\ref{td}), and taking the limit $\tau_{\text{sf}}\rightarrow\infty$, we find that Eq.~(\ref{afnn}) reduces to Eq.~(\ref{aest}), only with $2g^{\uparrow\downarrow}$ (where the factor of two corresponds to two F$\mid$N interfaces in the N$\mid$F$\mid$N trilayers) replaced by $g_{\text{eff}}^{\uparrow\downarrow}$:
\begin{equation}
\frac{1}{g_{\text{eff}}^{\uparrow\downarrow}}=\frac{1}{\tilde{g}^{\uparrow\downarrow}}+R_{N_1}+\frac{1}{\tilde{g}}\,,
\label{serie}
\end{equation}
where $R_{N_1}= (2e^2/h)(L/S \sigma) $ is the dimensionless resistance of the N$_1$ layer with conductivity $\sigma$. The right-hand side of Eq.~(\ref{serie}) is simply the inverse bare mixing conductance of the diffuse N$_1$ spacer in series with its two interfaces, one with F and the other with N$_2$ \cite{Bauer:prb03}. In particular, when N$_1$ is thick enough, the total mixing conductance $g_{\text{eff}}^{\uparrow\downarrow}$ is limited by the spacer separating F and N$_2$ \cite{Brataas:prl00,Brataas:epjb01}. The spin pumping into layer N$_1$ with subsequent spin-conserving diffusion and then spin absorption by the ideal spin sink N$_2$ is thus equivalent with spin pumping (\ref{Is}) across an effective scatterer separating the ferromagnet F from the ideal spin sink N$_2$.

\begin{figure*}[pth]
\includegraphics[width=0.7\linewidth,clip=]{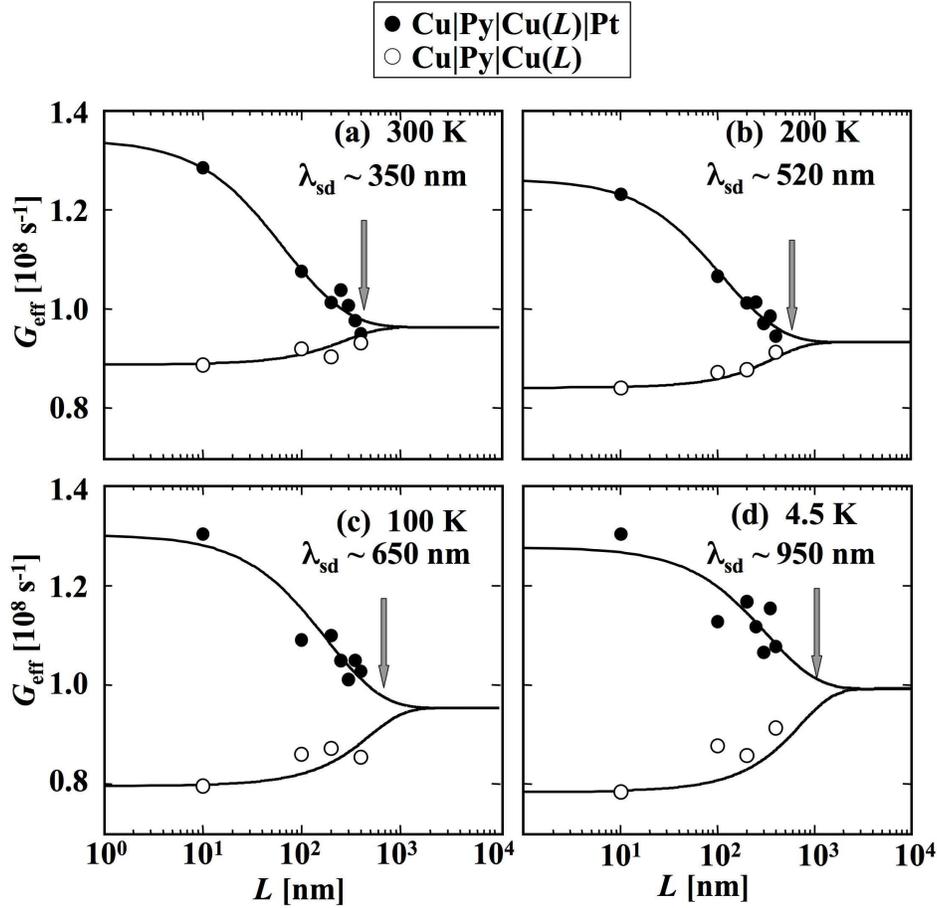}
\caption{Gilbert damping measured in Cu$\mid$Py(3~nm)$\mid$Cu($L$)$\mid$Pt(2~nm) and Cu$\mid$Py(3~nm)$\mid$Cu($L$) structures for several Cu-layer thicknesses $L$ at various temperatures. Solid lines are fits based on the spin-pumping theory. Adapted from \onlinecite{Ando:jmsj05}.}
\label{miz3}
\end{figure*}

The general trends in Fig.~\ref{miz2} can be understood as follows. Since Cu is a poor spin sink, a Py$\mid$Cu contact with a single Cu film only weakly increases the damping for all thicknesses. This enhancement saturates at $L\gg\lambda_{\text{sd}}$ and vanishes in the limit $L\ll\lambda _{\text{sd}}$. When a Pt film, a very good spin sink, is connected to the bilayer and the Cu spacer is thinner than the transport mean free path, $L\ll\lambda$, the spin accumulation is uniform throughout the Cu. The spin pumping is then partitioned: One fraction of the pumped spins is reflected back into the ferromagnet, while the remainder is transmitted to and subsequently relaxes in the Pt layer. Their ratio equals the ratio between the conductance $\tilde{g}_r^{\uparrow\downarrow}$ of the Py$\mid$Cu and the conductance $\tilde{g}$ of the Cu$\mid$Pt interfaces, and is of the order of unity. Since a significant fraction of the spin-pumping current is dissipated in Pt, a large magnetization damping is achieved. When $L$ is increased, less spins manage to diffuse through the Cu spacer, and, in the limit $L\gg\lambda_{\text{sd}}$, the majority of the spins scatter back into the ferromagnet before sensing the presence of the Pt layer. In the intermediate regime, the spin pumping into the Pt layer decays algebraically on the scale of the transport mean free path and exponentially on the scale of the spin-diffusion length.

The temperature dependence of the Gilbert damping in such trilayers has been measured recently by the same group \cite{Ando:jmsj05}, see Fig.~\ref{miz3}. The spin-diffusion length extracted from the data by means of an analysis similar to the one presented here increases as the temperature is lowered, indicating a reduced role of phonon scattering at low temperatures. The good agreement with results from electrical-transport experiments by \onlinecite{Jedema:nat01} establishes FMR experiments as an important tool to measure spin transport in magnetic heterostructures.

The dependence of the damping on the Cu-layer thickness $L$ in the Cu$\mid$Py$\mid$Cu$(L)$$\mid$Pt multilayers reflects the amount of accumulation in the normal metals. This spin accumulation, in turn, indicates that an excited ferromagnet (as in the FMR experiment discussed here) transfers spins into adjacent nonmagnetic layers according to the spin pumping (\ref{Is}). The concept of the spin battery discussed in section \ref{sb} relies on this effect.

\subsection{Enhanced Gilbert damping in spin valves: First-principles calculations vs experiment}
\label{abi}

In F$\mid$N$\mid$F structures, the presence of two ferromagnetic layers can make damping possible for each individual layer even in the absence of spin-flip relaxation in the system \cite{Berger:prb96}. The point is that \textit{if one ferromagnet is excited while the other is static,} the latter acts as a sink for transverse spin currents pumped by the former. In this subsection, we are assuming a sufficiently thick or disordered normal spacer, so that it does not support any persistent spin currents or, in other words, the static exchange interaction between the magnetic films vanishes, see Sec.~\ref{rkky}. Static exchange will be taken into account in Sec.~\ref{dec} in discussing coupled dynamics of two or more magnetic films. If the F$\mid$N$\mid$F magnetic structure is weakly excited from a collinear equilibrium state, then, defining the effective (complex-valued) ``spin-pumping efficiency" $\mathcal{A}_{{\rm F}\mid{\rm N}\mid{\rm F}}^{\uparrow\downarrow}$ by
\begin{equation}
\frac{1}{\mathcal{A}_{{\rm F}\mid{\rm N}\mid{\rm F}}^{\uparrow\downarrow}}=\frac{1}{\tilde{g}^{\uparrow\downarrow}_1}+\frac{2e^{2}}{h}\frac{L}{S\sigma}+\frac{1}{\tilde{g}^{\uparrow\downarrow}_2}\,,
\label{Af}
\end{equation}
we can summarize an analysis similar to that of the preceding section as follows. The total spin current $\mathbf{I}_s$ through the normal spacer consisting of the pumped and backflow components, Eq.~(\ref{Ist}), is given by the right-hand side of Eq.~(\ref{Is}) with $\mathcal{A}^{\uparrow\downarrow}$ replaced by $\mathcal{A}_{{\rm F}\mid{\rm N}\mid{\rm F}}^{\uparrow\downarrow}$ where $\mathbf{m}$ is the magnetization of the excited film (assuming one of the ferromagnets is static). Effective damping and gyromagnetic ratio are then given by Eqs.~(\ref{geff}) and (\ref{aeff}) with ${\mathcal{A}_{{\rm F}\mid{\rm N}\mid{\rm F}}^{\uparrow\downarrow}}$ substituted for $\mathcal{A}^{\uparrow\downarrow}+\mathcal{A}^{\prime\uparrow\downarrow}$. In Eq.~(\ref{Af}), $\sigma$ is the conductivity of the N spacer, $L$ is its thickness, and $S$ is the area of the trilayer. In the spirit of the theory discussed in Sec.~\ref{icm}, Eq.~(\ref{Af}) requires that the spacer (in series with the two interfaces) is diffuse. The transmission mixing conductance $t^{\uparrow\downarrow}$ is disregarded, assuming sufficiently thick magnetic films, $d\gg\lambda_{\text{sc}}$, or insulating substrate and cap for the F$\mid$N$\mid$F trilayer. Adding inverse spin-mixing conductances in series with the diffuse-spacer resistance in Eq.~(\ref{Af}) reflects partitioning of the pumped spin currents between the two magnetic layers, having disregarded here spin relaxation in the spacer.

\onlinecite{Urban:prl01} reported room-temperature (RT) observations of increased Gilbert damping for a system consisting of two epitaxially-grown Fe layers separated by a Au spacer layer. The complete structures were GaAs$\mid$(8,11,16,21,31)Fe$\mid$40Au$\mid$40Fe$\mid$20Au(001), where the integers represent the number of monolayers (ML's), and the samples differ in the thickness of the thinner Fe film. The interface magnetic anisotropies allowed \onlinecite{Urban:prl01} to separate the FMR fields of the two Fe layers with resonance-field differences that can exceed 5 times the FMR linewidths. Hence, the FMR measurements for thinner F layer can be carried out with a nearly static thick layer: The FMR linewidth of the thin F layer increases in the presence of the second layer. The difference in the FMR linewidths between the magnetic bilayer and single-layer structures is nearly inversely proportional to the thin-film thickness $d$, suggesting that the additional damping occurs due to its F$\mid$N interface. Secondly, the additional linewidth is linearly dependent on microwave frequency for both the in-plane (the saturation magnetization parallel to the film surface) and perpendicular (the saturation magnetization perpendicular to the film surface) configurations, strongly implying that the additional contribution to the FMR linewidth can be described strictly as an interface Gilbert damping \cite{Urban:prl01}.

\begin{figure}[pth]
\includegraphics[width=0.95\linewidth,clip=]{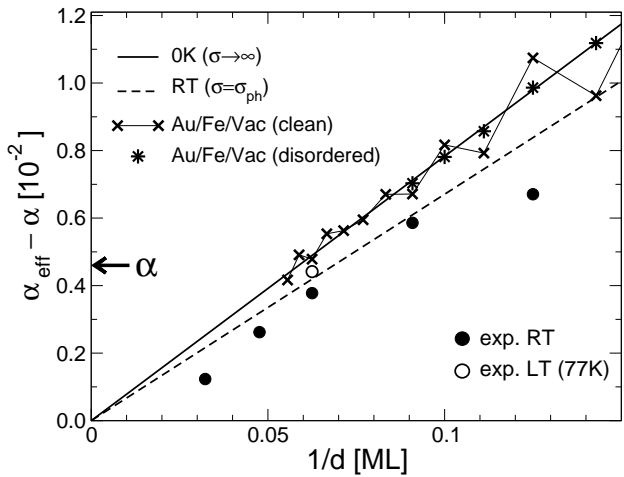}
\caption{Enhancement of the Gilbert damping coefficient for an Fe$\mid$Au$\mid$Fe trilayer as a function of $1/d$ where $d$ is the thickness of the excited Fe layer in monolayers (ML's). The filled circles ($\bullet$) are the room-temperature values measured by \onlinecite{Urban:prl01} and the open one ($\circ$) is a low-temperature (77~K) value from \onlinecite{Heinrich:jap03}. The theoretical prediction based on Eq.~(\ref{gc}) for 0~K (with $\sigma\rightarrow\infty$) is shown as solid and the RT-corrected (with phonon scattering) one as dashed lines. The results of 0~K calculations for a Au$\mid$Fe$\mid$vacuum system are given by crosses (\textbf{\textsf{x}}) and stars ($\ast$) for specular and disordered interfaces, respectively. The value of the Gilbert damping $\alpha$ for a single Fe film is marked with an arrow. From \onlinecite{Zwierzycki:prb05}.}
\label{urban}
\end{figure}

The magnetization of the thin ferromagnetic layer precesses in the external magnetic field, while the other static magnetic layer acts as a spin sink. No modification of the damping coefficient was measured for GaAs$\mid$Fe$\mid$Au structures without a second Fe layer. The latter finding is consistent with the prediction given by Eq.~(\ref{afn}) in the  $L\ll\lambda_{\text{sd}}$ limit, well fulfilled for thin Au films of \onlinecite{Urban:prl01}. In the presence of the second Fe layer, Eq.~(\ref{Af}) should be used: Neglecting $\mbox{Im}\mathcal{A}_{{\rm F}\mid{\rm N}\mid{\rm F}}^{\uparrow\downarrow}$ leads to $\gamma_{\text{eff}}=\gamma$ and the damping enhancement
\begin{equation}
\alpha_{\text{eff}}-\alpha=\frac{\hbar\gamma\mbox{Re}\mathcal{A}_{{\rm F}\mid{\rm N}\mid{\rm F}}^{\uparrow\downarrow}}{4\pi M_sdS}\,,
\label{gc}
\end{equation}
where $\alpha\approx0.0046$ is the dimensionless damping measured for a single Fe layer. Using $\gamma=2.1\mu_B/\hbar$ \cite{Heinrich:prl87} and the values of the interface and Sharvin conductances from Table~\ref{tabmix}, Eq.~(\ref{gc}) is compared with the experimental data in Fig.~\ref{urban} for various assumptions about $\sigma$ in (\ref{Af}). In the low-temperature limit and neglecting the residual resistivity of the Au layer, $\sigma\rightarrow\infty$, Eq.~(\ref{gc}) yields the solid line which is seen to overestimate the damping enhancement compared to the measured results. Using finite values of $\sigma$ will lead to lower values of $\mathcal{A}^{\uparrow\downarrow}_{{\rm F}\mid{\rm N}\mid{\rm F}}$ and indeed, it was found experimentally \cite{Heinrich:jap03} that lowering the temperature (increasing the conductivity) increases the damping by as much as about 20\% (open circle in Fig.~\ref{urban}). If one uses the RT conductivity due to phonon scattering in crystalline bulk Au, $\sigma_{\text{ph}}=0.45\times10^8~\Omega^{-1}$m$^{-1}$, the dashed line is obtained which, as expected, is closer to the RT measurements. Measurements of the sheet conductivity \cite{Heinrich:jap03} indicate that the Au layers used in the experiments have non-negligible residual resistances. [We note, however, that the conductivity entering Eq.~(\ref{Af}) does not include the interfacial-scattering contribution; the measurement of the sheet conductivity therefore does not give us a direct information about $\sigma_{\text{res}}$.] Assuming for example $\sigma_{\text{res}}\approx\sigma_{\text{ph}}$ would yield the 0~K (with $\sigma=\sigma_{\text{res}}$) and RT (with $\sigma^{-1}=\sigma_{\text{res}}^{-1}+\sigma_{\text{ph}}^{-1}$) lines in the close vicinity of the measured points.

The theoretical results represented by the straight lines in Fig.~\ref{urban} are based upon the asymptotic, single-interface value of $g^{\uparrow\downarrow}_r$ for Au$\mid$Fe from Table~\ref{tabmix}. This approximation needs to be relaxed in order to study possible size-dependent corrections in thin films. To estimate the variation which can result from finite-size effects, \onlinecite{Zwierzycki:prb05} carried out a series of calculations for a Au$\mid$Fe$\mid$vacuum system, using vacuum instead of GaAs as in the experiment \cite{Urban:prl01} for simplicity. The mixing conductance of the other, Fe$\mid$Au, interface in Eq.~(\ref{Af}) was kept at its asymptotic value (Table~\ref{tabmix}). The calculated thickness ($d$)-dependent mixing conductance $g^{\uparrow\downarrow}_r$ was then converted into the Gilbert damping via Eq.~(\ref{gc}). The results for perfect (specular) structures, marked in Fig.~\ref{urban} with black crosses (\textbf{\textsf{x}}), exhibit oscillations of non-negligible amplitude about the asymptotic values given by the solid line (arbitrarily taking the low-temperature regime, i.e., $\sigma\to\infty$ for reference). The introduction of interface disorder (two ML's of 50\%-50\% alloy) yields values for the damping [stars ($\ast$) in Fig.~\ref{urban}] essentially averaged back to the limit given by the single-interface calculations of Table~\ref{tabmix}. 

\onlinecite{Lubitz:jap03} also reported an increased magnetization damping in polycrystalline Cu$\mid$Fe$\mid$Cu$\mid$Fe$\mid$Cu multilayers as compared to Cu$\mid$Fe$\mid$Cu structures. The additional damping scaled as $1/d$ with magnetic-film thickness $d$, but was considerably larger than that for epitaxially-grown systems reported by \onlinecite{Urban:prl01}. Besides, this damping was rapidly reduced by increasing the thickness of the Cu spacer to only several nanometers, which was interpreted by \onlinecite{Lubitz:jap03} to be due to possibly a very short spin-flip length in their Cu. We note that, based on Eq.~(\ref{Af}), this could be alternatively explained by a short elastic scattering length in the polycrystalline Cu. \onlinecite{Lubitz:jap03} found a moderate increase by about 10\% in the additional damping on lowering the temperature to 77~K, which is roughly consistent with the decrease in the phonon contribution to the resistivity.

This subsection demonstrated that direct first-principles calculations can produce values of the damping coefficient in the same range as those measured experimentally in good-quality structures. Moreover, by taking into account various other sources of scattering in the Au spacer and/or quantum-size effects, the calculations can be brought into close agreement with observations. A more conclusive comparison with experiments would require a detailed knowledge of the microscopic structure of the experimental system, which is currently not available.

\section{Dynamic exchange interaction}
\label{dec}

\subsection{Magnetic bilayers}
\label{wsc}

The ground-state energy (free energy at finite temperatures) of more than one magnetic layer embedded into a nonmagnetic medium depends on the relative orientation of the magnetic moments. This is the essence of the static exchange coupling discussed in Sec.~\ref{rkky}. Disorder scrambles ballistic electron paths connecting magnetic layers and exponentially suppresses the static interaction as a function of the nonmagnetic spacer thickness between magnetic layers and the inverse mean free path. We can picture the dynamic coupling in terms of a localized magnetic moment that suddenly changes its direction, thus creating a nonequilibrium local spin accumulation. The latter partly precesses in the local-moment effective field and partly diffuses away. Other magnetic moments at distances $L$ experience this spin accumulation after a time $t_d=L^2/D$, where $D$ is the diffusion coefficient, as long as $t_d<\tau_{\text{sf}}$, the spin-flip time, or, equivalently, $L<\sqrt{D\tau_{\text{sf}}}$, the spin-diffusion length. These other moments start to move by the torque they experience by absorbing part of the diffusing spin accumulation, and, in turn, emit spin currents by themselves. When the magnetic dynamics are slow on the scale of the diffusion time $t_d$, the retardation of spin diffusion may be disregarded and the dynamic exchange coupling is practically instantaneous. This is the regime we treat in the present section. (A similar discussion applies to the ballistic transport regime.) The dynamic coupling between moving magnetizations by the exchange of nonequilibrium spin currents is affected by spin-conserving random scattering much less than the rapid suppression of the static coupling.

When the normal spacer is much thinner than the spin-diffusion length, the dynamic spin exchange is governed by $\mathcal{A}_{{\rm F}\mid{\rm N}\mid{\rm F}}^{\uparrow\downarrow}$, Eq.~(\ref{Af}), which does not depend on the N-spacer width $L$ for ballistic spacers. The coupling decays as $1/L$ for spacer widths larger than the scattering mean free path and is exponentially suppressed when $L$ becomes larger than the spin-diffusion length. The dynamic exchange interaction is less sensitive to disorder than its static counterpart because the latter relies on the orbital quantum interference between electron trajectories connecting magnetic moments/layers, whereas the former requires spin coherence on traversing the normal spacer. The dynamic exchange coupling has been first addressed perturbatively in the context of electron spin resonance by \onlinecite{Barnes:jpf74}, who already pointed out its long-range nature as compared to the static coupling. However, the dynamic correlations between an individual magnetic moment and proximate conduction electrons (which, as described in Sec.~\ref{gde}, lead to magnetization relaxation) were overlooked at that time. In the context of ferromagnetic-resonance experiments, dynamic exchange coupling has been studied by \onlinecite{Hurdequint:mmm912} and, more recently, by \onlinecite{Heinrich:prl03} and \onlinecite{Lenz:prb04}.

The spin-pumping picture of dynamic exchange coupling is impressively confirmed by recent experiments with sufficiently large normal spacers \cite{Heinrich:prl03}, as will be discussed in detail below. In the limit when the static coupling becomes appreciable, the spin-pumping--based circuit analysis leading to Eq.~(\ref{Af}) may not hold exactly, however, since the circuit model of Sec.~\ref{sm} assumes that the interfaces and normal spacers scramble the electron distribution in momentum space. A sizable static exchange is thus in principle inconsistent with the key assumption underlying the circuit model. A rigorous treatment of the dynamic coupling in the regime of significant static exchange coupling is difficult, as explained in Sec.~\ref{uns}. Here we rather heuristically assume that the dynamic coupling as described by spin-pumping theory still holds in the presence of a residual static exchange interaction, which turns out to be sufficient to make a connection with some recent experiments \cite{Lenz:prb04}. This picture can be justified by the general observation that in intermetallic heterostructures quantum size effects often modulate the semiclassical transport contribution only weakly.

In the following, we consider the collective dynamics of an F$\mid$N$\mid$F structure, i.e., two magnetic films separated by a normal metal. A possible static exchange through very thin spacers is taken into account phenomenologically by postulating a Heisenberg-type contribution (per cross-sectional unit area)
\begin{equation}
E_x=-J\mathbf{m}_1\cdot\mathbf{m}_2
\label{E}
\end{equation}
to the magnetic free-energy functional that determines effective fields (\ref{Heff}) experienced by the ferromagnets. $J$ is the Heisenberg coupling constant which is assumed to be small compared to the magnetic bulk exchange stiffness $A$ divided by the magnetic-film thickness $d$, $|J|\ll A/d$. This assumption is necessary in order to treat each individual magnetic layer as a macrospin pointing along a unit vector $\mathbf{m}_i$. $J$ depends in an oscillatory fashion on the spacer-layer thickness and favors either parallel $\left(J>0\right)$ or antiparallel $\left(J<0\right)$ orientation of the magnetic layers, as discussed in Sec.~\ref{rkky}. In nanostructured pillars, the magnetostatic interaction can also favor an antiparallel coupling of the form (\ref{E}), even at spacer layer thicknesses at which the exchange coupling vanishes. 

\begin{figure}[pth]
\includegraphics[width=0.95\linewidth,clip=]{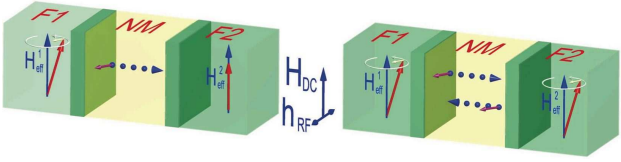}
\caption{\label{hein2} A cartoon of the dynamic-coupling phenomenon. In the left drawing, layer F$_1$ is at resonance and its precessing magnetic moment pumps spin current into the spacer, while F$_2$ is detuned from its FMR. In the right drawing, both films resonate at the same external field, inducing spin currents in opposite directions. The short arrows in N indicate the instantaneous direction of the spin angular momentum $\propto\mathbf{m}_i\times\mathbf{\dot{m}}_i$ carried away by the spin currents. Darker areas in F$_i$ around the interfaces represent the narrow regions in which the transverse spin momentum is absorbed. From \onlinecite{Heinrich:prl03}.}
\end{figure}

We consider magnetic films that are thinner than $A/|J|$ but thicker than $\lambda_{\text{sc}}$ [Eq. ~(\ref{lsc})], so that they completely absorb transverse spin currents. (Note that in typical metallic structures, $A/|J|\gg\lambda_F$ and $\lambda_{\text{sc}}\sim\lambda_F$.) A precessing magnetization vector $\mathbf{m}_i$ of ferromagnet F$_i$ pumps spin angular momentum at the rate (\ref{Is}) determined by the spin-pumping efficiency $\mathcal{A}_{{\rm F}\mid{\rm N}\mid{\rm F}}^{\uparrow\downarrow}$ [Eq.~(\ref{Af})] into the normal-metal spacer. We concentrate here on small-angle excitations of a collinear magnetic equilibrium configuration. Eq.~(\ref{Af}) applies to the typical situation that the magnetization dynamics are slow on the characteristic time scales for electron transfer across the spacer. When one ferromagnet is stationary, see the left drawing in Fig.~\ref{hein2}, the dynamics of the other film are governed by the LLG equation (\ref{llg}) but with the effective damping parameter enhanced with respect to the intrinsic value, as given by Eq.~(\ref{gc}). When both magnetizations are allowed to move, see the right sketch in Fig.~\ref{hein2}, the coupled LLG equations expanded to take into account the spin torques (\ref{Gllg}) read
\begin{align}
\frac{d\mathbf{m}_i}{dt}=&-\gamma_i\mathbf{m}_i\times\left(\mathbf{H}_{\text{eff},i}+\frac{J\mathbf{m}_j}{M_{s,i}d_i}\right)+\alpha_i\mathbf{m}_i\times\frac{d\mathbf{m}_i}{dt}\nonumber\\
&+\alpha_i^\prime\left(\mathbf{m}_i\times\frac{d\mathbf{m}_i}{dt}-\mathbf{m}_j\times\frac{d\mathbf{m}_j}{dt}\right)\,,
\label{Cllg}
\end{align}
where $\mathbf{H}_{\text{eff},i}$ are effective fields not including the exchange contribution (\ref{E}), $\alpha_i^\prime=\hbar\gamma_i\mbox{Re}\mathcal{A}_{{\rm F}\mid{\rm N}\mid{\rm F}}^{\uparrow\downarrow}/(4\pi M_{s,i}d_iS)$ is given by Eq.~(\ref{gc}), having disregarded $\mbox{Im}\mathcal{A}_{{\rm F}\mid{\rm N}\mid{\rm F}}^{\uparrow\downarrow}$, and $j=1(2)$ for $i=2(1)$. As a first simple example, consider the parallel equilibrium configuration, $\mathbf{m}_1^{(0)}=\mathbf{m}_2^{(0)}$, with zero static coupling, $J=0$, and matched resonance conditions: $\alpha_1=\alpha_2=\alpha$ and $\gamma_1\mathbf{H}_{\text{eff},1}=\gamma_2\mathbf{H}_{\text{eff},2}$, with magnetization-independent effective fields $\mathbf{H}_{\text{eff},i}$. After linearizing Eq.~(\ref{Cllg}) in terms of small deviations $\mathbf{u}_i(t)=\mathbf{m}_i(t)-\mathbf{m}_i^{(0)}$ of the magnetization direction $\mathbf{m}_i$ from its equilibrium value $\mathbf{m}_i^{(0)}$, we immediately see that $\mathbf{u}=(\mathbf{u}_1s_1+\mathbf{u}_2s_2)/(s_1+s_2)$, where $s_i=\gamma_iM_{s,i}d_i$, viz., the symmetric mode, is damped with the intrinsic Gilbert parameter $\alpha$, whereas the difference $\Delta\mathbf{u}=\mathbf{u}_1-\mathbf{u}_2$, viz., the antisymmetric mode, relaxes with enhanced damping constant $\alpha=\alpha+\alpha_1^\prime+\alpha_2^\prime$. This demonstrates that the dynamic interaction can lead to nontrivial collective magnetization dynamics even when the static interaction vanishes.

\begin{figure}[pth]
\includegraphics[width=0.95\linewidth,clip=]{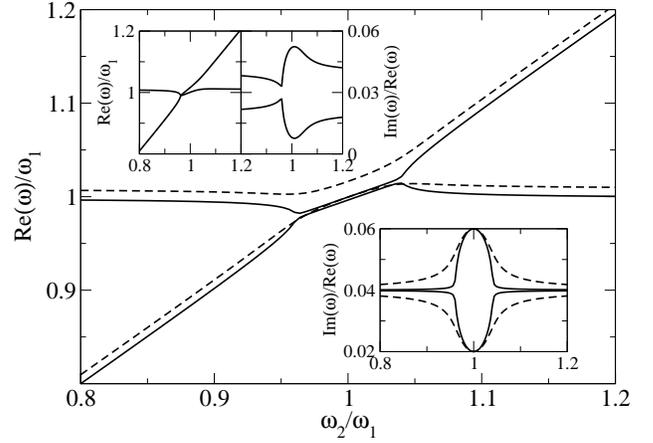}
\caption{\label{b1} The main panel displays the real part of the two frequencies $\omega$ that solve Eq.~(\ref{Sllg}) for the linearized dynamics of a symmetric magnetic bilayer close to the parallel magnetization configuration. Solid lines are obtained by setting the static coupling to zero, $\omega_x=0$, and dashed lines for $\omega_x=0.01\omega_1$. The lower inset shows the corresponding imaginary part of $\omega$. When $\omega_x=0$ and near the frequency crossing, $\omega_1\approx\omega_2$, the symmetric mode has a smaller imaginary part (less damping) and only slightly larger real part. For $\omega_x < 0$, the symmetric mode has a larger (real part of the) frequency than the antisymmetric mode, whereas the antisymmetric mode acquires a higher frequency for sufficiently large and positive $\omega_x$. In all cases, the antisymmetric mode remains to be stronger damped. The upper inset shows the results for $\omega_x=0.01\omega_1$ but setting $\alpha_1=0$ while keeping a finite $\alpha_2$. This inset illustrates the effect of a bilayer asymmetry.}
\end{figure}

Let us now analyze a more general case of a coupled magnetic bilayer undergoing a collective circular precession near a parallel equilibrium configuration $\mathbf{m}_1^{(0)}=\mathbf{m}_2^{(0)}=\mathbf{\hat{z}}$. For simplicity, we still focus on a nearly symmetric structure by setting $\gamma_i=\gamma$, $\gamma J/(M_{s,i}d_i)=\omega_x$, $\alpha_i=\alpha$, and $\alpha_i^\prime=\alpha^\prime$, but allow the effective fields $\gamma \mathbf{H}_{\text{eff},i}=\omega_i\mathbf{\hat{z}}$ to differ, $\omega_1\neq\omega_2$. Eq.~(\ref{Cllg}) then reduces to
\begin{align}
\frac{d\mathbf{m}_i}{dt}=&\omega_i\mathbf{\hat{z}}\times\mathbf{m}_i+\omega_x\mathbf{m}_i\times\mathbf{m}_j+\alpha\mathbf{m}_i\times\frac{d\mathbf{m}_i}{dt}\nonumber\\
&+\alpha^\prime\left(\mathbf{m}_i\times\frac{d\mathbf{m}_i}{dt}-\mathbf{m}_j\times\frac{d\mathbf{m}_j}{dt}\right)\,.
\label{Sllg}
\end{align}
The linearized equations of motion in the absence of driving force are solved by the form $\propto\exp(i\omega t)$ with two complex-valued natural frequencies $\omega$ and definite circular polarization. Solid lines in the main panel of Fig.~\ref{b1} show the real part of these $\omega$'s for various ratios $\omega_2/\omega_1$ after setting $\omega_x=0$, and dashed lines after setting $\omega_x/\omega_1=0.01$, i.e., introducing a finite static exchange coupling. In both cases, $\alpha=\alpha^\prime=0.02$. The lower inset shows the corresponding normalized imaginary part of $\omega$. For $\omega_x=0$ and resonance frequencies $\omega_1$ and $\omega_2$ well separated on the scale of the enhanced damping $\alpha^\prime$, viz., $|\omega_2/\omega_1-1|\gg2\alpha^\prime$ (assuming $\alpha^\prime\ll1$), the dynamics of two ferromagnets decouple and the spin pumping can be accounted for by simply adding $\alpha^\prime$ to the effective Gilbert parameter of each F layer. When, on the other hand, $|\omega_2/\omega_1-1|\lesssim2\alpha^\prime$, the spin pumping locks the collective dynamics to independent symmetric (acoustic) and antisymmetric (optic) normal modes with frequencies that are nearly degenerate and close to $(\omega_1+\omega_2)/2$. It then follows from Eq.~(\ref{Sllg}) that the symmetric mode is weakly damped, with $\mbox{Im}(\omega)/\mbox{Re}(\omega)$ close to $\alpha$, whereas the antisymmetric mode experiences an enhanced Gilbert damping $\alpha+2\alpha^\prime$.

\begin{figure}[pth]
\includegraphics[width=0.95\linewidth,clip=]{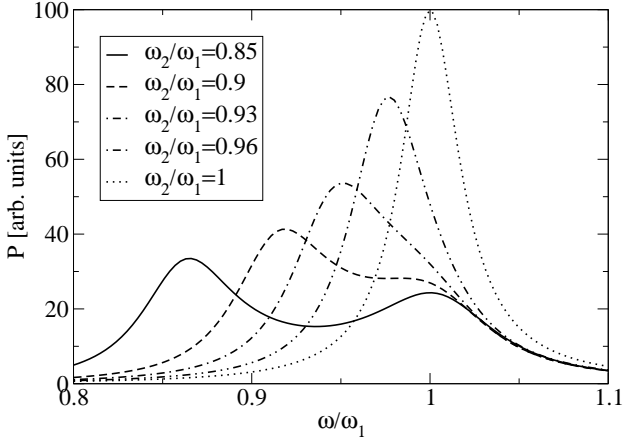}
\caption{\label{b2}Dissipation power $P$ of a symmetric bilayer in the presence of a uniform circularly-polarized rf driving field at frequency $\omega$. The parameters are the same as those used to generate the dashed lines in Fig.~\ref{b1}.}
\end{figure}

The presence of a static exchange interaction lifts the frequency of the antisymmetric mode by $2\omega_x$, not affecting the symmetric mode. When the frequencies $\omega_i$ are well separated, the static interaction lifts both of them by $\omega_x$. One of the modes then acquires some symmetric and the other some antisymmetric character even when $|\omega_2/\omega_1-1|\gtrsim2\alpha^\prime$, as illustrated by the damping of both modes (dashed lines in the lower inset of Fig.~\ref{b1}). In the upper inset of Fig.~\ref{b1}, the calculations are repeated for finite $\omega_x$, but setting the intrinsic damping of one of the films to zero, $\alpha_1=0$, while $\alpha_2=\alpha^\prime_i=0.02$. The dynamic locking into symmetric/antisymmetric modes close to the resonance crossing has disappeared. The modes are harder to synchronize when the participating modes have different amplitudes. A significant locking may be expected only in a symmetric bilayer with the individual films having similar resonant modes near parallel equilibrium axes. In particular, a bilayer in an antiparallel equilibrium configuration is not disposed to dynamic locking, since individual layers have excitations with opposite circular (or elliptic, in the presence of anisotropies) polarizations.

We now turn to a discussion of the consequences for the observable in FMR experiments, viz., the energy-dissipation power $P$ of the dynamically-locked collective dynamics as plotted in Fig.~\ref{b2} as a function of the frequency $\omega$ of the uniform and circularly-polarized transverse rf driving field. The system parameters are the same as those used to generate the dashed lines in Fig.~\ref{b1}. For the smallest ratio $\omega_2/\omega_1$ in Fig.~\ref{b2}, the two bilayer excitations are recognized as separated Lorentzian peaks with halfwidths close to $(\alpha+\alpha^\prime)\omega$, see Eq.~(\ref{P}). When the elementary bilayer excitations become locked at $|\omega_2/\omega_1-1|\lesssim2\alpha^\prime$, the rf radiation excites only the symmetric mode, and the two Lorentzians merge into a single sharper Lorentzian with halfwidth of only $\alpha\omega$. This narrowing is explained by the cancellation of spin currents and can be quite dramatic when $\alpha\ll\alpha^\prime$.

In FMR experiments, the applied magnetic field is swept, whereas the rf frequency $\omega$ is fixed by a resonant cavity. In Sec.~\ref{abi}, we discussed FMR experiments by \onlinecite{Urban:prl01} on magnetic bilayers in which one of the layers remains close to equilibrium while the other is resonantly excited. The collective magnetization dynamics were measured by \onlinecite{Heinrich:prl03} in the same system by making use of an accidental crossing of the resonance frequencies when the static field is reoriented relative to the crystal anisotropy axes. The resulting spectra could be quantitatively explained in terms of the dynamic coupling in the limit of a vanishing static exchange interaction. We summarize their findings in the following.

\begin{figure}[pth]
\includegraphics[width=0.95\linewidth,clip=]{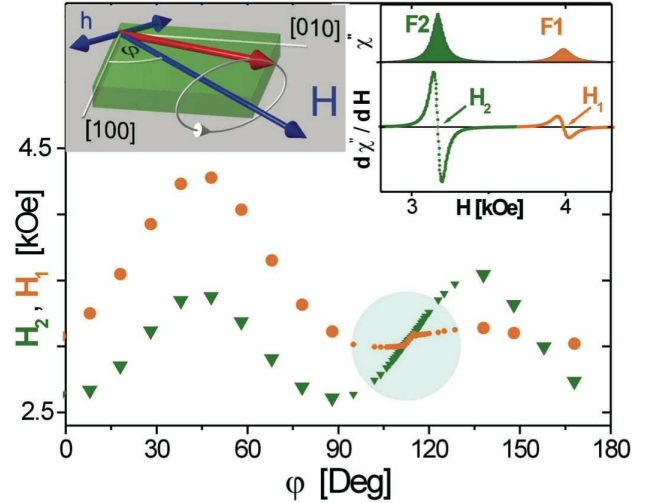}
\caption{\label{hein1} Dependence of the static FMR fields $H_1$ (circles) and $H_2$ (triangles) of the thin Fe film (F$_1$) and the thick Fe film (F$_2$), respectively, on the angle $\varphi$ of the static magnetic field with respect to the Fe [100] crystallographic axis. The sketch of the in-plane measurement in the left inset shows how the rf magnetic field (double-pointed arrow) drives the magnetization (on a scale grossly exaggerated for easy viewing). The right inset shows the measured absorption peaks for layers F$_1$ and F$_2$ at $\varphi=60$~Deg. The absorption power is given by the imaginary part of the susceptibility of the rf magnetization component along the rf driving field, which is denoted in the figure by $\chi^{\prime\prime}$. From \onlinecite{Heinrich:prl03}.}
\end{figure}

\begin{figure*}[pth]
\includegraphics[width=0.9\linewidth,clip=]{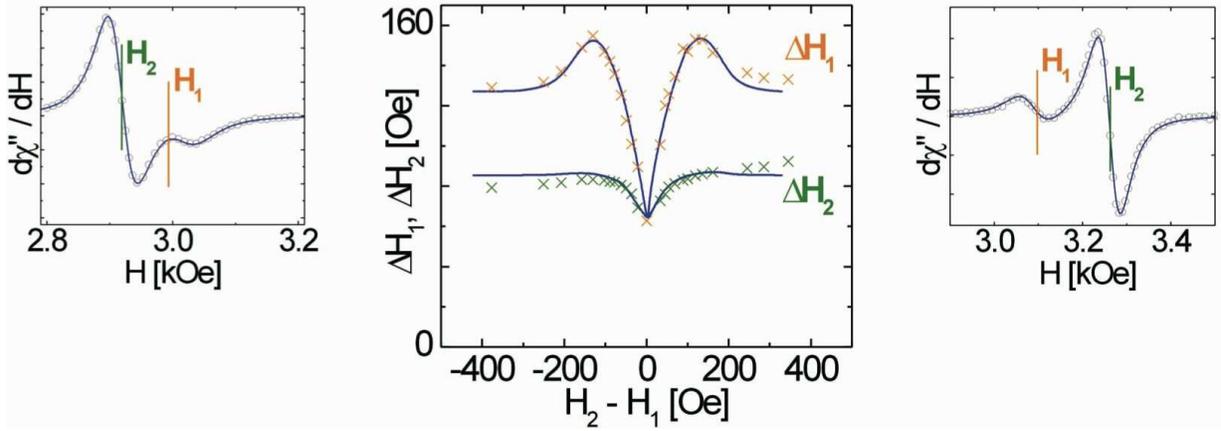}
\caption{\label{hein3} Comparison of theory (solid lines) with room-temperature measurements (symbols) close to and at the crossover of the FMR fields, marked by the shaded area in Fig.~\ref{hein1}. The left and right frames show FMR signals for the field difference, $H_2-H_1$, of -78~Oe and +161~Oe, respectively. The theoretical results are parametrized by the full set of magnetic parameters which were measured independently \cite{Urban:prl01}. The magnitude of the spin-pump current was determined by the linewidth at large separation of the FMR peaks and agrees well with that predicted theoretically, see Sec.~\ref{abi}. The middle frame displays the effective FMR linewidth of magnetic layers for the signals fitted by two Lorentzians as a function of the external field. At $H_1=H_2$, the FMR linewidths reached their minimum values at the level of intrinsic Gilbert damping of isolated films. The calculations in the middle frame did not take small variations of the intrinsic damping with angle $\varphi$ into account, which resulted in deviations between theory and experiment for larger $|H_1-H_2|$. Notice that $\Delta H_1$ first increases before attaining its minimum, which is due to a contribution of the antisymmetric collective mode. As a side comment, it should be noted that although fitting the absorption signal by two Lorentzians is a legitimate approach to comparing theoretical calculation with experimental curves, the analysis does not imply that the signal is always well approximated by the sum of two Lorentzians, which may not be the case very close to the FMR-field crossover. From \onlinecite{Heinrich:prl03}.}
\end{figure*}

The MBE-grown structure of \onlinecite{Heinrich:prl03} incorporates two ferromagnetic films, a thinner 16~ML (F$_1$) and a thicker 40~ML (F$_2$) Fe film, separated by 40~ML's of Au, grown on GaAs and capped with Au, i.e., the stacking order is GaAs$\mid$16Fe$\mid$40Au$\mid$40Fe$\mid$20Au(001). The uniaxial magnetic anisotropy in F$_1$ at the GaAs$\mid$Fe interface can be used to intentionally tune the resonance fields for F$_1$ and F$_2$ into a crossover by rotating the static magnetic field by an angle $\varphi$ with respect to the (001) crystal axis. In a finite interval of $\varphi$ near the crossover (shaded area in Fig.~\ref{hein1}), the two FMR fields clearly ``stick" to each other, a phenomenon explained above. When the resonance fields are identical, $H_1=H_2$, the rf magnetization components of F$_1$ and F$_2$ are moving in phase as depicted in the right drawing in Fig.~\ref{hein2}. For similar trajectories of F$_1$ and F$_2$, the total spin current through the spacer vanishes resulting in zero excess damping for both films, as follows from Eq.~(\ref{Cllg}). The locked collective motion is then hindered only by the intrinsic local damping. This is experimentally verified, as shown in Fig.~\ref{hein3}. For a theoretical analysis, \onlinecite{Heinrich:prl03} solved Eq.~(\ref{Cllg}), taking into account the ellipticity of the magnetic motion caused by the anisotropies. Using parameters derived from measurements on the uncoupled layers, they calculated the total FMR signal as a function of the difference between the resonance fields, $H_2-H_1$, without additional fitting parameters. The predictions are compared with the measurements in Fig.~\ref{hein3}. The remarkably good agreement between experiment and theory provides strong evidence that the dynamic exchange coupling not only contributes to the damping but leads to a new collective behavior of magnetic heterostructures. \onlinecite{Heinrich:prl03} additionally carried out measurements on samples with Au spacer thicknesses between 14 and 100 monolayers: The reported weak dependence of the FMR response on the spacer thickness proves the long range of the dynamic interaction.

\onlinecite{Lenz:prb04} investigated collective FMR dynamics in Ni$\mid$Cu$\mid$Ni and Ni$\mid$Cu$\mid$Co structures with Cu thicknesses down to 2~ML's. Such thin Cu spacers support a sizable static coupling between magnetic films. Like \onlinecite{Heinrich:prl03}, \onlinecite{Lenz:prb04} observed a sharp drop in the linewidth near an FMR-field crossing. Far from it, the optic-mode resonance is systematically broader than that of the acoustic mode, consistent with the spin-pumping mechanism. The difference between the optic and acoustic linewidths exhibits an oscillatory dependence on the Cu-spacer thickness, which roughly follows the predicted exchange-coupling constant $J$. As shown in the lower inset of Fig.~\ref{b1}, a stronger coupling $J$ would indeed result in a larger asymmetry in the two linewidths, away from the crossing, which could lead to oscillations in the linewidth difference. However, an inhomogeneous spread of the coupling strengths may also contribute to the broadening which depends nonmonotonically on the spacer thickness, providing an alternative explanation for the linewidth oscillations.

In the regime of very strong static exchange coupling, $|J|\gtrsim A/d$, opposite to what was assumed so far, magnetization gradients across the bulk magnetic layers alleviate the energy cost of the discontinuity in the magnetic orientation between two magnetic layers. This reduces the spin pumping through the spacer, and thus the additional broadening of the antisymmetric mode, because the adjacent magnetizations are better locked by the strong static exchange interaction.

\subsection{Magnetic superlattices}

Magnetic multilayers display a rich pattern of physical properties that have been well investigated (see, \text{e.g.}, \onlinecite{Camley:jp93} for a review). However, the relevance of the dynamic exchange coupling discussed in the previous subsection on the spin-wave dispersions and lifetimes appears to have not been recognized yet. In the following, we present a simple model description that should suffice to estimate the order of magnitude of the predicted effects that hopefully will stimulate new experiments. We also remark that superlattices can serve a theoretical purpose as toy models for describing certain features of \textit{bulk} magnetism.

\begin{figure}[pth]
\includegraphics[width=0.95\linewidth,clip=]{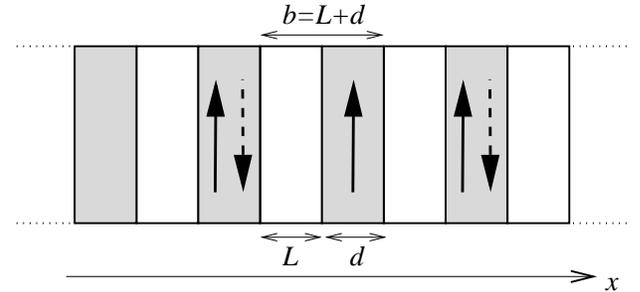}
\caption{A schematic view of the superlattice and its geometric parameters as considered in the text: An F$\mid$N bilayer is repeated along the $x$ axis, with either ferromagnetic or antiferromagnetic alignment of the consecutive magnetic layers. The system is translationally invariant along the two remaining axes.}
\label{sl}
\end{figure}

Consider a periodic stack (in the $x$ direction) of alternating F and N layers forming a two-component superlattice. We treat the model depicted in Fig.~\ref{sl}, in which an F$\mid$N bilayer forms the unit cell with thickness $b=L+d$, where the normal-metal spacer of width $L$ separates the magnetic films of thickness $d\gg\lambda_{\text{sc}}$. Translational invariance is assumed in the lateral directions. We consider here collective spin-wave excitations, taking both static and dynamic exchange couplings into account, in exactly the same fashion as in the previous subsection on magnetic bilayers, modeling each magnetic layer as a single macrospin.

The small-angle magnetization dynamics of a multilayer in an all-parallel configuration are described in terms of local deviations from the equilibrium: $\mathbf{u}_{i}(t)=\mathbf{m}_{i}(t)-\mathbf{m}^{(0)}$. For long-wavelength excitations, it may be approximated as a continuous function $\mathbf{u}(x,t)$ of the coordinate $x$ normal to the interfaces. For a uniaxial effective field $H_{\text{eff}}=H_{\text{eff}}\mathbf{m}^{(0)}$, the spin-wave dynamics obey the differential equation
\begin{align}
\partial_t\mathbf{u}=&\mathbf{m}^{(0)}\times\left[\omega_0\mathbf{u}-\omega_xb^2\partial_{xx}\mathbf{u}\right.\nonumber\\
&\hspace{1.3cm}\left.+\alpha\partial_t\mathbf{u}-\alpha^\prime b^2\partial_{xx,t}\mathbf{u}\right]\,,
\label{sle}
\end{align}
where we used the quantities defined for magnetic bilayers in Sec.~\ref{wsc}: $\omega_x=\gamma J/(M_sd)$, $\alpha^\prime=\hbar\gamma\mbox{Re}A_{{\rm F}\mid{\rm N}\mid{\rm F}}^{\uparrow\downarrow}/(4\pi M_sdS)$, and $\omega_0=\gamma H_{\text{eff}}$. The second term on the right-hand side of Eq.~(\ref{sle}) is due to the static exchange interaction mediated by quantum-well states of the spacer layers, and in the last term we recognize the dynamic coupling induced by the spin pumping. The second spatial derivatives reflect simply the difference of the spin currents through two consecutive normal spacers in the continuum limit. The static Heisenberg coupling can be interpreted as the superlattice equivalent of the bulk exchange-stiffness parameter $A$, which for the superlattice is given by $\tilde{A}=Jb^2/d$. Both $\omega_x$ and $\alpha^\prime$ depend on the normal-interlayer thickness $L$. It follows from Eq.~(\ref{sle}) that the small-momentum, $k\ll b^{-1}$, spin-wave excitations of the superlattice, propagating perpendicular to the interfaces, $\propto\exp\{i[kx-\omega(k)t]\}$, obey a dispersion relation
\begin{equation}
\omega(k)=\frac{\omega_0+(bk)^2\omega_x}{1+i\left[\alpha+(bk)^2\alpha^\prime\right]}\,.
\label{oak}
\end{equation}
When $k\rightarrow0$, $\omega(k)$ reduces to the Larmor frequency $\omega_0$ of the individual magnetic layers because the static and dynamic exchange couplings vanish when the consecutive magnetic layers move coherently in phase, as explained in Sec.~\ref{wsc}. Eq.~(\ref{oak}) holds for momenta comparable to $b^{-1}$ when $bk$ is replaced by $2\sin(bk/2)$.

The situation is very different for an antiferromagnetically-aligned superlattice, which is the lowest-energy state when, for example, $J<0$ and $H_{\text{eff}}=0$. In this case,
\begin{equation}
\omega(k)=\frac{-i\omega_x\left[2\alpha+(bk)^2\alpha^\prime\pm\sqrt{4\alpha^2-(bk)^2(1+\alpha^2)}\right]}{1+\alpha^2+4\alpha\alpha^\prime+\alpha^\prime(bk)^2}\,,
\label{obk}
\end{equation}
where plus and minus signs refer, respectively, to the modes with antisymmetric and symmetric dynamics in adjacent layers for overdamped motion, and to the right- and left-propagating modes when the real part of $\omega(k)$ is significant. Note that now $\omega_x<0$, so that $\mbox{Im}\,\omega>0$, as required for a stable configuration. In the absence of bulk magnetization damping, $\alpha=0$, Eq.~(\ref{obk}) reduces to
\begin{equation}
\omega(k)=\frac{\pm(bk)\omega_x}{1\pm i(bk)\alpha^\prime}\,,
\label{obks}
\end{equation}
with linear dispersion and damping at small $k$. Eqs.~(\ref{obk}) and (\ref{obks}) can also be generalized to large momenta by replacing $bk$ with $2\sin(bk/2)$. Notice that in Eqs.~(\ref{sle}), (\ref{oak}), and (\ref{obks}), the dynamic coupling modifies the damping similarly to the way the static coupling affects the excitation frequency of the magnetic superlattice. Crystal and shape anisotropies on top of the simple effective fields assumed above, might become important in real structures, and their inclusion is straightforward.

FMR experiments access the multilayer dynamics from the sample surface down to the microwave skin depth $\lambda_{\text{skin}}\sim100$~nm (which is even smaller in BLS). Therefore only modes with momenta $k\gtrsim\pi/\lambda_{\text{skin}}$ can be measured. Since the skin depth decreases when the temperature is lowered, both the FMR frequency and the damping in a ferromagnetically-aligned multilayer should grow roughly as $1/\lambda_{\text{skin}}^2\propto\tau$, the momentum scattering time in the normal skin-effect regime. Such studies can thus give information about the temperature-dependent scattering in superlattices; a weak temperature dependence could indicate that scattering is dominated by structural disorder. Inelastic neutron-scattering spectroscopy may be useful in elucidating the collective dynamics in thick multilayers, especially if supported by elastic neutron scattering \cite{Fitzsimmons:mmm04} to probe the magnetic profile in the superlattice.

\subsection{Large-angle motion in biased spin valves}
\label{bsv}

Perpendicular spin valves, i.e., F$_s$$\mid$N$\mid$F$_h$ trilayer pillar structures with layer thicknesses down to a few monolayers and submicron lateral dimensions, are ideal systems to study precession and switching phenomena in magnetic heterostructures. By attaching contacts on the outer sides an electric bias can be applied perpendicular to the interface planes. In many experiments, F$_s$ is a ``soft" ferromagnetic film with a magnetization that can change easily, whereas F$_h$ is a ``hard" magnetic layer whose magnetization is assumed to be stationary. The relevant variable is then the time-dependent magnetization of the soft layer. The soft layer can be excited by a current bias or an applied rf magnetic field (or both). In the second case (realized as, e.g., an isolated magnetic bilayer of Secs.~\ref{abi} and \ref{wsc}), F$_h$ can be pinned by an exchange bias or surface magnetic anisotropy \cite{Urban:prl01}. In the former case, the magnetization of layer F$_h$ may be also rendered less sensitive to a given spin torque simply by growing it much thicker than F$_s$ or by a resistance anisotropy \cite{Kovalev:prb02}. For a sufficiently thick spacer N, the static interaction between the ferromagnetic layers can be disregarded, while the dynamic coupling induced by the spin pumping may be still sizable, see Sec.~\ref{wsc}.

\onlinecite{Sloncz:mmm96} and \onlinecite{Berger:prb96} predicted that spin valves should display novel time-dependent effects. \onlinecite{Sloncz:mmm96} realized that a current flowing through a spin valve causes a spin transfer through the nonmagnetic spacer, inducing spin torques on the ferromagnets. \onlinecite{Berger:prb96} predicted that the two ferromagnets should interact even without an applied electric current, resulting in a significant contribution to the Gilbert damping of the magnetization dynamics. He further showed that an electric current can excite zero-momentum spin-waves in the ferromagnet, an idea that was later supported experimentally, see, e.g., \onlinecite{Tsoi:nat00}. The condition for the resonant spin-wave emission \cite{Berger:prb96} is similar to the criterion for magnetization reversal by \onlinecite{Sloncz:mmm96,Sloncz:mmm99}, who treated the Gilbert damping parameter as a phenomenological constant. \onlinecite{Berger:jap01} however found a dependence of the damping parameter in spin valves on the relative magnetization angle. Some of Berger's and Slonczewski's results as well as the underlying theoretical models are thus not consistent with each other. The spin-pumping concept unified the seminal work of these pioneers \cite{Tserkovnyak:prb031} as explained in the remainder of this subsection. We will calculate the critical current bias for the low-temperature magnetization instability and the configuration-dependent Gilbert damping parameter. The treatment here is limited to the macrospin model, i.e., to small systems in which the magnetic layers are monodomain ferromagnets characterized by two magnetization vectors. If combined with micromagnetic simulations \cite{Lee:natmat04}, the full range of the precession and switching dynamics can be studied in principle, however.

\begin{figure}[pth]
\includegraphics[width=0.95\linewidth,clip=]{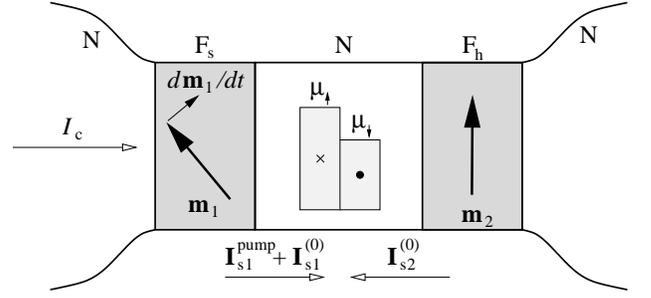}
\caption{\label{fnf}Schematic of a current-biased spin valve (a ``dynamic" version of Fig.~\ref{sv}). The symbols are explained in the text.}
\end{figure}

Consider the system sketched in Fig.~\ref{fnf}. The F$_s$$\mid$N$\mid$F$_h$ trilayer is sandwiched between two normal-metal contacts sustaining a charge-current bias $I_c$. The soft layer F$_s$ magnetization $\mathbf{m}_1$ then starts moving from its equilibrium direction at a critical value that depends on the applied magnetic field. Thermal activation facilitates current-induced magnetization switching \cite{Myers:prl02}, but we focus here on the low-temperature regime. The spin torque on the magnetization of F$_s$ in the presence of a spin current $\mathbf{I}_{s1}$ flowing from F$_s$ into the normal spacer is given by Eq.~(\ref{Gllg}). The spin current
\begin{equation}
\mathbf{I}_{s1}=\mathbf{I}_{s1}^{\text{exch}}+\mathbf{I}_{s1}^{\text{bias}}
\label{Ieb}
\end{equation}
consists of the dynamic-exchange current $\mathbf{I}_{s1}^{\text{exch}}$ induced by the spin pumping (\ref{Is}) and of the current $\mathbf{I}_{s1}^{\text{bias}}$ driven by an applied current bias. The former is responsible for a dynamic coupling between the ferromagnets, see Sec.~\ref{wsc}, and, as we discuss in the following, can be interpreted as a viscous friction term that stabilizes the relative magnetization configuration of the spin valve against the torques exerted by $\mathbf{I}_s^{\text{bias}}$ or an applied magnetic field. In high-density metallic systems, the applied voltages and spin accumulations are safely smaller than the Fermi energies, which means that we are in the linear-response regime and both spin currents may be calculated independently of each other. Spin pumping in the outward direction, i.e., into the external connectors, would only increase the intrinsic damping coefficient by a constant value, as discussed in Secs.~\ref{pss} and \ref{ds}, and can thus be added trivially to the intrinsic damping $\alpha$. Spin pumping into the normal spacer, which gives rise to $\mathbf{I}_{s1}^{\text{exch}}$, requires more attention.

We start by considering the spin current (\ref{Is}) pumped into the spacer by a time-dependent $\mathbf{m}_1(t)$ in the absence of an applied current bias, $I_c=0$. The following assumptions are convenient and realistic: (i) The magnetic films are sufficiently thicker than $\lambda_{\text{sc}}$, so that the transmission contribution to the spin-pumping parameter $\mathcal{A}^{\uparrow\downarrow}$ can be disregarded, (ii) the mixing conductance $\tilde{g}^{\uparrow\downarrow}$ is real-valued, (iii) two magnetic films including the interfaces have the same conductances, (iv) the normal spacer is ballistic, and (v) the spin-flip processes are restricted to the ferromagnetic layers, with the spin-diffusion length $\lambda_{\text{sd}}\gg\lambda_{\text{sc}}$. (Note that in this subsection $\lambda_{\text{sd}}$ denotes the spin-diffusion length inside the \textit{ferromagnets} for longitudinal spin transport along the magnetization direction). The spin currents out of the magnetic layers into the normal spacer are then given by $\mathbf{I}_{si}=\mathbf{I}_{si}^{\text{pump}}+\mathbf{I}_{si}^{(0)}$ with
\begin{align}
\label{Ipump}
\mathbf{I}_{si}^{\text{pump}}=&\frac{\hbar}{4\pi}\tilde{g}_r^{\uparrow\downarrow}\mathbf{m}_i\times\frac{d\mathbf{m}_i}{dt}\,,\\
\mathbf{I}_{si}^{(0)}=&-\frac{1}{4\pi}\left[\frac{2\tilde{g}^{\uparrow\uparrow}\tilde{g}^{\downarrow\downarrow}}{\tilde{g}^{\uparrow\uparrow}+\tilde{g}^{\downarrow\downarrow}}\mathbf{m}_i\left(\Delta\boldsymbol{\mu}_{si}\cdot\mathbf{m}_i\right)\right.\nonumber\\
&+\left.\tilde{g}^{\uparrow\downarrow}_r\mathbf{m}_i\times\Delta\boldsymbol{\mu}_{si}\times\mathbf{m}_i\right]\,,
\label{Iback}
\end{align}
where the last equation is derived using Eqs.~(\ref{IcR}), (\ref{IsR}) and a circuit-theory analysis discussed in Sec.~\ref{Mdc}, assuming vanishing charge current. Here, $\mathbf{m}_i$ is the $i$th-layer magnetization direction and $\Delta\boldsymbol{\mu}_{si}=\boldsymbol{\mu}_{sN}-\mu_{sFi}\mathbf{m}_i$ is the spin-accumulation difference across the N$\mid$F$_i$ interface. The magnetization-precession period is typically much longer than the electron dwell time in metallic spacers. Assuming weak spin-flip scattering in N, conservation of angular momentum then implies that $\mathbf{I}_{s1}+\mathbf{I}_{s2}=0$. It is now straightforward to calculate the exchange spin current which is given by $\mathbf{I}_{s1}^{\text{exch}}=\mathbf{I}_{s1}$ when $I_c=0$.

The longitudinal component of the spin accumulation $\boldsymbol{\mu}_{sN}$ can penetrate into the ferromagnets on the scale of the spin-diffusion length $\lambda_{\text{sd}}$, whereas the transverse component vanishes on the shorter scale of $\lambda_{\text{sc}}$ near the interface. The longitudinal spin accumulation  in the ferromagnet, $\mu_{sFi}$, and thereby $\mathbf{I}_{si}^{(0)}$, can be obtained for a given $\boldsymbol{\mu}_{sN}$ from the diffusion equation for the (longitudinal) spin transport in the ferromagnets, similarly to the normal-layer spin diffusion discussed in Sec.~\ref{ds}. To be specific, we take both the charge and spin currents to vanish on the outer boundaries of F$_s$ and F$_h$. It can then be shown \cite{Tserkovnyak:prb031} that the longitudinal spin current flowing into a ferromagnetic slab of thickness $d$ is governed by an effective conductance $g^\ast$ defined by
\begin{equation}
\frac{1}{g^\ast}=\frac{\tilde{g}^{\uparrow\uparrow}+\tilde{g}^{\downarrow\downarrow}}{2\tilde{g}^{\uparrow\uparrow}\tilde{g}^{\downarrow\downarrow}}+\frac{1}{g_{\text{sd}}\tanh(d/\lambda_{\text{sd}})}\,,
\label{gast}
\end{equation}
where $g_{\text{sd}}=(h/e^2)(S/\lambda_{\text{sd}})(2\sigma^\uparrow\sigma^\downarrow)/(\sigma^\uparrow+\sigma^\downarrow)$ and $\sigma^s$ is the spin-$s$ conductivity of the ferromagnetic bulk, so that
\begin{equation}
\mathbf{I}_{si}^{(0)}=-\frac{1}{4\pi}\left[g^\ast\mathbf{m}_i\left(\boldsymbol{\mu}_{sN}\cdot\mathbf{m}_i\right)+\tilde{g}^{\uparrow\downarrow}_r\mathbf{m}_i\times\boldsymbol{\mu}_{sN}\times\mathbf{m}_i\right]\,.
\label{backN}
\end{equation}
The transverse spin current is determined simply by $\tilde{g}^{\uparrow\downarrow}_r$, since we have taken $\lambda_{\text{sc}}$ to be the shortest relevant length scale in the problem. Note that $g^\ast\rightarrow0$ when $d\ll\lambda_{\text{sd}}$, \text{i.e.}, when the spin-flip relaxation vanishes, or when the ferromagnet is halfmetallic, so that it completely blocks the longitudinal spin flow for a vanishing charge flow. A new parameter
\begin{equation}
\nu=\frac{\tilde{g}_r^{\uparrow\downarrow}-g^\ast}{\tilde{g}_r^{\uparrow\downarrow}+g^\ast}
\end{equation}
characterizes the asymmetry of the absorption of transverse vs longitudinal spin currents. Putting together Eqs.~(\ref{Ipump}) and (\ref{backN}) and demanding conservation of angular momentum in the spacer (i.e., $\mathbf{I}_{s1}+\mathbf{I}_{s2}=0$), one arrives (after some algebra) at
\begin{equation}
\mathbf{I}_{s1}^{\text{exch}}=\frac{1}{2}\left[\mathbf{I}_{s1}^{\text{pump}}-\nu\left(\mathbf{I}_{s1}^{\text{pump}}\cdot\mathbf{m}_2\right)\frac{\mathbf{m}_2-\nu\mathbf{m}_1\cos\theta}{1-\nu^2\cos^2\theta}\right]\,,
\label{exch}
\end{equation}
where $\cos\theta=\mathbf{m}_1\cdot\mathbf{m}_2$. Since the normal spacer was taken to be ballistic, $\boldsymbol{\mu}_{sN}$ is uniform. The exchange spin current would otherwise be somewhat suppressed by the spacer diffuse scattering, which can be taken into account easily by solving the spin-diffusion equation as in Sec.~\ref{ds}, if necessary. Let us estimate typical values of $\nu$ for sputtered Cu$\mid$Co and Cu$\mid$Py systems at low temperatures, taking $d=5$~nm. The main difference between the two material combinations is the spin-diffusion length in the ferromagnets: Co has a relatively long $\lambda_{\text{sd}}\approx60$~nm, while $\lambda_{\text{sd}}\approx5$~nm is very short in Py \cite{Piraux:mmm96,Fert:mmm99,Bass:mmm99}. Using known values for spin-dependent conductivities \cite{Piraux:mmm96,Fert:mmm99,Bass:mmm99}, we obtain $g_{\text{sd}}/S\approx2.7$~nm$^{-2}$ for Co and 16~nm$^{-2}$ for Py. $2\tilde{g}^{\uparrow\uparrow}\tilde{g}^{\downarrow\downarrow}/(\tilde{g}^{\uparrow\uparrow}+\tilde{g}^{\downarrow\downarrow})/S\approx20$~nm$^{-2}$ for the Cu$\mid$Co interface, Table~\ref{tabmix}, we may expect the value for Cu$\mid$Py to be similar. With $\tilde{g}_r^{\uparrow\downarrow}/S\approx27$~nm$^{-2}$ for Cu$\mid$Co, Table~\ref{tabmix}, and 15~nm$^{-2}$ for Cu$\mid$Py \cite{Bauer:prb03}, one finds $\nu\approx0.98$ for Cu$\mid$Co and $\nu\approx0.33$ for Cu$\mid$Py.

\begin{figure}[pth]
\includegraphics[width=0.95\linewidth,clip=]{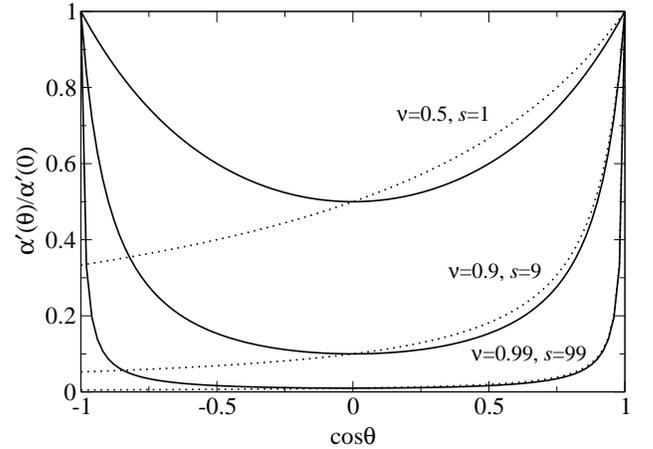}
\caption{\label{cone}Solid lines display the precession-cone angle dependence of the Gilbert damping parameter, Eq.~(\ref{or}), and the dotted lines show the extrapolated small-angle approximation (\ref{B}). The lowest lines are representative for Co and the uppermost ones for Py, assuming magnetic-film thickness of 5~nm. Fe and Ni are expected to be intermediate between these two cases.}
\end{figure}

The magnetization dynamics in the absence of an applied bias are determined by substituting $\mathbf{I}_s^{\text{exch}}$ into equation (\ref{Gllg}), which thus becomes inconsistent with a constant effective Gilbert parameter. We now analyze the configuration dependence of the damping in more detail, which can be measured, in principle, by the FMR line width broadening at high rf intensities (and therefore finite ``precession cones"). For $\mathbf{m}_1$ precessing around $\mathbf{m}_2$,
\begin{equation}
\mathbf{m}_1\times\mathbf{I}^{\text{exch}}_{s1}\times\mathbf{m}_1=\frac{1}{2}\left(1-\frac{\nu\sin^2\theta}{1-\nu^2\cos^2\theta}\right)\mathbf{I}_{s1}^{\text{pump}}\,.
\label{pr}
\end{equation}
The angular dependence of the additional Gilbert damping parameter due to the exchange spin current then reads \cite{Tserkovnyak:prb031}
\begin{equation}
\frac{\alpha^\prime(\theta)}{\alpha^\prime(0)}=1-\frac{\nu\sin^2\theta}{1-\nu^2\cos^2\theta}\,,
\label{or}
\end{equation}
where $\alpha^\prime(0)=\gamma\hbar\tilde{g}^{\uparrow\downarrow}_r/(8\pi M_sdS)$ is the damping enhancement in a collinear configuration, see Eq.~(\ref{gc}). For small angles, $\theta\approx0$, Eq.~(\ref{or}) can be rewritten as
\begin{equation}
\frac{\alpha^\prime(\theta)}{\alpha^\prime(0)}\approx\frac{1}{1+s(1-\cos\theta)}\,,
\label{B}
\end{equation}
where
\begin{equation}
s=\frac{\nu}{1-\nu}\,.
\label{snu}
\end{equation}
Eq.~(\ref{B}) was also obtained by \onlinecite{Berger:jap01} in the limit of small precession angles, however, for a definition of $s$ with $s\propto\tau_{\text{sf}}$ differing from Eq.~(\ref{snu}). Expressions (\ref{or}) and (\ref{B}) are plotted in Fig.~\ref{cone}.

As mentioned above, $\nu$ is close to 0.98 for cobalt, so that the lower solid line in Fig.~\ref{cone} represents the damping for Co according to Eq.~(\ref{or}). $s=333$ is found by \onlinecite{Berger:jap01} for Cu$\mid$Co with Co 1.5~nm thick, which is remarkably similar to the estimate based on $\nu$ for this thickness when substituted into Eq.~(\ref{snu}). The damping is thus predicted to be significantly reduced for precession angles deviating only slightly from the collinear configurations (we expect this conclusion to be also true for Fe and Ni). Modeling the magnetization dynamics with a constant damping parameter is therefore not allowed for sufficiently thin magnetic layers. For permalloy, on the other hand, the magnetization damping is expected to remain significant at all angles, see the upper solid line in Fig.~\ref{cone}. This implies that the magnetization reversal by an applied magnetic field should be faster in Py than in Co spin valves.

If $\mathbf{m}_1$ rotates around an axis perpendicular to $\mathbf{m}_2$, i.e., only the relative angle $\theta$ changes, then $\mathbf{I}_{s1}^{\text{pump}}\perp\mathbf{m}_2$ and Eq.~(\ref{exch}) reduces to $\mathbf{I}_{s1}^{\text{exch}}=\mathbf{I}_{s1}^{\text{pump}}/2$. The corresponding damping $\alpha^\prime(0)$ thus has an angle-independent enhancement with respect to the intrinsic Gilbert damping.

An applied current bias is an additional instrument to control the magnetization dynamics. When the conductance parameters of the spin valve are mirror symmetric, the bias-induced spin transfer $\mathbf{I}_{s1}^{\text{bias}}$ is coplanar with the magnetization directions and can be written as
\begin{equation}
\mathbf{I}_{s1}^{\text{bias}}=I_s^{\text{bias}}(\cos\theta)\frac{\mathbf{m}_1+\mathbf{m}_2}{2\cos\theta/2}\,,
\end{equation}
for a given applied bias. This follows from expanding the spin current as
\begin{equation}
\mathbf{I}_{s1}^{\text{bias}}=f_{11}(\cos\theta)\mathbf{m}_1+f_{22}(\cos\theta)\mathbf{m}_2+f_{12}(\cos\theta)\mathbf{m}_1\times\mathbf{m}_2
\end{equation}
and noting that $\mathbf{I}_{s1}^{\text{bias}}(\mathbf{m}_1,\mathbf{m}_2)=\mathbf{I}_{s1}^{\text{bias}}(\mathbf{m}_2,\mathbf{m}_1)$, by symmetry, which implies that $f_{11}=f_{22}$ and $f_{12}=0$. The electric current necessary to induce a given spin current $\mathbf{I}_s^{\text{bias}}$ depends on $\theta$ and can be calculated readily by the circuit theory summarized in Sec.~\ref{Mdc}. Eqs.~(\ref{Gllg}), (\ref{Ieb}), (\ref{exch}) and the form of the bias current completely determine the dynamics of $\mathbf{m}_1(t)$. The exchange induced by the spin pumping causes relaxation toward an equilibrium configuration, while the bias current can either relax or excite the magnetization, depending on the sign of the current, as discussed below.

Near the parallel configuration, $\theta\approx0$, Eq.~(\ref{exch}) simplifies to $\mathbf{I}_{s1}^{\text{exch}}=\mathbf{I}_{s1}^{\text{pump}}/2$. Let $\mathbf{m}_1$ circularly precess around a fixed $\mathbf{m}_2$ with the frequency $\omega$ (determined by effective fields): $\mathbf{m}_1\times\mathbf{\dot{m}_1}=\omega\mathbf{m}_1\times\mathbf{m}_2\times\mathbf{m}_1$. The total projected spin current has the Gilbert form:
\begin{equation}
\mathbf{m}_1\times\mathbf{I}_{s1}\times\mathbf{m}_1=\left(\frac{\hbar\tilde{g}_r^{\uparrow\downarrow}}{8\pi}+\frac{I_s^{\text{bias}}}{2\omega}\right)\mathbf{m}_1\times\frac{d\mathbf{m}_1}{dt}\,.
\end{equation}
Instability is reached when the effective Gilbert-damping coefficient becomes negative. The critical bias is thus given by
\begin{equation}
I_{s,\text{crit}}^{\text{bias}}=-\left(\frac{\tilde{g}_r^{\uparrow\downarrow}}{4\pi}+\frac{2\alpha M_sSd}{\hbar\gamma}\right)\hbar\omega\,,
\label{Ics}
\end{equation}
where $\alpha$ is the intrinsic Gilbert constant. Neglecting the first term in the brackets, one obtains a result analogous to that of \onlinecite{Sloncz:mmm96,Sloncz:mmm99}, while neglecting the second term leads to a condition similar to the resonant spin-wave emission criterion \cite{Berger:prb96}. The spin-pumping contribution (the first term) is comparable with the intrinsic damping (the second term) for transition-metal films with thickness $d$ of several nanometers, but dominates for very thin films. When the instability is reached, the trajectory of $\mathbf{m}_1(t)$ can become very complicated and could possibly lead to a magnetization reversal to a different (meta)stable configuration. A more complete discussion of spin torques and macrospin switching dynamics in asymmetric spin valves can be found in \onlinecite{Manschot:apl04,Brataas:prp05}. Nonlinear large-angle dynamics and field-induced switching in the presence of spin pumping were discussed by \onlinecite{Kim:mmm05} within the macrospin model. Nonuniform dynamics have been studied by \onlinecite{Brataas:cm05}. Experimental evidence for a spin-pumping--induced damping in current-driven spin valves has been reported by \onlinecite{Krivorotov:sc05}. The importance of the spin-pumping contribution to the critical current (\ref{Ics}) has been investigated experimentally by \onlinecite{Sun:jap05} for Co$\mid$Cu$\mid$Co spin valves. 

\section{Linear-response approach}
\label{lra}

\subsection{Heterostructures}
\label{lsp}

This subsection is devoted to an alternative description of the spin emission by a dynamic ferromagnetic magnetization embedded in a conducting nonmagnetic matrix, as in Fig.~\ref{sc}, which was put forward by \onlinecite{Simanek:prb031} and further elaborated by \onlinecite{Simanek:prb032,Mills:prb03,Simanek:cm04}. In Sec.~\ref{pump}, we formulated such spin emission as a scattering pumping process, which requires the concept of waveguide leads for electron states that are reflected by or transmitted through a ferromagnetic layer between normal reservoirs. The time dependence of the scattering matrix caused by a moving magnetization leads to the pumping of spin currents, and the corresponding loss of angular momentum by the ferromagnet to an increased viscous damping of the magnetization dynamics. Such language is standard in the field of mesoscopic transport phenomena \cite{Beenakker:rmp97}, but the magnetism community is more accustomed to linear-response susceptibilities rather than scattering matrices. Of course, the final result should not depend on the formalism used, as verified in the following, but one may argue that the scattering theory has distinct advantages over the linear-response approach for the present problem.

For a linear-response formulation, we proceed from a Hamiltonian for conducting electrons experiencing a ferromagnetic exchange field as given by Eqs.~(\ref{Ht}), (\ref{Hf}). When the exchange field $V_x=-\hbar\Omega/2$ is uniform inside the ferromagnetic volume $V$ and vanishes outside, we may rewrite Eq. (\ref{Hf}) as
\begin{equation}
H^\prime(t)=-\frac{\Omega}{M_s}\int_V d^3r\left[\mathbf{s}(\mathbf{r})\cdot\mathbf{M}(\mathbf{r},t)\right]\,.
\label{HI}
\end{equation}
Here, $\mathbf{s}(\mathbf{r})$ is the spin-density operator for conduction electrons that are polarized by the exchange field with strength $\Omega$ along the magnetization direction $\mathbf{m}=\mathbf{M}/M_s$. $\mathbf{M}$ is a collective property that is treated as a classical time-dependent potential. The Hamiltonian (\ref{HI}) follows, e.g., from a mean-field approximation of the $s-d$ model. The exchange is then viewed as an external potential (provided by the $d$ electrons) that does not depend on the $s$-electron distribution. The spin-density--functional formulation of the magnetization dynamics in itinerant ferromagnets, see, e.g., \onlinecite{Qian:prl02,Capelle:prl01}, leads to the Hamiltonian Eq.~(\ref{HI}) in the local-density approximation. The effective field (\ref{HM}) due to the interaction (\ref{HI}) then reads
\begin{equation}
\mathbf{H}^\prime_{\text{eff}}(\mathbf{r},t)=\frac{\Omega}{M_s}\left\langle\mathbf{s}(\mathbf{r})\right\rangle_t\,.
\label{Hp}
\end{equation}
In the $s-d$ model, this field corresponds to the reaction torque by the nonequilibrium conduction-electron-spin distribution inside the ferromagnet, which is excited by the moving magnetization direction.

\onlinecite{Simanek:prb031} suggested to calculate the reaction torque on the ferromagnetic magnetization directly by using Eq.~(\ref{Hp}) in the LLG equation of motion. This appears to be very different from the spin-pumping picture that relies on the spin currents that are emitted at the interface to the normal metal. The two approaches are however related by the continuity equation for electron spin dynamics, see, e.g., \onlinecite{Capelle:prl01}, according to which the spin current equals the torque exerted on the magnetization $\mathbf{M}$ by the effective field (\ref{Hp}), up to a term given by the time derivative of the average spin density, $\langle\mathbf{\dot{s}}\rangle_t$, times volume. When $\langle\mathbf{s}\rangle_t$ follows the magnetization adiabatically, the difference between the spin current and torque is thus proportional to $\mathbf{\dot{m}}$. In the case of the $s-d$ model, this has a physical meaning, corresponding to the torque required to change the angular momentum of the $s$-electron spin density. For Stoner ferromagnets, in which the conduction electrons are identical to the ones that make up the magnetization, we have to calculate the spin flow emitted by the ferromagnet as an additional term in the LLG equation, while the reaction torque on the magnetization by the conduction electrons has no obvious physical meaning. When treating itinerant (Stoner) ferromagnets within the density-functional theory, the spin currents induced by the Kohn-Sham Hamiltonian (\ref{HI}) in principle differ from the physical ones inside or very close to the ferromagnet. However, this difference vanishes asymptotically as a function of distance from the ferromagnet \cite{Capelle:prl01}. The emitted spin currents are thus similar to those in the $s-d$ model, and their evaluation can thus be mapped on calculating corresponding reaction torques.

A uniform small-angle dynamics (of a possibly large $\Omega$) only weakly perturbs the system. The induced spin imbalance $\left\langle\delta\mathbf{s}(\mathbf{r})\right\rangle_t$ can therefore be expressed in terms of the linear-response susceptibility of the unperturbed system
\begin{equation}
\chi_{s_as_{a^\prime}}(t)=\frac{i}{\hbar V}\Theta(t)\int_V d^3rd^3r^\prime\left\langle\left[s_a(\mathbf{r},t),s_{a^\prime}(\mathbf{r}^\prime,0)\right]\right\rangle ,
\label{chi}
\end{equation}
where $\Theta(t)$ is the Heaviside step function. The effective field due to the induced nonequilibrium spin density is then given by
\begin{align}
\delta\mathbf{H}^\prime_{\text{eff}}(t)=&\frac{\Omega}{M_sV}\int_Vd^3r\left\langle\delta\mathbf{s}(\mathbf{r})\right\rangle_t=\frac{\Omega}{M_s}\sum_{aa^\prime}\int_{-\infty}^\infty dt^\prime\nonumber\\
&\times\mathbf{\hat{a}}\chi_{s_as_{a^\prime}}(t-t^\prime)\delta m_{a^\prime}(t^\prime)\,,
\label{aH}
\end{align}
where $a,~a^\prime \in \{ x,y,z \}$ are the indices of the Cartesian axes and $\mathbf{\hat{a}}$ stands for the corresponding unit vectors. Suppose now for simplicity that the system is invariant under spin rotations about the $z$ axis, and consider small-angle magnetization dynamics near this axis, $\delta\mathbf{m}=\mathbf{m}-\mathbf{\hat{z}}$. Substituting Eq.~(\ref{aH}) into the LLG equation (\ref{llg}) then yields the following lowest-order dynamic term:
\begin{equation}
-\gamma\mathbf{m}\times\delta\mathbf{H}^\prime_{\text{eff}}(t)=\frac{\gamma\Omega^2}{M_s}\left(\Lambda_1\frac{d\mathbf{m}}{dt}+\Lambda_2\mathbf{m}\times\frac{d\mathbf{m}}{dt}\right)\,,
\label{LL}
\end{equation}
which is the most general adiabatic term for axially-symmetric systems \cite{Mills03}. Here,
\begin{align}
\Lambda_1=&\left.-i\frac{d\chi_{s_xs_y}(\omega)}{d\omega}\right\vert_{\omega=0}\,,\nonumber\\
\Lambda_2=&\left.-i\frac{d\chi_{s_xs_x}(\omega)}{d\omega}\right\vert_{\omega=0}
\end{align}
are real numbers. $\Lambda_1$ renormalizes the effective gyromagnetic ratio $\gamma_{\text{eff}}$ in Eq.~(\ref{llg}) and $\Lambda_2$ is a Gilbert-like damping parameter
\begin{equation}
\alpha_{\text{eff}}=\left.-i\frac{\gamma_{\text{eff}}\Omega^2}{M_s}\frac{d\chi_{s_xs_x}(\omega)}{d\omega}\right\vert_{\omega=0}\,.
\label{alr}
\end{equation}
Eq.~(\ref{alr}) can also be obtained by equating the energy dissipated into the itinerant degrees of freedom by moving magnetization and the work done by an rf magnetic field applied against the effective viscous Gilbert term, at a steady magnetic precession, since $\lim_{\omega\to0}\mbox{Re}\chi_{s_xs_y}(\omega)/\omega=0$ in the case of spin-rotational symmetry around the $z$ axis. For thermally-equilibrated $s$-electron subsystem, Eq.~(\ref{alr}) implies $\alpha_\text{eff}\gamma_\text{eff}>0$, as required by the LLG phenomenology, see Sec.~\ref{llgt}.

The damping constant $\alpha$ can generally be expressed in terms of the response function of the total magnetization (e.g., of the $s$ and $d$ electrons in the $s-d$ model) by inverting Eq.~(\ref{P}). In the present discussion, this problem is self-consistently reduced to the simpler task of calculating the spin response of $s$ electrons to a time-dependent exchange field aligned with the $d$-electron magnetization.

The similarity between the torque (\ref{LL}) and the spin-pumping current (\ref{Is}) should not be surprising in the light of the above discussion. Indeed, \onlinecite{Simanek:prb032} explicitly demonstrated that the low-frequency linear-response and spin-pumping pictures lead to identical $\alpha_{\text{eff}}$ for a $\delta$-function magnetic layer embedded into a clean normal metal (which corresponds to an ideal spin sink for emitted spin currents in the spin-pumping language). However, evaluating the linear-response correlation functions becomes very tedious for more realistic models, see, e.g., \onlinecite{Mills:prb03}. It is also not obvious how to treat the current-induced spin transfer and the spin pumping on the same footing by the linear-response formalism. On the other hand, the formulation in terms of the susceptibilities is complementary to the spin-pumping approach. It appeals to the intuition of many researchers in the magnetism community. Furthermore, it could be helpful to obtain insights into problems that are hard to solve within the scattering theory. Examples of these are magnetic bilayers coupled by a strong static exchange interaction (see Sec.~\ref{uns}), strongly-correlated systems (Sec.~\ref{eei}), and the bulk damping of the magnetization dynamics (Sec.~\ref{sdm}).

\subsection{Bulk damping}
\label{sdm}

In the previous subsection, we considered coupling of the magnetization to itinerant electrons via an exchange interaction (\ref{HI}). In particular, it was reasserted that when a ferromagnetic film is inserted into a nonmagnetic metal, the magnetization dynamics causes an emission of spins. In the presence of a spin sink outside the ferromagnet, this pumping leads to a net angular-momentum loss that is equivalent to an excess damping of the magnetization dynamics, as discussed in Sec.~\ref{gde}. An analogous mechanism could be effective in bulk ferromagnets provided that spin-relaxation decay channels exist that dissipate a spin accumulation created by the pumping. This happens in the presence of momentum-scattering mechanisms, such as lattice impurities and phonons, and the accompanying (or band-structure) relativistic spin-orbit interaction. The inductive coupling of the magnetization to the conduction electrons that causes dissipation by eddy currents is less important in thin layered structures and is disregarded in the following. For simplicity, the discussion is restricted in the following to a mean-field $s-d$ model.

A mechanism of bulk magnetization damping bearing similarity with the spin-pumping picture in heterostructures was proposed a long time ago by \onlinecite{Mitchell:pr57}: It involves a transfer of the angular momentum (and energy) of a nonequilibrium ferromagnetic configuration to the itinerant electrons via the exchange interaction, with a subsequent spin-orbit relaxation to the lattice. The consequences of such process for macrospin (long-wavelength) dynamics have been worked out for the $s-d$ model by \onlinecite{Heinrich:pss67}. The $s-d$ picture has been resurrected recently by \onlinecite{Sinova:prb04} in order to address the magnetization relaxation in the ferromagnetic semiconductor (Ga,Mn)As, in which ferromagnetism originates from the free-hole--mediated exchange interaction between dilute, substitutional spin-5/2 Mn atoms \cite{Ohno:mmm99}. It is possible to reproduce and generalize the results of \onlinecite{Heinrich:pss67,Sinova:prb04} by calculating the magnetization damping in terms of the conduction-electron-spin dynamics in a time-dependent exchange field similarly to Sec.~\ref{lsp} \cite{Tserkovnyak:apl04}.

Consider an $sp-d$ model of a conducting ferromagnet, where the spin $\mathbf{S}$ of the itinerant $s$ or $p$ states (either electrons or holes) experiences an exchange field of magnitude $\Omega$ along the uniform magnetization direction $\mathbf{m}$ of the localized $d$ orbitals, as in Eq.~(\ref{HI}):
\begin{equation}
H(t)=H_0-\Omega\mathbf{m}(t)\cdot\mathbf{S}\,.
\label{H}
\end{equation}
Here, $H_0$ is a one-particle Hamiltonian reflecting the host band structure. In (Ga,Mn)As, substitutional Mn are paramagnetic acceptors that strongly interact with the free holes. Although the exchange field can be highly nonuniform on atomic scales, it is customary to start with a simplifying assumption that it may be smeared out. As before, we treat the magnetization $\mathbf{m}$ as a classical and, on the relevant length scales of the carrier dynamics, spatially-uniform variable. A suitable model for the valence bands of dilute $p$-doped semiconductor (e.g., GaAs, Si, or Ge) is the spherical Luttinger Hamiltonian for spin-$3/2$ holes:
\begin{equation}
H_0=\frac{1}{2m_e}\left[\left(\gamma_1+\frac{5}{2}\gamma_2\right)p^2-2\gamma_2\left(\frac{\mathbf{p}\cdot\mathbf{S}}{\hbar}\right)^2\right]\,,
\label{H0}
\end{equation}
where $m_e$ is the free-electron mass and $\gamma_i$ are the so-called Luttinger parameters \cite{Luttinger:pr56}. The spin-orbit term couples the hole momentum $\mathbf{p}$ with its spin $\mathbf{S}$. The four-band model (\ref{H0}), is valid when the hole Fermi energy is sufficiently smaller than the spin-orbit splitting of the semiconductor host between the spin-$3/2$ and spin-$1/2$ valence bands. Eq.~(\ref{H0}) neglects corrections for lattices with broken inversion symmetry. Spin-$1/2$ electron systems with vanishing spin-orbit coupling are recovered by setting $\gamma_2=0$. Suppose the magnetization of the localized orbitals varies slowly in time (being uniform at all times), so that the time-dependent $\mathbf{m}$ modulates the Hamiltonian (\ref{H}) adiabatically. This means that the system equilibrates on time scales faster than the motion of $\mathbf{m}$ and all the quantities parametrizing the carrier Hamiltonian stay constant. The observation of ferromagnetic resonance indicates that time-dependent long-range ferromagnetic order indeed exists in thin films of magnetic transition metals \cite{Heinrich:ap93} and semiconductors \cite{Goennenwein:apl03,Rappoport:prb04}. Eq.~(\ref{alr}) may then be taken as a microscopic definition of the Gilbert-damping parameter. In the following, $\alpha$ is formally evaluated for electron and hole systems with an emphasis on its dependence on spin dephasing. (We will drop the subscripts ``eff" on $\alpha$ and $\gamma$ from the previous subsection.)

\begin{figure}[pth]
\includegraphics[width=0.95\linewidth,clip=]{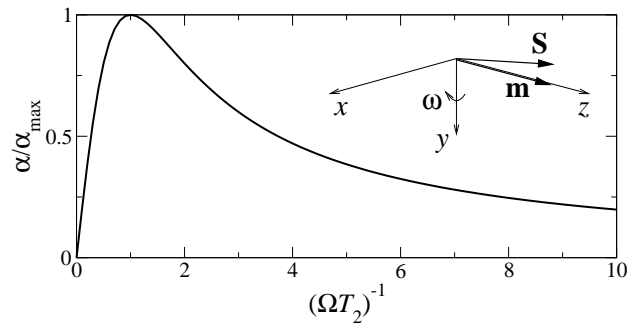}
\caption{\label{semi} Gilbert damping, Eq.~(\ref{a}), in units of $\alpha_{\max}=\gamma s_0/(2M_s)$ as a function of the normalized spin-relaxation rate. Inset: Geometry of the model.}
\end{figure}

We initially focus on a system without spin-orbit coupling in the band structure, $\gamma_2=0$. Suppose the transverse spin-density dynamics in the exchange field $\Omega$ is described by the Bloch equation
\begin{equation}
\frac{d\mathbf{s}}{dt}=\Omega\mathbf{s}\times\mathbf{m}(t)-\frac{\mathbf{s}-s_0\mathbf{m}(t)}{T_2}\,,
\label{T2}
\end{equation}
where the last term is a phenomenological relaxation due to impurities, parametrized by the transverse dephasing time $T_2$. Let us assume that $\mathbf{m}(t)$ moves slowly on the scales of $\Omega^{-1}$ and $T_2$. It is then convenient to transform Eq.~(\ref{T2}) into a frame of reference (for spin variables) that moves together with $\mathbf{m}(t)$. When, for example, $\mathbf{m}$, at a given instant, rotates with frequency $\omega$ around the $y$ axis, see inset of Fig.~\ref{semi}, it is stationary in the rotating frame at the cost of a new (Larmor) term $\omega\mathbf{s}\times\mathbf{\hat{y}}$ on the right-hand side of Eq.~(\ref{T2}), which corresponds to a magnetic field along the $y$ axis. Since the motion is slow, it is sufficient to solve for $\mathbf{s}$ as the instantaneous stationary state in the moving frame of reference. If $\mathbf{m}$ points along the $z$ axis in the rotating frame, spin polarization along the $x$ axis then exerts a damping torque on $\mathbf{m}$ that corresponds to
\begin{equation}
\alpha=-\frac{\gamma\tilde{\chi}_{s_xs_y}\Omega}{M_s}=\frac{\Omega T_2}{1+(\Omega T_2)^2}\frac{\gamma s_0}{M_s}\,,
\label{a}
\end{equation}
where $\tilde{\chi}_{s_xs_y}$ is the stationary (real-valued) response function in the rotating frame. (We recall that $\gamma$ is the gyromagnetic ratio of $d$ orbitals.) The calculation of the time-dependent linear response in the laboratory frame, Eq.~(\ref{alr}), is thus simplified to that of the static response in the rotating frame. Such a transformation is however not possible  in general, in particular for Hamiltonians that are not spin-rotationally invariant, such as Eq.~(\ref{H}). Eq.~(\ref{a}) is plotted in Fig.~\ref{semi}. The equilibrium spin density $s_0$ can be calculated from the specific form of the static Hamiltonian. $\alpha$ vanishes at both small and large spin-relaxation limits.

The damping parameter (\ref{a}) depends only on the ratio of spin-relaxation rate and exchange energy. The low spin-relaxation rate regime, $\alpha\propto T_2^{-1}$, is analogous to the spin-pumping damping of thin films in contact with a spin-sink conductor: The moving magnetization ``pumps" spins into the itinerant carriers at a constant rate, which are then relaxed with a probability $\propto T_2^{-1}$ before exchanging spins with the ferromagnet again. The difference is that now the spins are pumped into the internal conduction electrons of the ferromagnet rather than those of an externally attached metal. The other limit, $\alpha\propto T_2$, can be understood by noting that for a very viscous dynamics of $\mathbf{s}(t)$, $\mathbf{s}(t)\approx \mathbf{s}_0(t)-T_2\mathbf{\dot{s}}_0(t)$ in the laboratory frame. $\alpha\propto T_2$ then follows from the torque $\propto(\mathbf{s}-\mathbf{s}_0)\times \mathbf{m}$. This is analogous to the ``breathing Fermi-surface" damping mechanism discussed by \onlinecite{Kunes:prb02} in the regime of fast relaxation: The itinerant carriers try to lower their energy by rearranging themselves in the field of slowly varying magnetization direction but lag behind with a short delay time determined by the relevant relaxation processes. In the presence of an anisotropic spin-orbit interaction in the ferromagnet's crystal field, the breathing Fermi surface gives an additional contribution to damping \cite{Kunes:prb02}, see also a short discussion in Sec.~\ref{uns}. In the present model described by equation (\ref{T2}), the ``breathing" takes place in spin space.

It is interesting to note that Eq.~(\ref{a}) reduces to the result obtained by \onlinecite{Heinrich:pss67} for the long-wavelength magnon lifetime due to the $s-d$ interaction with spin-$1/2$ conduction electrons in the random-phase approximation:
\begin{equation}
\alpha=\frac{\Omega T_2}{1+(\Omega T_2)^2}\frac{\gamma\left[\Omega m^\ast k_F/(4\pi^2 \hbar)\right]}{M_s}\,,
\label{rpa}
\end{equation}
where $m^\ast$ is the band-structure mass and $k_F$ the Fermi wave vector. Here, it was assumed that $\hbar\Omega\ll E_F$ (the Fermi energy). The quantity in the square brackets is just the total carrier spin density. Eq.~(\ref{rpa}) was used by \onlinecite{Ingvarsson:prb02} in order to explain the measured damping in thin permalloy films, which scaled linearly with the film resistivity $\rho$. A linear relation between the two quantities is expected when $\Omega\gg T_2^{-1}$, so that $\alpha\propto T_2^{-1}$, and using Drude formula $\rho\propto\tau^{-1}$ ($\tau$ is the transport mean free time) and assuming $\tau/T_2={\rm const}$ which depends on material and scattering-impurity type but not on scattering rate \cite{Abrikosov:zetf62}. Unlike for ferromagnetic semiconductors, however, the use of the $s-d$ model for itinerant ferromagnets like transition metals is questionable, since the separate treatment of magnetic and conduction electrons is unphysical.

Let us turn to the magnetization relaxation in hole-doped magnetic semiconductor (Ga,Mn)As. Eq.~(\ref{a}) may be used to obtain a rough estimate of the damping coefficient: The largest achievable value of $\alpha_{\max}=\gamma s_0/(2M_s)$ occurs when the holes are fully polarized, giving $\alpha_{\max}\sim 0.1-0.3$, roughly one third the ratio of the (spin-$3/2$) hole to the substitutional (spin-$5/2$) Mn concentrations. For realistic samples with a spin polarization of the order of unity, therefore, $\alpha_{\max}\sim 0.1$. The damping $\alpha$ is further suppressed by the factor $\alpha/\alpha_{\max}=2\Omega T_2[1+(\Omega T_2)^2]^{-1}<1$. For clean bulk samples of GaAs, the spin-relaxation time is $\sim 100$~fs \cite{Hilton:prl03}. For approximately 5\% Mn doping, $\hbar\Omega\sim0.1$~eV \cite{Dietl:prb01}, so that $\Omega T_2\sim10$. This  corresponds to the $\alpha \propto T_2^{-1}$ regime with $\alpha\sim0.01$. Reduced spin-relaxation times should thus result in a larger damping. Experimentally, the impurity scattering is likely to be the easiest parameter to vary in order to tune $\alpha$ to a desired value. The strong spin-orbit coupling $\gamma_2$, however, makes the validity of the phenomenological equation (\ref{T2}), and thus the result (\ref{a}), questionable for the hole system. Besides, a strong crystal anisotropy would require a further refinement of the analysis. Returning to our basic equation (\ref{alr}) and inserting the response function for a noninteracting system yields
\begin{align}
\alpha=&\frac{\gamma\Omega^2}{M_sV}\lim_{\omega\rightarrow0}\frac{\pi}{\omega}\sum_{ij}\left|\langle i|S_x|j\rangle\right|^2\left[f_{\text{FD}}(\varepsilon_i)-f_{\text{FD}}(\varepsilon_j)\right]\nonumber\\
&\times\delta(\hbar\omega+\varepsilon_i-\varepsilon_j)\,,
\label{aij}
\end{align}
where $i$, $j$ label the one-particle eigenstates of the sample with volume $V$. When the wave vector $\mathbf{k}$ is conserved, $\sum_{ij}/V=\sum_{\sigma\sigma^\prime}\int d^3k/(2\pi)^3$, where $\sigma$, $\sigma^\prime$ label spin states. For a perfect crystal, therefore, $\alpha$ vanishes, unless there is a finite Fermi-surface area with spin degenerate states. Introducing lattice defects leads to a finite $\alpha$, see, e.g., \onlinecite{Sinova:prb04} (these authors did not include important vertex corrections in the response function, however). 

In the above analysis, we have assumed a coherent motion of the ferromagnetic magnetization without specifying the source of excitation. An FMR magnetic field (with a large dc and small rf components), for example, can be included explicitly into the Hamiltonian (\ref{H}) of the itinerant carriers. The results for the Gilbert damping will stay unaffected, however, as long as the exchange energy $\hbar\Omega$ is much larger than the carrier Zeeman splitting in the applied field and the ferromagnetic magnetization is mainly carried by the localized orbitals. (The energy pumped into the carrier-magnetization dynamics by the rf field must be taken into account otherwise.) Inhomogeneities in the bulk magnetization are not important on the length scales set by the transverse spin-relaxation rate and the precession frequency in the exchange field. The bulk spin dynamics discussed in this subsection have no effect on the spin pumping into adjacent conductors discussed in Sec.~\ref{pump}, as long as the transverse spin-relaxation rate is small compared to the exchange precession frequency. In the opposite, rather unrealistic, limit, the $s$-electron spin dynamics are locked to the $d$-electron magnetization motion, suppressing spin leakage (pumping) into adjacent normal conductors.

\section{Miscellaneous}
\label{misc}

\subsection{Quantum-size effects}
\label{qse}

\subsubsection{Ultrathin magnetic layer}
\label{uml}

As explained in Sec.~\ref{pump}, the spin pumping by a magnetic layer in contact with normal metals is governed by the parameter $\mathcal{A}^{\uparrow\downarrow}=g^{\uparrow\downarrow}-t^{\prime\uparrow\downarrow}$, Eq.~(\ref{A}). When the magnetic-film thickness $d$ exceeds the spin-coherence length (\ref{lsc}), $d\gg\lambda_{\text{sc}}$, $t^{\prime\uparrow\downarrow}$ vanishes and $\mathcal{A}^{\uparrow\downarrow}$ is given simply by the interfacial spin-mixing conductance $g^{\uparrow\downarrow}$. In this subsection, on the other hand, we focus on the regime in which $d$ is smaller or of the order of $\lambda_{\text{sc}}$, i.e., for thicknesses of a few monolayers in the case of transition-metal ferromagnets. In this limit, the coherence between up- and down-spin states in the ferromagnet leads to a thickness dependence of $\mathcal{A}^{\uparrow\downarrow}$, and thus of the spin pumping and magnetization torque (\ref{tR}). The decoherence of the orbital wave function due to inelastic scattering processes is disregarded, assuming its characteristic length scale is much longer than $\lambda_{\text{sc}}$.

As in most of this review, we discuss in the following layered structures. A large lateral area $S$ renders mesoscopic phenomena such as the Coulomb blockade irrelevant, and we focus on quantum-coherence effects due to small layer thicknesses. For a study of magnetization dynamics in magnetic nanoparticles in contact with large reservoirs, see, e.g., \onlinecite{Waintal:prl03,Waintal:prl05}.

The linear-response framework \cite{Simanek:prb031}, see also Sec.~\ref{lra}, has been used to calculate the enhanced Gilbert damping of finite-thickness ferromagnetic films embedded into a conducting medium by \onlinecite{Mills:prb03}. For an idealistic model of isotropic band structure with spin-dependent exchange potential, he found that ultrathin films display oscillatory damping (as a function of thickness) due to quantum-size effects. However, by first-principles scattering matrices based on realistic electronic structures computed in the local spin-density approximation, \onlinecite{Zwierzycki:prb05} have shown that quantum-size oscillations are much smaller than those obtained by \onlinecite{Mills:prb03}, and small amounts of disorder suppress the remaining oscillations even further. \onlinecite{Zwierzycki:prb05} also found that the spin-pumping torque is of the Gilbert-damping form, with only a very small correction to the gyromagnetic ratio.

\begin{figure}[pth]
\includegraphics[width=0.95\linewidth,clip=]{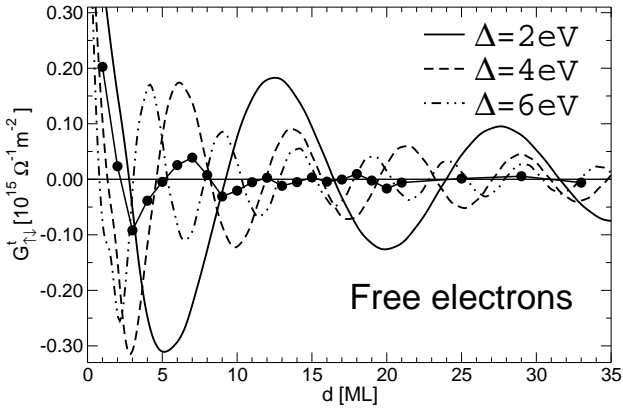}
\caption{The real part of $G^t_{\uparrow\downarrow}=(e^{2}/h)t^{\uparrow\downarrow}/S$ as a function of the magnetic-layer thickness $d$ calculated from first principles ($\bullet$) for realistic multiband electronic structures and that calculated for an isotropic free-electron gas with Fermi energy $\varepsilon_F=7$~eV (chosen to obtain the correct value of the Sharvin conductance in Cu) and various choices of the exchange splitting $\Delta$. From \onlinecite{Zwierzycki:prb05}.}
\label{free}
\end{figure}

The above observations have been made for copper/cobalt and gold/iron heterostructures, of which we will restrict our discussion to the former. Realistic band-structure and simple-model calculations on Cu$\mid$Co$\mid$Cu(111) trilayers are compared in Fig.~\ref{free}, where the real part of $G_{\uparrow\downarrow}^t=(e^{2}/h)t^{\uparrow\downarrow}/S$ is plotted as a function of thickness $d$ of the magnetic layer (measured in atomic monolayers). Solid circles in figure~\ref{free} show the result of the \textit{ab initio} calculation \cite{Zwierzycki:prb05} without impurities and for specular ($\mathbf{k}_\parallel$-conserving) interfaces. The smooth solid lines represent the calculations for the isotropic free-electron model. For exchange splittings $\Delta=2,~4,~6$~eV, the amplitude of oscillation is much larger and the decay is much slower in the model than in the more realistic first-principles calculations. As might be expected, increasing the exchange splitting from 2 to 6~eV leads to a shorter period and more rapid decay of the oscillations. In order to mimic the parameter-free \textit{ab initio} result, an exchange splitting in the range of 10~eV would be needed (not shown in the figure), however. Such a large value can be justified neither on theoretical nor experimental grounds. This discrepancy illustrates the difficulty of quantitatively representing the complex electronic structure of transition metals by simple model Hamiltonians.

\begin{figure}[pth]
\includegraphics[width=0.95\linewidth,clip=]{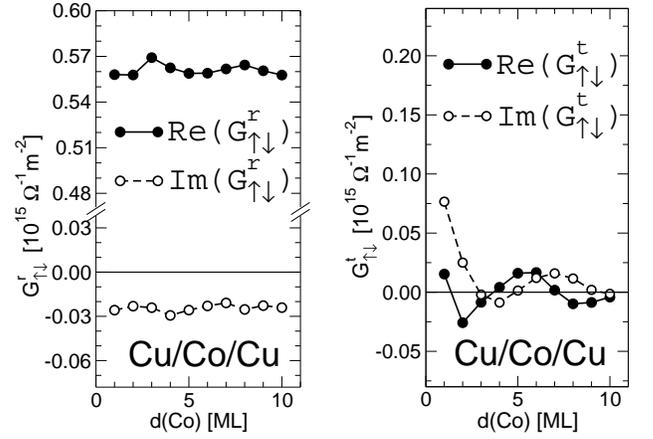}
\caption{Spin-mixing conductances of a Cu$\mid$Co$\mid$Cu(111) structure with disordered interfaces as a function of the Co-layer thickness. From \onlinecite{Zwierzycki:prb05}.}
\label{gccd}
\end{figure}

Figure \ref{gccd} shows $G_{\uparrow\downarrow}^r=(e^2/h)g^{\uparrow\downarrow}/S$ and $G_{\uparrow\downarrow}^t$ calculated from first principles in the presence of disorder modeled by a monolayer of 50\% alloy added on each side of the magnetic layer. The thickness $d$ in this case is defined as that of the remaining clean ferromagnetic layer. The disorder strongly quenches the amplitudes of the oscillations as a function of $d$, so that $G_{\uparrow\downarrow}^r$ is practically constant at the level of its asymptotic (i.e., single-interface) value. For $G_{\uparrow\downarrow}^t$, the oscillations do not vanish completely, but their amplitude is substantially reduced to values that are negligibly small compared to $\mbox{Re}G_{\uparrow\downarrow}^r$ for all but the thinnest magnetic layers. Diffusive scattering in the bulk of the magnetic layer, which has not been considered here explicitly, should have similar effects.

Both $G_{\uparrow\downarrow}^r$ and $G_{\uparrow\downarrow}^t$ are governed by (i) the matching of the normal-metal and ferromagnetic-metal states described by the scattering matrix of the isolated interface and (ii) the phases accumulated by the electrons on their quantum-coherent propagation through the magnetic layer. The interface determines the amplitudes of the oscillations and, the asymptotic value of $G_{\uparrow\downarrow}^r$, whereas the bulk term is responsible for the oscillation period. It is instructive to interpret the transmission and reflection coefficients of the finite-size magnetic layer in terms of multiple scattering between the two interfaces. The task is simplified by the simple Fermi surface of Cu, which corresponds to only one left- and right-going state at the Fermi energy for each value of $\mathbf{k}_\parallel$ and spin. The sums over states in Eqs.~(\ref{g}) and (\ref{t}) therefore reduce to integrations over the two-dimensional Brillouin zone (2DBZ) involving the complex-valued functions $r^\sigma(\mathbf{k}_\parallel)$ and $t^\sigma(\mathbf{k}_\parallel)$. To lowest order in the number of scattering processes and dropping explicit reference to $\mathbf{k}_\parallel$, 
\begin{align}
\label{ts}
t^\sigma&\approx t_{F\to N}^\sigma\Lambda^\sigma t_{N\to F}^\sigma\\
r^\sigma&\approx r_{N\to N}^\sigma+t_{F\to N}^\sigma\Lambda^\sigma r_{F\to F}^\sigma\Lambda^\sigma t_{N\to F}^\sigma\,,
\label{rs}
\end{align}
where $t_{N\to F}^\sigma=(t_{1}^\sigma,\ldots,t_n^\sigma)^T$ is a vector of transmission coefficients between a single propagating state in the normal metal and a set of states in the ferromagnet, $\Lambda^\sigma$ is a diagonal matrix of phase factors $e^{ik_{j\perp}^\sigma d}$ ($j$ is an index of the states in the ferromagnet), $r_{N\to N}^\sigma$ is a scalar reflection coefficient for states incoming from the normal metal and $r_{F\to F}^\sigma$ is a square matrix describing reflection at the ferromagnetic side. The set of states in the ferromagnet consists of both propagating and evanescent states. The contribution of the latter decreases exponentially with the thickness of the layer.

Let us first analyze the thickness dependence of $t^{\uparrow\downarrow}$. Substituting Eq.~(\ref{ts}), the summation in Eq.~(\ref{t}) is carried out over terms containing phase factors $e^{i(k_{i\perp}^\uparrow-k_{j\perp}^\downarrow)d}$. Because of the large differences between majority and minority Fermi surfaces of the ferromagnet, this typically leads to rapidly oscillating terms that to a large extent cancel in the 2DBZ integration over $\mathbf{k}_\parallel$. In the spirit of the theory of interlayer exchange coupling \cite{Bruno:prb95, Stiles:prb02}, long-range contributions must originate from the vicinity of points for which $\nabla_{\mathbf{k}_\parallel}(k_{i\perp}^\uparrow-k_{j\perp}^\downarrow)=0$, corresponding to a constant phase in the integrand of Eq.~(\ref{t}). The damped oscillations around zero are therefore caused by the stationary-phase Fermi-surface calipers.

By substituting Eq.~(\ref{rs}) into $g^{\uparrow\downarrow}$, Eq.~(\ref{g}), two thickness-independent contributions can be identified. The first, coming from integrating the $\delta_{nn^\prime}$ term in Eq.~(\ref{g}), is just the number of transport channels in the normal metal, i.e., the dimensionless Sharvin conductance. The second contribution comes from the $r_{N\to N}^\uparrow r_{N\to N}^{\downarrow\ast}$ term and provides an interface-specific correction to the first contribution. The thickness-dependent contributions contain, to lowest order, phase factors $e^{i(k_{i\perp}^\sigma+k_{j\perp}^\sigma)d}$ and $e^{-i(k_{i\perp}^\sigma+k_{j\perp}^\sigma)d}$. Just as in the case of $t^{\uparrow\downarrow}$, their 2DBZ integrals have an oscillatory character, with periods governed by different Fermi-surface calipers. These oscillations now occur around the constant value set by the first two contributions. The asymptotic value of $g^{\uparrow\downarrow}$ is clearly the reflection mixing conductance of a single F$\mid$N interface.

In metallic films, the electronic structure of all but the outermost atomic layers is practically identical to that of the bulk material. The period of oscillations of $g^{\uparrow\downarrow}$ and $t^{\uparrow\downarrow}$ as a function of the magnetic-layer thickness $d$ is thus a bulk property of the magnetic layer. The amplitudes, on the other hand, involve the interfacial scattering coefficients introduced in Eqs.~(\ref{ts}) and (\ref{rs}). When the quantum-size oscillations are small, $\mathcal{A}^{\uparrow\downarrow}\approx g^{\uparrow\downarrow}$, where $g^{\uparrow\downarrow}$ is a property of the N$\mid$F interface instead of the entire structure. Furthermore, $g^{\uparrow\downarrow}$ can be estimated by the Sharvin conductance of the normal metal, Eq.~(\ref{gg2}). The results of single-interface calculations are listed in Table~\ref{tabmix} for clean and disordered interfaces of Cu$\mid$Co and Au$\mid$Fe material combinations. The disorder in Table~\ref{tabmix} was modeled by 2 monolayers of 50\% alloy instead of a single monolayer in the present subsection. In spite of this difference, the values are practically identical to the asymptotic ones seen in Fig.~\ref{gccd} for the Cu$\mid$Co combination.

\subsubsection{Ultrathin normal spacer}
\label{uns}

Let us now turn to a discussion of the magnetization dynamics of two monodomain magnetic layers separated by an ultrathin normal-metal spacers with quantum-well states that penetrate into and couple the ferromagnets. As explained in Sec.~\ref{rkky}, the free energy $F$ of the system depends on the relative angle between the two magnetizations even in the absence of magnetic anisotropies. The dependence of $F(\mathbf{M}_i)$, $i=1,2$, on the magnetic configuration corresponds to nonlocal effective fields (\ref{Heff}) exerting torques on the magnetizations. In the following, we comment on the dynamic component of the exchange interaction in time-dependent problems, cf. the semiclassical dynamic exchange interaction discussed in section \ref{dec}.

Consider for simplicity an $s-d$ model with noninteracting $s$ electrons, where the magnetic $d$ orbitals are coupled to itinerant electrons via a mean-field exchange interaction. The transverse component of effective field (\ref{HM}) entering the LLG equation (\ref{llg}) for the $i$th magnetic moment is then given by
\begin{align}
\mathbf{H}_i(t)=&-\frac{1}{M_{si}V_i}\partial_{\mathbf{m}_i}\left\langle H(\mathbf{m}_i)\right\rangle_t\nonumber\\
=&-\frac{1}{M_{si}V_i}\partial_{\mathbf{m}_i}\sum_\kappa\varepsilon_\kappa(\mathbf{m}_i)n_\kappa(\mathbf{m}_i,t)\,,
\label{Hi}
\end{align}
where $H(\mathbf{m}_i)$ is the Hamiltonian for itinerant electrons, parametrized by the magnetization directions $\mathbf{m}_i$, and $\langle\rangle_t$ denotes the (quantum-mechanical) expectation value at time $t$. The sum on the second line of Eq.~(\ref{Hi}) runs over all eigenstates of $H(\mathbf{m}_i)$, where $\kappa$ labels both the orbital and spin degrees of freedom, $n_\kappa$ is the occupation number corresponding to the many-body state at time $t$ and $\varepsilon_\kappa$ is the energy of the $\kappa$th eigenstate. Setting the many-body ensemble at time $t$ to its equilibrium value determined by $\mathbf{m}_i(t)$ reproduces the Landau-Lifshitz (LL) definition (\ref{Heff}), leading to dissipationless trajectories [assuming there are no other sources of damping, i.e., $\alpha=0$ in Eq.~(\ref{llg})]. Such an approximation thus captures only the static exchange component.

In reality, $\langle\rangle_t$ lags behind its instantaneous equilibrium value. The corresponding nonequilibrium component of $n_\kappa(\mathbf{m}_i,t)$ in momentum space reflects the ``breathing Fermi surface" discussed by \onlinecite{Kunes:prb02} in the context of transition-metal bulk magnetization damping in the presence of crystal anisotropy. \onlinecite{Heinrich:jap03} conjectured that such mechanism may also play a role in the magnetic dynamics of F$\mid$N$\mid$F structures: The modulation of the exchange energy stored in the normal spacer may cause an additional damping through the time lag in the itinerant-electron response.

For a spin-rotationally--invariant system, the effective field (\ref{Hi}) reduces to
\begin{align}
\mathbf{H}_i^{(1)}(t)=-\frac{1}{M_{si}V_i}\sum_\kappa\varepsilon_\kappa\partial_{\mathbf{m}_i}n_\kappa(\mathbf{m}_i,t)\,.
\label{Hii}
\end{align}
The bulk damping coefficient (\ref{a}) determined by the transverse spin-relaxation time $T_2$ in a model discussed in Sec.~\ref{sdm}, where the average electron spin density lagged behind that corresponding to the instantaneous equilibrium, can be understood to arise from such an effective field. Spin-orbit interaction in the presence of a crystal field in bulk ferromagnets and/or the exchange coupling through the normal spacer modulate $\varepsilon_\kappa(\mathbf{m}_i)$ via the time-dependent $\mathbf{m}_i$, leading to an additional dynamic contribution to the effective field, as follows from Eq.~(\ref{Hi}). This contribution has a particularly simple form in the limit of short relaxation times $\tau$ \cite{Korenman:prb72,Kunes:prb02,Heinrich:jap03}:
\begin{equation}
\mathbf{H}_i^{(2)}(t)=-\frac{\tau}{M_{si}V_i}\sum_\kappa\delta(\varepsilon_\kappa-\varepsilon_F)\frac{\partial\varepsilon_\kappa}{\partial\mathbf{m}_i}\left(\frac{\partial\varepsilon_\kappa}{\partial\mathbf{m}_i}\cdot\frac{d\mathbf{m}_i}{dt}\right)\,,
\end{equation}
where $\varepsilon_F$ is the Fermi energy and assuming low temperature. It is clear that this effective field results in a (tensor) Gilbert damping.

In disordered structures and/or thick spacer layers with vanishing static exchange coupling between the magnetic films, Eq.~(\ref{Hi}) should in principle reproduce the dynamic exchange coupling discussed in Sec.~\ref{dec}. In other words, the spin pumping captures the semiclassical component of the time-dependent exchange coupling between ferromagnetic films when the static contribution vanishes.

\subsection{Spin-orbit coupling}
\label{mca}

The derivation of the spin-pumping current in Sec.~\ref{pump} relies on Eqs.~(\ref{s}) and (\ref{u}) which relate the scattering matrix in spin space, $\hat{s}_{nn^\prime,ll^\prime}$, for given channel and lead indices, $n$, $n^\prime$ and $l$, $l^\prime$, to the magnetization direction $\mathbf{m}$. For systems isotropic in spin space, the scattering matrix depends on $\mathbf{m}$ only through the simple projection (\ref{u}). In transition-metal ferromagnets, this is a good approximation, since their exchange splitting is by far the largest relevant energy scale. A large spin-orbit coupling in the electronic structure, such as in $p$-doped magnetic semiconductors like (Ga,Mn)As, on the other hand, can considerably modify the spin-pumping currents.

In ferromagnets with spin-orbit interaction, the rotating-frame analysis of Sec.~\ref{rfa} becomes tedious by the need to apply the transformation to the orbital as well as spin degrees of freedom. Crystal anisotropies or even the presence of planar interfaces would make such an approach impractical. The adiabatic-pumping formalism of Sec.~\ref{psp} still applies, however. One may in general calculate the tensor current (\ref{Ip}) in terms of the emissivity (\ref{em}), when the dependence of the full scattering matrix on the magnetization direction is known. Of course, the $2\times2$ matrices have to be generalized to $(2S+1)\times(2S+1)$ for spin-$S$ carriers. In the case of spin-orbit coupling in the nonmagnetic leads, however, such tensor currents are in general different from spin currents, since spin is not a good quantum number for transverse quantum channels. This complicates the matter beyond the scope of this article. In the simplest case of weak spin-orbit coupling within the ferromagnets, Eq.~(\ref{s}) holds approximately near a given $\mathbf{m}$ with weakly $\mathbf{m}$-dependent scattering coefficients $s^{\uparrow(\downarrow)}$ for spin-$1/2$ carriers. The adiabatic spin-pumping current is then given by Eq.~(\ref{Is}) in terms of a possibly $\mathbf{m}$-dependent $\mathcal{A}^{\uparrow\downarrow}$. 

For ferromagnets with a strong spin-orbit coupling, the most general form for the adiabatic spin-pumping current is
\begin{equation}
I_{s,a}=\sum_{aa^\prime}\mathcal{A}_{aa^\prime}(\mathbf{M})\frac{dM_{a^\prime}}{dt}\,,
\label{Isa}
\end{equation}
with constraints on the form of the $3\times3$ tensor $\mathcal{A}_{aa^\prime}(\mathbf{M})$, $a,a^\prime=1,2,3$, for a given crystal symmetry. In particular, $I_{s,a}$ may now have a component along the magnetization direction. It is always the case, for example, when the ferromagnetic exchange field varies in magnitude as well as in direction. By conservation of the total angular momentum, the spin-pumping current (\ref{Isa}) is accompanied by a torque on the magnetization, and also transferred into the orbital angular momentum as well as lattice torque. Note that in the presence of magnetic anisotropies, the LLG equation of motion is a tensor equation in which the scalar Gilbert parameter $\alpha$ is replaced by a $3\times3$ tensor \cite{Mills03}.

\subsection{Inhomogeneous magnetization dynamics}
\label{nmd}

We have so far restricted our attention to spatially-uniform magnetization dynamics. Recall, for example, Eq.~(\ref{afn}) for spin-pumping--enhanced Gilbert damping in F$\mid$N heterostructures. Several authors recently discussed possibilities to access also nonuniform spin-wave modes. \onlinecite{Polianski:prl04} showed that for sufficiently large perpendicular-current densities, a single thin-film ferromagnet sandwiched between diffuse normal-metal contacts becomes unstable with respect to transverse (to the layering direction) excitations of finite-wavelength spin waves when the source and drain contacts are asymmetric, and only for one direction (determined by the asymmetry) of the current bias. Their calculation is based on the magnetoelectronic circuit theory and adiabatic spin pumping. \onlinecite{Stiles:prb04} considered the magnetic instability in diffuse N$\mid$F$\mid$N structures in the limit of thicker F layers that can undergo longitudinal (along the layering direction) as well as transverse magnetization dynamics, allowing to relax the contact-asymmetry condition. \onlinecite{Ozyilmaz:prl04} reported an experimental study of the current-induced excitations in Cu$\mid$Co$\mid$Cu nanopillars, qualitatively confirming the theoretical ideas. \onlinecite{Ozyilmaz:prb05,Brataas:cm05} investigated current-driven spin-wave instabilities also in spin valves, in which they compete with the coherent macrospin motion. The detailed discussion of the current-induced instabilities is beyond the scope of the present review. We would nonetheless like to outline how the Gilbert damping (\ref{afn}) is affected by a weak transverse spin-wave excitation in a single magnetic film in contact with a diffuse metal.

Consider a nonuniform transverse excitation of the thin-film magnetization in the limit of small amplitudes, so that spin waves at different wavelengths do not couple. It is then sufficient to study excitations at a single wavelength with the magnetization direction deviating from its equilibrium value $\mathbf{m}_0$ as
\begin{equation}
\mathbf{m}(\mathbf{r})-\mathbf{m}_{0}=\delta\mathbf{m}\cos\left(\mathbf{q}_\perp\cdot\boldsymbol{\rho}+\omega t+\varphi\right)\,,
\end{equation}
where $\varphi$ is an arbitrary phase. The spatial-position vector $\mathbf{r}$ is decomposed here into a coordinate along the layering direction, $x$, and a transverse vector in the interface plane, $\boldsymbol{\rho}$. The wave vector $\mathbf{q}_\perp$ of transverse spin waves is parallel to the F$\mid$N interface and the amplitude $\delta\mathbf{m}$ does not depend on $x$. The derivation of the effective Gilbert damping for transverse spin waves \cite{Polianski:prl04} closely resembles that in section \ref{ds} for the macrospin dynamics. Let us consider a normal layer capping one side of the ferromagnet. The presence of two normal layers sandwiching a magnetic film thicker than its coherence length $\lambda_{\text{sc}}$ simply doubles the effect. For an unbiased structure, there is no charge current or voltage imbalance as long as the dynamics are slow on the characteristic spin-relaxation time scales, $\omega\ll\tau_{\text{sf}}^{-1}$, which is assumed in the following. In order to find the enhanced Gilbert damping, the diffusion equation for spins in the normal metal (\ref{de}) must be solved with boundary condition (\ref{bc}). In contrast to section \ref{ds}, the spin-pumping current now depends on the transverse coordinate $\boldsymbol{\rho}$. The solution of the diffusion equation for the spin accumulation in the normal-metal film with thickness $L$ as a function of distance $x$ from the F$\mid$N interface is [cf. Eq.~(\ref{sn})]
\begin{align}
\boldsymbol{\mu}_s(x,\boldsymbol{\rho})=&\cosh\left[(x-L)/\lambda_{\text{sd}}^{\text{(eff)}}\right]/\sinh\left[L/\lambda_{\text{sd}}^{\text{(eff)}}\right]\nonumber\\
&\times\frac{2\lambda_{\text{sd}}^{\text{(eff)}}}{\hbar\mathcal{N}SD}\cos\left( \mathbf{q}_\perp\cdot\boldsymbol{\rho}+\omega t+\varphi \right)\mathbf{I}_s\,,
\end{align}
where the effective spin-diffusion length for the transverse mode is defined by
\begin{equation}
\lambda_{\text{sd}}^{\text{(eff)}}=\frac{\lambda_{\text{sd}}}{\sqrt{1+\left(\lambda_{\text{sd}}\mathbf{q}_\perp\right)^2}}
\end{equation}
which reduces to the usual $\lambda_{\text{sd}}$, Eq.~(\ref{sd}), for uniform dynamics. By a calculation similar to that of section \ref{ds}, the wave-vector dependence of the enhanced Gilbert damping that generalizes Eq.~(\ref{afn}) reads \cite{Polianski:prl04}: 
\begin{equation}
G_{\text{eff}}(\mathbf{q}_\perp)-G=\left[1+\tilde{g}_r^{\uparrow\downarrow}\frac{R_{\text{sd}}\lambda_{\text{sd}}^{\text{(eff)}}/\lambda_{\text{sd}}}{\tanh\left(L/\lambda_{\text{sd}}^{\text{(eff)}}\right)}\right]^{-1}\frac{\hbar\gamma^2\tilde{g}_r^{\uparrow\downarrow}}{4\pi V}\,.
\end{equation}
Thick layers with $L\gg\lambda_{\text{sd}}^{\text{(eff)}}$ are thus the best spin sinks for a given material composition and spin-wave wave vector.

There is a crossover in the behavior of the enhanced Gilbert damping when the wavelength is comparable to the spin-diffusion length: In the long-wavelength limit, $\lambda_{\text{sw}}\gg2\pi\lambda_{\text{sd}}$ ($\lambda_{\text{sw}}=2\pi/q_\perp$), the result reduces to that of a uniformly-precessing ferromagnet; for short-wavelength excitations, $\lambda_{\text{sw}}\ll2\pi\lambda_{\text{sd}}$, the damping depends on the wavelength corresponding to the reduced effective spin-diffusion length $\lambda_{\text{sd}}^{\text{(eff)}}\approx\lambda_{\text{sw}}/(2\pi)$. In the latter regime, numerical estimates for transition-metal ferromagnets in contact with simple normal metals, in the spirit of Sec.~\ref{ds}, give
\begin{equation}
G_{\text{eff}}(\lambda_{\text{sw}})-G\sim\frac{\hbar\gamma^2\tilde{g}_r^{\uparrow\downarrow}/(4\pi V)}{1+\left[4\lambda/\lambda_{\text{sw}}\tanh(2\pi L/\lambda_{\text{sw}})\right]^{-1}}\,,
\label{Gsw}
\end{equation}
where $\lambda$ is the transport mean free path. We thus find that a normal metal is always a good spin sink in the limit $\lambda_{\text{sw}}\ll\lambda<L$, independently of the spin-relaxation rates, in stark contrast to the long-wavelength result (\ref{sp11}). This can be understood referring to the discussion of the dynamic exchange coupling in Sec.~\ref{wsc} and noticing that the pumping and backflow reabsorption of spins are separated in space by distances of the order or larger than $\lambda$, corresponding to regions of the magnetic layer with dynamics that are incoherent upon averaging over various diffuse paths when $\lambda_{\text{sw}}\ll\lambda$. The damping of each magnetic region is therefore affected by the spin pumping with vanishing spin backflow, rendering the normal metal a good spin sink. Consequently, the efficiency of the normal metal as a spin sink increases for shorter-wavelength spin-wave excitations and longer mean free paths. In particular, the Gilbert damping can be enhanced even when there is no enhancement of the Gilbert damping for long-wavelength excitations. This general conclusion can also be extended to magnetic films that are inhomogeneous or disordered on length scales shorter than the transport mean free path in nonmagnetic buffers. In addition, the normal metal may become an efficient spin sink for both short- and long-wavelength spin waves, when the frequency of these excitations becomes larger than the normal-metal scattering rate, a regime not explicitly treated in this review. (The interfacial spin-pumping current can still be evaluated by the adiabatic formalism, as long as the frequency remains much smaller than the ferromagnetic exchange energy.) Finally, we remark that quite generally, in the limit when the normal metal becomes effectively an ideal spin sink, the mixing conductance that determines the strength of the spin-pumping current should not be renormalized (in the sense of the Sec.~\ref{icm} discussion). In particular, it is $g^{\uparrow\downarrow}$ and not $\tilde{g}^{\uparrow\downarrow}$ that enters Eq.~(\ref{Gsw}) when the denominator on the right-hand side is close to unity. In the intermediate spin-sink regime, one has to extend the discussion of Sec.~\ref{icm} to laterally-inhomogeneous systems.

\subsection{Electron-electron interactions}
\label{eei}

The appropriate framework for describing metallic magnetism, including the 3$d$ transition metals, is band theory \cite{Kubler00}, treating the electron-electron interaction in a mean-field approximation. For qualitative purposes, this comes down to a simple Stoner or $s-d$ model Hamiltonian having parabolic free-electron dispersion for the conduction electrons with parametrized masses and exchange splittings. The local--spin-density approximation (LSDA) of density-functional theory is a very successful framework for treating itinerant electron systems from first principles. Electron-correlation effects are taken into account by the exchange-correlation potentials, be it in the local approximation. This approach has been taken by, e.g., \onlinecite{Zwierzycki:prb05} in calculating the scattering matrix and the conductance parameters.

\onlinecite{Simanek:prb031,Simanek:prb032} raised the question of possible enhancement of spin pumping in metallic F$\mid$N heterostructures by electron-electron interactions in the normal metals. A potential candidate would be Pd as a normal metal with an interaction-enhanced magnetic susceptibility. Pd is ``nearly" ferromagnetic, causing, e.g., giant moments around magnetic impurities. A ferromagnetic film in contact with such a material can thus induce magnetic moments renormalizing the exchange potential felt by electrons at the F$\mid$N contact. \onlinecite{Simanek:prb031,Simanek:prb032} considered a problem of an ultrathin ``$\delta$-function" magnetic layer embedded in a nonmagnetic material with a large Stoner enhancement of the spin susceptibility. Treating the ferromagnetic exchange field as a perturbation felt by normal-metal electrons, one can obtain a significant enhancement of its effective mean-field profile by electron-electron interactions. This in turn can considerably increase the spin-mixing conductances that govern the spin pumping \cite{Simanek:prb032}. Such perturbative analysis, however, overestimates the effect of electron correlations on spin-mixing conductances and thereby spin pumping in transition-metal heterostructures. It has been explained in Sec.~\ref{uml} that mixing conductance $g^{\uparrow\downarrow}$ (computed nonperturbatively by density-functional theory) of even the thinnest (and more so the thicker) magnetic films is close to the normal-metal Sharvin conductance, regardless of the possible Stoner enhancement \cite{Zwierzycki:prb05}. The mixing conductance that determines the strength of the spin pumping is thus not expected to be correlated with the normal-metal spin susceptibility, as already emphasized in Sec.~\ref{pss}.

In order to understand that a Stoner enhancement does not directly affect the spin-pumping enhancement, it is convenient to perform the rotating-frame analysis of Sec.~\ref{rfa}, that is also valid in the presence of electron-electron interactions, which are invariant to the rotating-frame transformation. The precessing ferromagnet changes the polarization in the normal metal along the axis of rotation by an amount which corresponds to the spin-imbalance potential $\mu_s=\hbar\omega$ that is determined by the precession frequency $\omega$, irrespective of the materials under consideration. In particular, the steady-state spin-accumulation $\mu_s$  due to the spin pumping does not depend on the magnetic susceptibility. Since it is $\mu_s$ that drives the nonequilibrium spin flow and not the spin density (that \textit{is} affected by the Stoner enhancement), the arguments given before in this article remain valid for interacting systems treated within the mean-field picture.

The electron-electron effects such as an enhanced normal-metal susceptibility are implicitly and nonperturbatively taken into account in self-consistent \textit{ab initio} calculations of heterostructures. First-principles calculations of the scattering matrix \cite{Zwierzycki:prb05}, fully include the electron-correlation effects which are difficult to capture by a perturbative formalism. Although many researchers in the field of solid-state magnetism are not familiar with scattering theory, it appears to be the most natural language to pursue the study of coupled transport and magnetization dynamics. 

Whereas electron-electron effects beyond the mean-field LSDA are probably small for transition metals, this does not mean that they can be neglected in other systems. For example, a suppression of spin pumping by the correlation effects in one-dimensional metals (``Luttinger liquids") has been predicted by \onlinecite{Bena:prb04}. These authors found that the spin pumping by a moving ferromagnet into a Luttinger liquid through a tunnel barrier is given by the same expression as for noninteracting electrons, viz., Eq.~(\ref{Is}), but with parameters $\mathcal{A}_r^{\uparrow\downarrow}$ and $\mathcal{A}_i^{\uparrow\downarrow}$ vanishing as a power of temperature at low temperatures with an exponent characteristic for the Luttinger-liquid zero-bias anomaly in tunneling density of states. Correlation effects might also become important if one wishes to describe spin pumping close to the ferromagnetic critical temperature, to understand significant deviations from the macrospin model for the magnetization, and to quantitatively model the spin-pumping parameter for interfaces with strongly-correlated materials.

\section{Summary and outlook}
\label{sum}

In this review, we presented a coherent picture of the nonlocal magnetization dynamics in heterostructures of ferromagnets and nonmagnetic conductors. It is based on the assumption of semiclassical transport in the bulk materials that is valid for diffuse and chaotic systems, as well as a separation of time scales of the electronic and magnetic degrees of freedom. Interfaces are treated as sharp quantum-mechanical boundary conditions for electron distribution functions and nonequilibrium transport. Except for the phenomenological treatment of spin-flip scattering processes, the theory is derived from first principles. The main subject in this context was the concept of spin pumping due to moving magnetization vectors. The magnetization dynamics affected by the spin-transfer torque in the presence of an electrical bias should be treated on equal footing with the spin pumping. The crucial material parameter is the spin-mixing conductance that can be computed from \textit{ab initio} electronic band structures. 

Several phenomena can be predicted or explained based on the basic formalism. One of them is the increased Gilbert damping of thin magnetic films in good electrical contact with normal metals that efficiently dissipate spin angular momentum. In more complex magnetic structures, we predict an interplay between the spin pumping and magnetization torques that is truly nonlocal, i.e., depends on the entire spin-coherent volume of the sample (determined by the spin-flip diffusion length). Novel collective effects appear when different magnetic elements in a spin-coherent circuit or device resonate at nearby frequencies. A moving ferromagnet can be used as a source that pumps spin currents into normal metals or semiconductors, that leads to a spin accumulation determined by the ferromagnetic-resonance frequency.

Although some basic principles are rather subtle, the final formalism is easy to use. It can often be mapped on an equivalent circuit model that is governed by a few parameters that can be either fitted to experiments or computed from first principles. Such calculations can be used, e.g., to estimate and optimize critical magnetization-switching currents \cite{Manschot:apl04}. We focused attention on quasi-one-dimensional configurations and the macrospin model for the magnetization. Generalization of the formalism to include spin-wave excitations in the ferromagnet or inhomogeneous spin currents have also been illustrated (Sec.~\ref{nmd}). Integration of micromagnetic simulations with the transport equations based on boundary conditions at the interfaces as described here might be necessary to improve the accuracy of first-principles modeling.

In the review, we had mainly metallic heterostructures with transition-metal ferromagnets in mind. But since the approach is quite general, we should by no means exclude other materials. We already speculated in Sec.~\ref{sdm} that the formalism can be used to understand the Gilbert damping in magnetic semiconductors. There is little doubt that a modeling of the current-induced switching of (Ga,Mn)As observed by \onlinecite{Ohno:prl04} requires the concept of magnetization torque. The spin pumping into carbon nanotubes as investigated by \onlinecite{Bena:prb04} is partly suppressed by correlation effects. The spin pumping by ferromagnetic superconductors \cite{Brataas:prl04} is entangled with Cooper-pair pumping and depends on the spin-pairing symmetry of the superconducting state. Spin dynamics in heterostructures of high-density magnets in contact with doped semiconductors \cite{Bauer:prl04,Bauer:cm05} are another promising playground for the formalism described here. Besides, it is possible to generalize the spin-pumping picture to other symmetry-broken heterostructures with adiabatically-varying order parameters \cite{Tserkovnyak:prb05}.

A critical parameter in the nonlocal magnetization dynamics is the spin-flip diffusion length in the normal-metal components, that can be of the order of microns even at room temperature, see, e.g., \onlinecite{Jedema:nat02}. The nonlocal effects in structures based on transition metals could therefore be robust and observable at ambient temperatures. We therefore believe that the formalism discussed here should be useful to understand, compute, and design magnetic-device operations. An example is the possibility of ``$\alpha$ engineering" based on the increase of the Gilbert damping in ultrathin magnetic films, e.g., by an order of magnitude by just evaporating a few monolayers of Pt on top of it. We also expect that with decreasing size of magnetic circuits and devices, the dynamic coupling discussed in this review will become even more relevant. It might lead to effects like cross talk between magnetic elements and excess noise. On the other hand, proper engineering of phenomena like the nonlocal dynamic locking might lead to an increased stability against external perturbations as well.

Theoretical challenges for the future include a proper treatment of spin-orbit interactions, the coupling of magnetic degrees of freedom to the lattice, and effects beyond the semiclassical regime.

\acknowledgments

We are grateful to G. A. Fiete, B.~Heinrich, P. J. Kelly, and M.~Zwierzycki for contributing to this review and to J.~Foros and M. D. Stiles for valuable discussions. This work was supported in part by the Harvard Society of Fellows, the Norwegian Research Council Grant No. 162742/V00, Nanomat Grants Nos. 158518/143 and 158547/431, Dutch FOM Foundation, DARPA Award No. MDA972-01-1-0024, and NSF Grants Nos. DMR 02-33773 and PHY 01-17795.

\end{document}